\documentclass{article}
\author{Daniel Rosin} \date{\today} \title{A fast algorithm for the Frobenius problem in three variables}
\newcommand{\version}{final}
\usepackage{ifthen}
\usepackage{etoolbox}
\usepackage[utf8]{inputenc}
\usepackage{mathtools}
\usepackage{amssymb}
\usepackage{amsthm}
\usepackage{thmtools}
\usepackage{mdframed}
\usepackage[parfill]{parskip}
\usepackage[nottoc,numbib]{tocbibind}
\usepackage{url}
\usepackage{appendix}
\usepackage[ singlelinecheck=false ]{caption}
\usepackage{float}
\usepackage{titlesec}
\titlespacing*{\section}{0pt}{3ex}{3ex}
\titlespacing*{\subsection}{0pt}{3ex}{3ex}
\usepackage{letltxmacro}
\usepackage{hyperref}
\hypersetup{colorlinks=true, 
linkcolor=blue}
\usepackage[nameinlink]{cleveref}
\usepackage[\version]{crossreftools}
\usepackage[left,\version]{showlabels}
\usepackage{makecell}

\declaretheorem[name=Theorem,
refname={theorem,theorems},
Refname={Theorem,Theorems}]{theo}

\declaretheorem[name=Lemma,
refname={lemma,lemmas},
Refname={Lemma,Lemmas}]
{lemm}
\declaretheorem[name=Definition,
refname={definition,definitions},
Refname={Definition,Definitions}]
{defi}
\declaretheorem[name=Assumption,
refname={assumption,assumptions},
Refname={Assumption,Assumptions}]
{assu}
\declaretheorem[name=Example,
refname={example,examples},
Refname={Example.,Examples.}]
{exam}
\newenvironment{frameit}
{\begin{mdframed}[innertopmargin=10pt,innerbottommargin=10pt, everyline=true]
}{\end{mdframed}}
\newenvironment{proveit}[1]
{\begin{proof}[\textbf{Proof of \cref{#1}:}\nopunct]}
{\end{proof}}
\newcommand{\am}{a\textsuperscript{-1}}
\newcommand{\azerom}{\azero\textsuperscript{-1}}
\newcommand{\amtwo}{a\textsubscript{2}\textsuperscript{-1}}
\newcommand{\azero}{a\textsubscript{0}}
\newcommand{\aone}{a\textsubscript{1}}
\newcommand{\atwo}{a\textsubscript{2}}
\newcommand{\athree}{a\textsubscript{3}}

\newcommand{\ac}
{a\textsubscript{c}}

\newcommand{\attwo}{\tilde{a}\textsubscript{2}}
\newcommand{\atthree}{\tilde{a}\textsubscript{3}}

\newcommand{\flooraoneazero}{\floorfrac{\aone}{\azero}}
\newcommand{\flooraoneoa}{\floorfrac{\aone}{\oa}}

\newcommand{\hs}{h\textsuperscript{s}}
\newcommand{\hone}{h\textsubscript{1}}
\newcommand{\hj}{h\textsubscript{j}}

\newcommand{\hop}{h\textsubscript{\(\psi\)}}
\newcommand{\hopp}{h\textsubscript{\(\psi + 1\)}}
\newcommand{\hops}{h\textsubscript{\(\psi\)}\textsuperscript{s}}
\newcommand{\hopps}{h\textsubscript{\(\psi + 1\)}\textsuperscript{s}}
\newcommand{\hm}{h\textsubscript{-1}}
\newcommand{\hms}{h\textsubscript{-1}\textsuperscript{s}}

\newcommand{\hjs}{h\textsubscript{j}\textsuperscript{s}}

\newcommand{\itilde}{\tilde{\text{\i}}}

\newcommand{\ilone}{\check{\text{\i}}\textsubscript{1}}
\newcommand{\iltwo}{\check{\text{\i}}\textsubscript{2}}
\newcommand{\ilk}{\check{\text{\i}}\textsubscript{k}}
\newcommand{\ilkp}{\check{\text{\i}}\textsubscript{k+1}}

\newcommand{\ilibjp}
{\check{\text{\i}}
\textsubscript{\((\ibj+1\))}}
\newcommand{\ilibjpp}
{\check{\text{\i}}
\textsubscript{\((\ibjp+1\))}}
\newcommand{\ilitp}
{\check{\text{\i}}
\textsubscript{\((
\itilde + 1)\)}}
\newcommand{\ilvjpp}
{\check{\text{\i}}
\textsubscript{\(\vjp + 1\)}}
\newcommand{\ihone}{\hat{\text{\i}}\textsubscript{1}}
\newcommand{\ihk}{\hat{\text{\i}}\textsubscript{k}}
\newcommand{\ihkp}{\hat{\text{\i}}\textsubscript{k+1}}

\newcommand{\ihibj}
{\hat{\text{\i}}
\textsubscript{\(\ibj\)}}
\newcommand{\ihibjp}
{\hat{\text{\i}}
\textsubscript{\(\ibjp\)}}
\newcommand{\ihit}
{\hat{\text{\i}}
\textsubscript{\(
\itilde\)}}
\newcommand{\ihvjp}
{\hat{\text{\i}}
\textsubscript{\(\vjp\)}}
\newcommand{\ib}{\text{\(\bar{\text{\i}}\)}}
\newcommand{\ibone}{\text{\(\bar{\text{\i}\textsubscript{1}}\)}}

\newcommand{\ibj}{\text{\(\bar{\text{\i}\textsubscript{j}}\)}}
\newcommand{\ibjp}{\text{\(\bar{\text{\i}}\textsubscript{j+1}\)}}
\newcommand{\ibjm}{\text{\(\bar{\text{\i}}\textsubscript{j-1}\)}}
\newcommand{\ibos}{\bar{\text{\i}}\textsubscript{\(\os\)}}
\newcommand{\ibosm}{\bar{\text{\i}}\textsubscript{\(\os-1\)}}

\newcommand{\khl}{\hat{k}\textsubscript{l}}
\newcommand{\kll}{\check{k}\textsubscript{l}}
\newcommand{\li}{l\textsubscript{i}}
\newcommand{\lone}{l\textsubscript{1}}

\newcommand{\ldip}{\text{\(\dot{l}\textsubscript{i+1}\)}}

\newcommand{\Lone}{\text{L\textsubscript{1}}}
\newcommand{\Lj}{\text{L\textsubscript{j}}}
\newcommand{\Ljp}{\text{L\textsubscript{j + 1}}}
\newcommand{\Ljm}{\text{L\textsubscript{j - 1}}}
\newcommand{\Los}{L\textsubscript{\(\os\)}}
\newcommand{\Losm}{L\textsubscript{\(\os-1\)}}

\newcommand{\mk}{m\textsubscript{k}}
\newcommand{\mzero}{m\textsubscript{0}}
\newcommand{\nb}{\text{\(\bar{n}\)}}
\renewcommand{\ni}{n\textsubscript{i}}
\newcommand{\nzero}{n\textsubscript{0}}
\newcommand{\none}{n\textsubscript{1}}
\newcommand{\ntwo}{n\textsubscript{2}}
\newcommand{\nim}{n\textsubscript{i-1}}

\newcommand{\nm}{n\textsubscript{\(m\)}}

\newcommand{\nk}{n\textsubscript{k}}
\newcommand{\nkm}{n\textsubscript{k -1}}
\newcommand{\nnb}{n\textsubscript{\nb}}
\newcommand{\nt}{\tilde{n}}
\renewcommand{\nu}{\ddot{n}}
\newcommand{\nlone}{\check{n}\textsubscript{1}}
\newcommand{\nltwo}{\check{n}\textsubscript{2}}
\newcommand{\nli}{\check{n}\textsubscript{i}}
\newcommand{\nlip}{\check{n}\textsubscript{i+1}}
\newcommand{\nlib}{\check{n}\textsubscript{\ib}}
\newcommand{\nlibp}{\check{n}\textsubscript{\ib+1}}
\newcommand{\nlibip}{\check{n}\textsubscript{\ib+i+1}}
\newcommand{\nlj}{\check{n}\textsubscript{j}}
\newcommand{\nljp}{\check{n}\textsubscript{j+1}}
\newcommand{\nhone}{\hat{n}\textsubscript{1}}
\newcommand{\nhtwo}{\hat{n}\textsubscript{2}}
\newcommand{\nhi}{\hat{n}\textsubscript{i}}
\newcommand{\nhip}{\hat{n}\textsubscript{i+1}}
\newcommand{\nhib}{\hat{n}\textsubscript{\ib}}
\newcommand{\nhibi}{\hat{n}\textsubscript{\ib+i}}

\newcommand{\nhibm}{\hat{n}\textsubscript{\ib-1}}
\newcommand{\nhj}{\hat{n}\textsubscript{j}}
\newcommand{\nhjp}{\hat{n}\textsubscript{j+1}}
\newcommand{\nhoa}{\hat{n}\textsubscript{\(\oa\)}}

\newcommand{\niseq}
{\text{\((\ni)_{i=0}^m\)}}
\newcommand{\ninbseq}
{\text{\((\ni)_{i=0}^\nb\)}}
\LetLtxMacro{\plusminus}{\pm}
\renewcommand{\pm}{p\textsuperscript{-1}}

\newcommand{\um}{u\textsubscript{-1}}

\newcommand{\uj}{u\textsubscript{j}}
\newcommand{\ujp}{u\textsubscript{j+1}}
\newcommand{\ujm}{u\textsubscript{j - 1}}

\newcommand{\uone}{u\textsubscript{1}}

\newcommand{\us}{u\textsuperscript{s}}
\newcommand{\ujs}{u\textsubscript{j}\textsuperscript{s}}

\newcommand{\uos}{u\textsubscript{\(\os\)}}
\newcommand{\uosm}{u\textsubscript{\(\os-1\)}}
\newcommand{\useq}{(u(x))\textsubscript{x\inZ}}

\newcommand{\vone}{v\textsubscript{1}}

\newcommand{\vj}{v\textsubscript{j}}
\newcommand{\vjp}{v\textsubscript{j+1}}
\newcommand{\vjpp}{v\textsubscript{j+2}}
\newcommand{\vjm}{v\textsubscript{j-1}}

\newcommand{\vt}{v\textsubscript{\(\tau\)}}
\newcommand{\vtp}{v\textsubscript{\(\tau+1\)}}
\newcommand{\vtm}{v\textsubscript{\(\tau-1\)}}

\newcommand{\vos}{v\textsubscript{\(\os\)}}
\newcommand{\vosm}{v\textsubscript{\(\os-1\)}}

\newcommand{\vbone}{\bar{v}\textsubscript{1}}
\newcommand{\vbj}{\text{\(\bar{v}\textsubscript{j}\)}}
\newcommand{\vbjp}{\text{\(\bar{v}\textsubscript{j+1}\)}}
\newcommand{\vbjpp}{\text{\(\bar{v}\textsubscript{j+2}\)}}
\newcommand{\vbjm}{\bar{v}\textsubscript{j-1}}
\newcommand{\vbt}{\text{\(\bar{v}\textsubscript{\(\tau\)}\)}}
\newcommand{\vbtp}{\text{\(\bar{v}\textsubscript{\(\tau+1\)}\)}}
\newcommand{\vbtm}{\text{\(\bar{v}\textsubscript{\(\tau-1\)}\)}}

\newcommand{\vbos}{\bar{v}\textsubscript{\(\os\)}}
\newcommand{\vbosm}{\bar{v}\textsubscript{\(\os-1\)}}
\newcommand{\vones}{v\textsubscript{1}\textsuperscript{s}}
\newcommand{\vjs}{v\textsubscript{j}\textsuperscript{s}}

\newcommand{\vbones}{\bar{v}\textsubscript{1}\textsuperscript{s}}
\newcommand{\vbjs}{\bar{v}\textsubscript{j}\textsuperscript{s}}

\newcommand{\xb}{\bar{x}}
\newcommand{\xone}{x\textsubscript{1}}
\newcommand{\xtwo}{x\textsubscript{2}}
\newcommand{\xthree}{x\textsubscript{3}}
\renewcommand{\xi}{x\textsubscript{i}}

\newcommand{\xt}{\tilde{x}}

\newcommand{\xlone}{\text{\(\check{x}\textsubscript{1}\)}}

\newcommand{\xli}{\text{\(\check{x}\textsubscript{i}\)}}
\newcommand{\xlip}{\text{\(\check{x}\textsubscript{i+1}\)}}
\newcommand{\xlj}{\text{\(\check{x}\textsubscript{j}\)}}
\newcommand{\xhone}{\text{\(\hat{x}\textsubscript{1}\)}}
\newcommand{\xhi}{\text{\(\hat{x}\textsubscript{i}\)}}
\newcommand{\xhip}{\text{\(\hat{x}\textsubscript{i+1}\)}}
\newcommand{\xhj}{\text{\(\hat{x}\textsubscript{j}\)}}
\newcommand{\xhp}{\text{\(\hat{x}\textsubscript{\(p\)}\)}}
\newcommand{\xhqmp}{\text{\(\hat{x}\textsubscript{\(q - p\)}\)}}
\newcommand{\oa}{\alpha}
\newcommand{\oaone}{\alpha\textsubscript{1}}

\newcommand{\oaj}{\alpha\textsubscript{j}}
\newcommand{\oajp}{\alpha\textsubscript{j+1}}

\newcommand{\oaos}{\alpha\textsubscript{\(\os\)}}
\newcommand{\oaosm}{\alpha\textsubscript{\(\os-1\)}}

\newcommand{\oab}{\bar{\alpha}}
\newcommand{\oabone}{\bar{\alpha}\textsubscript{1}}
\newcommand{\oabj}{\bar{\alpha}\textsubscript{j}}

\newcommand{\oabos}{\bar{\alpha}\textsubscript{\(\os\)}}
\newcommand{\oabosm}{\bar{\alpha}\textsubscript{\(\os-1\)}}

\newcommand{\ob}{\beta}
\newcommand{\oG}{\Gamma}
\newcommand{\oD}{\Delta}
\newcommand{\oDone}{\Delta\textsubscript{1}}

\newcommand{\oDi}
{\Delta\textsubscript{i}}
\newcommand{\oDk}
{\Delta\textsubscript{k}}

\newcommand{\oDt}{\tilde{\Delta}}
\newcommand{\oDu}{\ddot{\Delta}}
\newcommand{\oDdi}
{\dot{\Delta}\textsubscript{i}}
\newcommand{\oDdk}
{\dot{\Delta}\textsubscript{k}}
\newcommand{\oDdone}
{\dot{\Delta}\textsubscript{1}}
\newcommand{\oDbi}
{\bar{\Delta}\textsubscript{i}}
\newcommand{\oDbk}
{\bar{\Delta}\textsubscript{k}}
\newcommand{\od}{\delta}

\newcommand{\odltwo}{\check{\delta}\textsubscript{2}}

\newcommand{\odi}{\delta\textsubscript{i}}
\newcommand{\odk}{\delta\textsubscript{k}}

\newcommand{\odij}{\delta\textsubscript{i,j}}

\newcommand{\odijp}{\delta\textsubscript{i,j + 1}}

\newcommand{\odkj}{\delta\textsubscript{k,j}}
\newcommand{\odkjp}{\delta\textsubscript{k,j + 1}}

\newcommand{\oddi}{\dot{\delta}
\textsubscript{i}}
\newcommand{\oddione}{\dot{\delta}
\textsubscript{i,1}}
\newcommand{\oddij}{\dot{\delta}
\textsubscript{i,j}}
\newcommand{\oddijp}{\dot{\delta}
\textsubscript{i,j+1}}
\newcommand{\oddkj}{\dot{\delta}
\textsubscript{k,j}}
\newcommand{\oddkjp}{\dot{\delta}
\textsubscript{k,j+1}}
\newcommand{\oddiop}{\dot{\delta}
\textsubscript{\(i,\psi\)}}
\newcommand{\odbi}{\bar{\delta}
\textsubscript{i}}
\newcommand{\odbione}{\bar{\delta}
\textsubscript{i,1}}
\newcommand{\odbij}{\bar{\delta}
\textsubscript{i,j}}
\newcommand{\odbijp}{\bar{\delta}
\textsubscript{i,j+1}}
\newcommand{\odbkj}{\bar{\delta}
\textsubscript{k,j}}
\newcommand{\odbkjp}{\bar{\delta}
\textsubscript{k,j + 1}}
\newcommand{\odbiop}{\bar{\delta}
\textsubscript{\(i,\psi\)}}

\newcommand{\odloddijpp}{\check{\delta}
\textsubscript{\(\dot{\delta}\textsubscript{i,j+1}+1\)}}
\newcommand{\odloddkjpp}{\check{\delta}
\textsubscript{\(\dot{\delta}\textsubscript{k,j+1}+1\)}}
\newcommand{\odlodbijp}{\check{\delta}
\textsubscript{\(\bar{\delta}\textsubscript{i,j+1}\)}}
\newcommand{\odlodbijpp}{\check{\delta}
\textsubscript{\(\bar{\delta}\textsubscript{i,j+1}+1\)}}
\newcommand{\odlodbkjpp}{\check{\delta}
\textsubscript{\(\bar{\delta}\textsubscript{k,j+1}+1\)}}
\newcommand{\odhone}{\hat{\delta}
\textsubscript{1}}

\newcommand{\odhoddijp}{\hat{\delta}
\textsubscript{\(\dot{\delta}\textsubscript{i,j+1}\)}}
\newcommand{\odhodbijp}{\hat{\delta}
\textsubscript{\(\bar{\delta}\textsubscript{i,j+1}\)}}
\newcommand{\odhodbkjp}{\hat{\delta}
\textsubscript{\(\bar{\delta}\textsubscript{k,j+1}\)}}

\newcommand{\oed}{\dot{\epsilon}}
\newcommand{\oeb}{\bar{\epsilon}}
\newcommand{\oo}{\theta}
\newcommand{\ooone}{\theta\textsubscript{1}}
\newcommand{\ooj}{\theta\textsubscript{j}}
\newcommand{\oojp}{\theta\textsubscript{j+1}}

\newcommand{\ooosm}{\theta\textsubscript{\(\os-1\)}}

\newcommand{\om}{\mu}
\newcommand{\omone}{\om\textsubscript{1}}
\newcommand{\omtwo}{\om\textsubscript{2}}
\newcommand{\omi}{\om\textsubscript{i}}
\newcommand{\omip}{\om\textsubscript{i + 1}}
\newcommand{\omipp}{\om\textsubscript{i + 2}}
\newcommand{\omippp}{\om\textsubscript{i + 3}}
\newcommand{\omipppp}{\om\textsubscript{i + 4}}
\newcommand{\omk}{\om\textsubscript{k}}
\newcommand{\omkm}{\om\textsubscript{k-1}}
\newcommand{\omiseq}
{(\omi)\textsubscript{i\inN}}
\newcommand{\os}{\sigma}
\newcommand{\ot}{\text{\(\tau\)}}
\newcommand{\op}{\psi}
\newcommand{\ovj}{\varphi\textsubscript{j}}
\newcommand{\ovjp}{\varphi\textsubscript{j+1}}
\newcommand{\ovone}{\varphi\textsubscript{1}}
\newcommand{\ovbj}{\bar{\varphi}\textsubscript{j}}
\newcommand{\ovbone}{\bar{\varphi}\textsubscript{1}}
\newcommand{\ovbop}{\bar{\varphi}\textsubscript{\(\psi\)}}
\newcommand{\ovbopp}{\bar{\varphi}\textsubscript{\(\psi + 1\)}}
\newcommand{\ovop}{\varphi\textsubscript{\(\psi\)}}
\newcommand{\ovopp}{\varphi\textsubscript{\(\psi + 1\)}}
\newcommand{\ovot}{\varphi\textsubscript{\(\tau\)}}
\newcommand{\intref}[1]
{\text{\{\Cref{#1}\}}}

\newcommand{\inZ}{\text{\(\in\mathbb{Z}\)}}
\newcommand{\inN}{\text{\(
\in\mathbb{N}\)}}

\newcommand{\abs}[1]{\mid #1\mid}

\newcommand{\floorfrac}[2]{\biggl\lfloor\frac{#1}{#2}\biggl\rfloor}
\newcommand{\ceilfrac}[2]{\biggl\lceil\frac{#1}{#2}\biggl\rceil}
\renewcommand{\mod}[2]{mod(#1,\;#2)}
\LetLtxMacro{\oldsum}{\sum}
\renewcommand{\sum}[2]{\oldsum\limits_{#1}^{#2}}

\LetLtxMacro{\oldbullet}{\bullet}
\renewcommand{\bullet}{\text{\(\oldbullet\)\;}}

\newcommand{\inputdraft}[1]
{\expandafter\ifstrequal\expandafter{\version}{draft}{\input{#1}}{}}

\begin{document}
\maketitle
\begin{abstract}
\noindent Given the function \(\oG (X)=\aone\xone+\atwo\xtwo+\athree\xthree\), where the set \(\{\aone,\atwo,\athree\}\), denoted \(A\), consists of positive integers and the set \(\{\xone,\xtwo,\xthree\}\), denoted \(X\), consists of non-negative integers, the Frobenius problem in three variables is to find the greatest integer, which is not in the codomain of \(\oG\). The fastest known algorithm for solving the three variable case of the Frobenius problem was invented by H. Greenberg in 1988 whose worst case time complexity is a logarithmic function of \(A\). In 2017 A. Tripathi presented another algorithm for solving the same problem. This article presents an algorithm whose foundation is the same as Tripathi's. However, this algorithm is significantly different from Tripathi's and we show that its worst case time complexity is also a logarithmic function of \(A\).
\end{abstract}
\tableofcontents
\section{Introduction}\label{IN}
Imagine a monetary system with coins of denominations three and five. What is the greatest integer amount which cannot be supplied exactly? The answer is seven. We can easily realize this by noting that seven can not be supplied but eight, nine and ten can since:
\begin{align*}
5+3 = & 8\\
3*3 = & 9\\
5*2 = & 10
\end{align*}
Any amount greater than ten can be supplied by adding a number of three-unit coins to one of the amounts above. This is one example of the so-called Frobenius problem. 
Given a set of integers \(A=\{\aone,\atwo,...\;\ac\}\) the Frobenius problem is to find the greatest integer \(g(A)\) which cannot be represented as a non-negative linear combination of \(A\). In the money supply problem the different denominations correspond to the elements in \(A\) and \(g(A)\) is the greatest amount which can not be supplied. J. Sylvester, who was the first mathematician to study the Frobenius problem, discovered a closed form formula in 1882 solving the problem when the set \(A\) has two elements. (This problem was later named after F. Frobenius probably because he popularized the problem through his lectures.)
\begin{frameit}
\begin{theo}\label{IN1}
Sylvester's theorem: Given \(A = \{\aone, \atwo\}\)
\[
g(A) = \aone\atwo-\aone-\atwo
\]
\end{theo}
\end{frameit}
In the general formulation of the Frobenius problem the number of variables is arbitrary. However, in this article we are presenting an algorithm solving the Frobenius problem in three variables. Below follows a formal definition of this problem. Note that \(\aone\) is defined to be greater than one. However, this is done without any major loss of generality since \(g(A)\) is always minus one if \(\aone\) equals one.
\begin{frameit}
\textbf{\Cref{IN2}: }

Given the set of integers \(A=\{\aone,\atwo,\athree\}\) where \(1 < \aone < \atwo < \athree\) the Frobenius problem in three variables is to find the greatest integer \(g(A)\) which is not in the codomain of \(\oG\) defined by:
\[\oG(X)=\aone\xone + \atwo\xtwo + \athree\xthree \text{ where }\xi\text{ are non-negative integers for all } i\]
\end{frameit}
The value of \(\oG\) is a multiple of \(gcd(\aone,\atwo, \athree)\) which implies that there is no greatest integer outside the codomain of \(\oG\) in the case \(gcd(\aone, \atwo, \athree )>1\), i.e. the Frobenius number does not exist in this case. It is also well-known that the Frobenius number exists if \(gcd(\aone, \atwo, \athree)=1\), i.e. the Frobenius number exists iff this is the case. A classic specialization of the Frobenius problem in three variables is the so-called Chicken McNuggets problem. Given that a McDonald's restaurant sells Chicken McNuggets in boxes of 6, 9 and 20 pieces, the problem is to find the largest amount of nuggets which cannot be served. The answer is 43.

A well-known theorem, discovered by Johnson \cite{Johnson1960}, allows any Frobenius number in three variables to be computed from a Frobenius number of a set with three pairwise coprimes. This implies that, for the three variable case, one can assume that all integers of the set \(A\) are pairwise coprimes without loss of generality. This is also assumed throughout this article. This assumption also guarantees that \(g(A)\) exists.  Below follows Johnson's theorem for the three variable case. (It can also be formulated for the general case.)

\begin{frameit}
\begin{theo}\label{IN3}
Johnson's theorem
\begin{gather*}
\atwo=gcd(\atwo,\athree)\attwo,\;\athree=gcd(\atwo,\athree)\atthree\\
\implies\\
g(\aone,\atwo,\athree)=gcd(\atwo,\athree)g(\aone,\attwo,\atthree)+(gcd(\atwo\athree)-1)\aone
\end{gather*}
\end{theo}
\end{frameit}
A closed form formula for the Frobenius problem is only known for the two variable case and Curtis\cite{Curtis1990} showed that the Frobenius number for more variables cannot be expressed by means of a finite set of polynomials. In addition to this, Ramírez-Alfonsín\cite{Alfonsin1996} showed that the problem in general is NP-hard. However, several algorithms exist solving the Frobenius problems when the number of variables are greater than two. These can be divided into two categories: three variable specific and general. General ones can be used to find \(g(A)\) for any number of variables. E.S. Selmer and  Ö. Beyer invented a three variable specific algorithm which is based on continued fractions. It was later improved by Ø. Rødseth whose result was even further elaborated by J.L. Davison \cite{Davison1994}. The worst case time complexity of this algorithm is O(\aone) and it was the fastest three variable specific algorithm for about 10 years until H. Greenberg \cite{Greenberg1988}
invented one whose worst case time complexity is \(O(\log \aone)\). There are several algorithms for the general case. E.g. A. Nijenhuis\cite{Beihoffer2005} developed an algorithm in which the Frobenius problem is transformed into a shortest path problem of a directed weighted graph and solved by Dijkstra's algorithm. This and other general algorithms can be used to solve the three variable case but none of them are as fast as Greenberg's, i.e. Greenberg's is the fastest algorithm known today for this case.

Several of the algorithms finding the Frobenius number, including the one presented here, are based on a theorem by Brauer and Shockley \cite{ShockleyBrauer1962}.
\begin{frameit}
\begin{theo}\label{IN14}
Brauer and Shockley's theorem

Let \([k]\) be the residue class modulo \(\aone\) containing \(k\) and \(\mk\) be the smallest integer in the codomain of \(\oG\) belonging to \([k]\).
\[g(A) = \underset{0 < k <\aone}{max}\;\mk - \aone\]
\end{theo}
\end{frameit}
Brauer and Shockley's theorem follows from the fact that \(\mk\) minus \(\aone\) is the greatest integer in \([k]\), which is not in the codomain of \(\oG\), since all integers in \([k]\) greater than \(\mk\) are in the codomain of \(\oG\). \(g(A)\) then equals the greatest \(\mk\) minus \(\aone\). However, \(\mzero\) can be disregarded since it always equals zero and therefore never will be the greatest \(\mk\), since the assumption that \(\aone\) is greater than one implies that \(g(A)\) is always greater than one.

A. Tripathi \cite{Tripathi2017} published in 2017 a three variable specific algorithm based on Brauer and Shockley's theorem. The time complexity of his algorithm is, to the best of the author's knowledge, unknown. However, Tripathi's algorithm and the algorithm presented in this article are founded on the same theorem presented below. Note that different notation is used in Tripathi's formulation of this theorem and that mod does not denote the modulo operator in this article. Instead, mod is a function with two integer variables, \(a\) and \(b\) (\(b \neq 0\)), whose value is \(a\) reduced modulo \(b\). Deriving the algorithm presented here, we use a few well-known lemmas of the modulo function. Below we list these as well.
\begin{frameit}
Given \(a,b,c\inZ\) and \(b\neq 0\)
\begin{lemm}\label{IN8}
\(0\leq a< b\iff \text{\mod{a}{b} = a}\)
\end{lemm}
\begin{lemm}\label{IN9}
\begin{align*}
\mod{a+c}{b} = & \mod{\mod{a}{b}+\mod{c}{b}}{b}\\
= & \mod{a+\mod{c}{b}}{b}
\end{align*}
\end{lemm}
\begin{lemm}\label{IN10}
\(gcd(a,b) = gcd(\mod{a}{b},b)\)
\end{lemm}
\begin{lemm}\label{IN11}
Given \(b > 1\)
\[
gcd(a,b)=1\implies \mod{a}{b} > 0
\]
\end{lemm}
\begin{lemm}\label{INT10}
\[
\mod{a}{b} = 
a - \floorfrac{a}{\abs{b}}\abs{b}
\]
\end{lemm}
\textbf{\Cref{IN4}:}
\begin{itemize}
\item
\amtwo\ is the multiplicative inverse of \atwo\ modulo \aone\ where \(0 < \amtwo <\aone\).
\item
\(\azero = \mod{-\amtwo\athree}{\aone}\)
\item
\(F(n,r) = \atwo\mod{\azero n-r}{\aone} + \athree n\)
\end{itemize}
Note that the assumption that \aone\ and \atwo\ are coprimes guarantees the existence of \amtwo.

\textbf{\Cref{IN5}: }
Tripathi's theorem
\[g(A)=
\underset{0<r<\aone}{max}\bigl(\underset{0 \leq n<\aone}{min}F(n,r)\bigl)-\aone\]
\end{frameit}
Tripathi's theorem is proven by showing that the minimum of \(F(n, r)\), for a fixed value of r, gives a distinct \(\mk\) for each value of \(r\) in the interval \([0, \aone)\). The greatest of these minima minus \(\aone\) then gives us \(g(A)\)
according to the theorem by Brauer and Shockley. Note that the minimum of \(F(n, r)\), for a fixed value of r, does not in general belong to \([r]\). However, when \(r\) is fixed to zero it always does. That is why we can exclude this minimum.

As already mentioned, the foundation of the algorithm presented here is \cref{IN5} (Tripathi's theorem). In \cref{FM} we show how to minimize \(F(n, r)\), for a fixed value of \(r\), using a sequence denoted \ninbseq. This sequence is defined by two functions, denoted \(f\) and \(h\), and an integer constant, denoted \nb, which is defined based on these functions. Below we formally define \(f\), \(h\), \nb\ and \ninbseq\ and present the theorem for minimizing \(F(n, r)\) by means of these.
\begin{frameit}
\textbf{\Cref{FM1}: }
\begin{itemize}
\item
\(h(n) = \mod{\azero n}{\aone}\)
\item
\(f(n) = \atwo h(n) + \athree n\)
\end{itemize}

\textbf{\Cref{IN13}: }\\
\nb\ is the smallest integer \(n\geq 0\) such that \(f(n+1)>\aone\atwo\).

\textbf{\Cref{FM21}: (slightly modified)}\\
\ninbseq\ consists of all integers in the interval \([0,\nb]\) where \(\nb\) is greater than zero. These integers appear exactly once and there are no other elements in the sequence. The elements appear in ascending order based on the value of \(h\) at the element.

\textbf{\Cref{FM2}: (slightly modified)}
\begin{gather*}
\underset{n}{min}\;F(n,r)=
\begin{cases}
f(\ni)-\atwo r &\text{if } h(\nim)<r\leq h(\ni)\\ 
\aone\atwo-\atwo r &\text{if } h(\nnb) < r
\end{cases}\\
\text{where } 0 < i \leq \nb
\end{gather*}
\end{frameit}
The function \(h\) (see \cref{FM1}) defines a sequence belonging to a sequence type, which here is called arithmetic reduced modulo (ARM) sequence. Deriving the algorithm presented here, we use the characteristics of this sequence type extensively. Therefore, in \cref{ARM} we summarise the propositions concerning ARM sequences important here. If you are not well familiar with ARM sequences the  author strongly recommends that you read that section before continuing reading this introduction.

In \cref{FF} we show how the Frobenius number can be computed without computing the minima of \(F\) for all values of r by cherry picking specific values. However, the computations in that section are based on knowing what \nb\ is. Therefore, in \cref{FN} we derive an algorithm for finding \nb\ which is substantially faster than finding \nb\ by simply computing \(f(n)\) from \(n\) equal to zero until \(f(n)\) becomes greater than \(\aone\atwo\). This analysis starts with concluding that \(\nb\) equals zero iff:
\[f(1) = \athree + \atwo\azero > \aone\atwo\]
After having concluded that we assume that the opposite is true and show that \(f(n)\) is, in this case, decreasing within the interval \([\nli, \nhi]\) for any \(i\) if \(h\) is decreasing, where \(\nli\) is a lower border of \(h\) and  \(\nhi\) is an upper border. This implies that the smallest integer \(n\), such that \(f(n)\) is greater than \(\aone\atwo\), equals \(\nli\) for some \(i\). This in turn implies that \(\nb\) equals \(\nhi\) for some positive integer \(i\). When \(h\) is increasing then \(f(n)\) is increasing within the interval \([\nli, \nhi]\) for any \(i\). We can also show that \(f(\nhi)\) is greater than \(f(n)\) for all \(n\) in the interval \([\nlip, \nhip)\). This gives us two cases when \(h\) is increasing. Either \(n\), such that \(f(n)\) is greater than \(\aone\atwo\), is not greater than \(\nhone\) or it equals \(\nhi\) for some \(i\) greater than one. All in all, to find \(\nb\) we want to find the smallest \(i\) such that \(f(\nhi)\) is greater than \(\aone\atwo\) when \(h\) is increasing and the smallest \(i\) such that \(f(\nli)\) is greater than \(\aone\atwo\) when \(h\) is decreasing. Below follows a lemma stating how this can be done by means of the border sequence of \(\hs\) together with the definition of the function defining this sequence, which we denote \(e\). (The constant \(\oo\) is derived from the set \(A\).)
\begin{frameit}
\textbf{\Cref{FM7}: }
\begin{align*}
e(i)=\mod{\oab i}{\oa} \text{ where }\left\{
\begin{aligned}
2\azero<\aone:\quad
&\oa=\azero\\
&\oab=\mod{\aone}{\oa}\\
\\
2\azero>\aone:\quad
&\oa=\aone-\azero\\
&\oab=\oa-\mod{\aone}{\oa}
\end{aligned}\right.
\end{align*}
\textbf{\Cref{FM9}: }
\begin{gather*}
\begin{aligned}
2\azero<\aone:\quad & f(\nhi)<\aone\atwo\\
2\azero>\aone:\quad & f(\nlip)<\aone\atwo\\
\end{aligned}
\iff
\frac{e(i)}{i} > \oo
\text{ where }0<i<\oa
\end{gather*}
\end{frameit}
As \cref{FM9} shows \(f(\nhi)\) is greater than \(\aone\atwo\), when \(h\) is increasing, if the ratio \(e(i)/i\) is less than \(\oo\) and the same condition determines whether \(f(\nli)\) is greater than \(\aone\atwo\), when \(h\) is decreasing, i.e. we can find \(\nb\) if we find the smallest integer \(i\) fulfilling this condition. In \cref{ARM} an algorithm is described finding the first element in an ARM sequence such that the ratio of the element and its index is less than a constant. Using this algorithm allows us to derive an algorithm 
finding \(\nb\) which worst case time complexity is \(O(\log \azero)\). This implies that finding \(\nb\) using this algorithm is substantially faster than finding \(\nb\) by simply computing \(f(n)\) from \(n\) equal to zero until \(f(n)\) becomes greater than \(\aone\atwo\). The algorithm finding the first element in an ARM sequence, such that the ratio of the element and its index is less than a constant, is based on that this element can be computed by means of closed form formulas when a specific condition apply and if this condition does not apply then this problem can be transformed into an equivalent problem of the border sequence of the ARM sequence. This transformation can be repeated until we have formulated a problem where the condition is satisfied, i.e. we can use the border sequence sequence of the ARM sequence to solve this problem. This ARM sequence is in our case defined by the function \(e\) so the algorithm finding \(\nb\) is based on the diff-mod sequence of \(e\) and the first pair in this sequence is \((\oab, \oa)\). Once this diff-mod sequence has been generated, \nb\ can be computed by means of closed form formulas. However, in most cases we do not have to generate the entire diff-mod sequence. The closed form formulas for computing \nb\ are based on a specific pair in the diff-mod sequence, denoted \((\oabos, \oaos)\), and its predecessor.
After generating the diff-mod sequence until the pair \((\oabos, \oaos)\), \nb\ can be computed by means of closed form formulas. Further down we describe the complete algorithm
presented here and list all formulas for computing \(\nb\).

Equipped with an efficient algorithm for computing \nb, we derive in \cref{FF} formulas for computing the Frobenius number. In total six formulas are derived for as many mutually exclusive cases. Jointly these cover all possible cases. First we derive a formula for the case where \(\athree + \atwo\azero\) is greater than \(\aone\atwo\). (Further down we describe the complete algorithm presented here and list the formulas for all cases.) Then we take a closer look at \cref{IN5} (Tripathi's theorem) to see how we can cherry pick values of \(r\) to compute the Frobenius number. Doing that gives us the following formula for doing that.
\begin{frameit}
\textbf{\Cref{FF3}: }
\(\oDi=h(\ni)-h(\nim)
\text{ where }0<i\leq \nb\)

\textbf{\Cref{FF4}: }
\[g(A)=max\Bigl(\underset{i}{max}\;\athree \ni+\atwo(\oDi-1),\atwo(\aone-h(\nnb)-1)\Bigl)-\aone\text{ where }0<i\leq \nb\]
\end{frameit}
The formula given by \cref{FF4}, consists of a maximum of two expressions minus \(\aone\). We label these expressions A and B:
\begin{align*}
A:&\quad\underset{i}{max}\;\athree \ni+\atwo(\oDi - 1)\\
B:&\quad\atwo(\aone - h(\nnb) - 1)
\end{align*}
Then we analyse the case where \(h\) is increasing and \(\oab\) is less than \(\oo\) and show that \(\oDi\) is constant for all \(i\). This implies that A is given by \(i\) such that \(\ni\) equals \(\nb\). This implies that the Frobenius number is given by either expression A and B. Then we show that A will always be greater than B for the remaining cases, i.e. B can be neglected for these cases.

A is a maximum of a number of candidates. It is trivial to realize that the greater \(\ni\) and \(\oDi\) are the stronger the corresponding candidate is to be the maximum of A. Therefore, we analyse how the size of \(\ni\) and \(\oDi\) are correlated and prove a lemma which will help us greatly to understand this correlation.
\begin{frameit}
\textbf{\Cref{FF6}: }
\begin{itemize}
\item
\(\oDdi\text{ is the smallest integer }\oD>0\text{ such that }\hm (\oD)\leq\ni.\)
\item
\(\oDbi\text{ is the smallest integer }\oD>0\text{ such that }\hms (\oD) \leq \nb - \ni.\)
\end{itemize}
\textbf{\Cref{FF7}: (shortened)}
\[
\oDi = min(\oDdi,\oDbi)
\]
\end{frameit}
The lemma above implies that \(\oDbi\) increases or stays the same when \(\ni\) increases since \(\oDbi\) is the index of the first element in an ARM sequence lower than a limit which decreases when \(\ni\) increases. Equivalent reasoning yields that \(\oDdi\) decreases or stays the same when \(\ni\) increases. We also show that \(\oDi\) equals \(\oDbi\) when \(\ni\) equals one and that \(\oDi\) equals \(\oDdi\) when \(\ni\) equals \(\nb\). This implies that \(\oDi\) increases or stays the same when \(\ni\) increases as long as \(\oDi\) equals \(\oDbi\). However, \(\oDi\) will equal \(\oDdi\) at one point and \(\oDi\) will then equal \(\oDdi\) from that point onwards and decrease or stay the same. Deeper analysis shows that the number of values of \(i\), which potentially can give the maximum of A, can be reduced to two in all cases except for one case where there is only one possible option. To derive formulas for the remaining cases, for which \(\oab\) is greater than \(\oo\), we use an algorithm described in \cref{ARM}. This algorithm finds the first element in an ARM sequence not greater than a limit. It is based on that this element will be one of the local minima of the sequence if the limit is not greater than the greatest local minima. This is the case if the modulus of the border sequence of the ARM sequence is greater than the limit and then the problem can be transformed into the problem of finding the first element of the border sequence less than the limit. This transformation can be repeated as long as the limit is not greater than the greatest local minima, i.e. we can again use the border sequence sequence of the ARM sequence to solve this problem. Here we use this approach to compute \(\oDdi\) and \(\oDbi\). Therefore, we have to compute the diff-mod sequence of the sequence defined by \(\hm\). The first pair of this diff-mod sequence equals \((\azerom, \aone)\) and we denote the pairs in this sequence \((\ovbj, \ovj)\). In addition, we need the diff-mod sequence of \(\hms\) to compute \(\oDbi\). However, each pair in this sequence is given by the diff-mod sequence of \(\hm\) (see \cref{FFA5}) so we do not need to compute it. Often we do not have to generate the entire diff-mod sequence of \(\hm\) either. The closed form formulas for computing \(g(A)\) are based on two consecutive pairs which moduli are defined by:
\[\ovopp \leq \nb < \ovop\]
Below we summarize the complete algorithm presented here and list formulas yielding \(g(A)\) for all cases.
\begin{frameit}
\textbf{Complete algorithm for computing the Frobenius number:}
\begin{enumerate}
\item Compute \(\azero = \mod{-\amtwo\athree}{\aone} \quad \intref{IN4}\)
\item If \(\athree + \atwo\azero > \aone\atwo\) then:
\[g(A) = \aone\atwo - \atwo -\aone \quad \intref{FF1}\]
\item else if \(\athree + \atwo\azero < \aone\atwo\) then:
\begin{enumerate}
\item
compute (see \cref{FM7,FM10}):
\begin{align*}
2\azero<\aone:\quad
&\oa=\azero\\
&\oab=\mod{\aone}{\oa}\\
&\ob=\atwo\oa+\athree\\
\\
2\azero>\aone:\quad
&\oa=\aone-\azero\\
&\oab=\oa-\mod{\aone}{\oa}\\
&\ob=\atwo\oa-\athree\\
\\
&\oo = \frac{\aone\athree}{\ob}
\end{align*}
\item if \(2\azero < \aone\) and \(\oab < \oo\):
\begin{align*}
\intref{FM13} & \quad
\nb = \ceilfrac{\aone\atwo}{\ob} - 1\\
\intref{FF5}& \quad g(A) = max(
\begin{aligned}[t]
&\athree\nb + \atwo(\azero - 1),\\
&\atwo(\aone - \nb\azero -1)) - \aone
\end{aligned}
\end{align*}
\item else if \(2\azero > \aone\) and \(\oab < \oo\):
\begin{align*}
\intref{FM13} & \quad \nb = \floorfrac{\aone}{\oa}\\
\intref{FF13} & \quad g(A) = max(
\begin{aligned}[t]
&\athree\nb + \atwo(\mod{\aone}{\oa} - 1),\\
&\athree(\nb - 1) + \atwo(\oa - 1)) - \aone
\end{aligned}
\end{align*}
\item else if \(\oab > \oo\)
\begin{enumerate}
\item compute the diff-mod sequence (see \cref{IN7}) starting with the pair \((\oab, \oa)\) until the pair \((\oabos, \oaos)\), where \(\os\) is the smallest integer \(j\) such that \(\oabj\) is less than \(\ooj\), where \(\ooj\) is defined by:
\begin{align*}
\intref{FN2} \quad 2\oabj < \oaj:\quad & \oojp=\frac{\oaj\ooj}{\oajp-\ooj}\\
2\oabj > \oaj:\quad & \oojp = \frac{\oaj\ooj}{\oajp+\ooj}\\
& \ooone = \oo
\end{align*}
\item compute \(\nb\) (see \cref{FN11}):
\begin{align*}
\nb= \hms(\oabos + \azero) \text{ if } 2\oabosm < \oaosm\\
\nb = \hms\left(\oaosm - \oaos \ceilfrac{\oaosm}{\oaos+\ooosm}+\azero\right)\text{ if }2\oabosm > \oaosm
\end{align*}
\item compute the diff-mod sequence starting with the pair \((\azerom, \aone)\) until the pair \((\ovbopp, \ovopp)\) defined by:
\[\ovopp \leq \nb < \ovop \quad \intref{FF14}\]
\item compute \(g(A)\):
\begin{align*}
\intref{FF25}\quad & \text{ if } \nb = \ovop -1:\\
\\
& \quad g(A) = max(
\begin{aligned}[t]
&\athree \nb + \atwo(h(\ovbop) - 1),\\
&\athree(\ovbop -1) + \atwo (\hs(\ovop - \ovbop) - 1))
- \aone
\end{aligned}\\
\\
\intref{FF29} \quad &\text{ else if } \nb < \ovop -1\text{ and }2\ovbop < \ovop:\\
\\
&\quad g(A) = max(
\begin{aligned}[t]
&\athree \nb + \atwo(h(\ovbop) - 1),\\
&\athree(\ovbop -1) + \atwo (\hs(\oed \ovbop - \ovbopp) - 1)) - \aone
\end{aligned}\\
\\
&\quad \text{where }\oed = \floorfrac{\nb - \ovopp + 1 + \ovbopp}{\ovopp}\\
\\
\intref{FF33}\quad &\text{ else if } \nb < \ovop -1 \text{, } 2\ovbop > \ovop \text{ and } \nb = \oeb \ovopp + \ovbopp - 1:\\
\\
&\quad g(A) = \athree\nb + \atwo(h(\ovbopp + (\oeb - 1)\ovopp) - 1) - \aone\\
\\
\intref{FF33}\quad &\text{ else if } \nb < \ovop -1 \text{, } 2\ovbop > \ovop \text{ and } \nb > \oeb \ovopp + \ovbopp - 1:\\
\\
& \quad g(A) = max(
\begin{aligned}[t]
&\athree \nb + \atwo(h(\ovbopp + \oeb \ovopp) - 1),\\
&\athree(\oeb \ovopp + \ovbopp - 1
) +\\
&\atwo (h(\ovbopp + (\oeb - 1)\ovopp) - 1)) - \aone\\
\end{aligned}\\
\\
&\quad \text{where } \oeb = \floorfrac{\nb + 1 - \ovbopp}{\ovopp} \text{ for the last two cases}
\end{align*}
\end{enumerate}
\end{enumerate}
\end{enumerate}
\end{frameit}
Note that the algorithm above covers all possible cases since:
\begin{itemize}
\item
\(\athree + \atwo\azero \neq \aone\atwo \quad \intref{FM24}\)
\item
\(\athree + \atwo\azero > \aone\atwo\text{ if }2\azero = \aone \quad \intref{FM25}\)
\item
\(\oab \neq \oo \quad \intref{FM9}\)
\item
\(2\oabosm \neq \oaosm \quad \intref{FN11}\)
\item
\(\oab < \oo\) if \(2\ovbop = \ovop \quad \intref{FF24}\)
\end{itemize}
You might also have noted that the formula for the case \(\athree + \atwo\azero > \aone\atwo\) is the same as the one given by Sylvester's theorem (\cref{IN1}) for solving the two variable case. Basically, this implies that \(g(\aone, \atwo, \athree)\) equals \(g(\aone, \atwo)\) for this case.

We finalize this introduction by giving an overview of the entire article. In \cref{ARM} we summarise the propositions concerning ARM sequences important here. In \cref{FM} we show how to minimize \(F(n, r)\), for a fixed value of \(r\), by means of the sequence \ninbseq. In \cref{FN} we derive an algorithm for finding \nb. In \cref{FF} we derive formulas for computing the Frobenius number. 
In \cref{EX} we provide examples. In \cref{TC} we finally analyse the time complexity of the algorithm presented here and conclude that it is is a logarithmic function of \(A\), which makes the time complexity comparable to Greenberg's algorithm, the fastest algorithm for solving the Frobenius problem known today.
\section{ARM sequences}\label{ARM}
The function \(h\) (see \cref{FM1}) defines a sequence belonging to a type of sequence, which here is called arithmetic reduced modulo (ARM) sequence. Deriving the algorithm presented here, we use the characteristics of this type of sequence extensively. In this section we summarise these characteristics. Most of them are well-known and presented here without proof. In addition to this, we present two algorithms solving subproblems of the algorithm presented in this article. The first algorithm finds the first element in an ARM sequence not greater than a specific limit. The second one finds the first element in an ARM sequence such that the ratio of the element and its index is not greater than a specific limit. Proofs of these algorithms are given. We also analyse the time complexity of these algorithms.

The sequence defined by \(h\) and all other ARM sequences presented here meet specific criteria. ARM sequences have two parameters, here called difference and modulus. The criteria is that these parameters are positive coprimes and that the difference is less than the modulus.  Below follows a formal definition of the ARM sequences discussed here.
\begin{frameit}
\begin{defi}\label{BA1}\ \\
The xth element in an ARM sequence, denoted \(u(x)\), is defined by:
\[u(x) = \mod{px}{q}\text{ where } p \text{ (difference) and } q \text{ (modulus) are parameters.}\]
\end{defi}
\begin{assu}\label{BA4}\ \\
In this article we assume that:
\[0<p<q \text{ and } gcd(p,q)=1\]
\end{assu}
\end{frameit}
Specializations of \(u(x)\) can be grouped into pairs, where two specializations make a pair if they have the same modulus \((q-parameter)\) and the sum of their differences \((p-parameters)\) equals the modulus. Note that specializations whose difference equals half the modulus do not have any other specialization to pair up with. These sequences will be discussed later on. Looking at two sequences, which form a pair, one will have a difference which is less than half the modulus and the other one will have a difference which is greater. Here we call the sequences of a pair siblings.
We also define two categories of ARM sequences, where the specializations whose difference is less than half the modulus are called increasing and the ones whose difference is greater are called decreasing. This means that a sibling pair consists of an increasing sequence and a decreasing one. ARM sequences are periodic, with cycle length equal to the modulus, and the naming comes from the fact that their cycles can be divided into subsequences which all either are increasing or decreasing. We start with describing increasing sequences. In this case, a cycle can be divided into subsequences which all are increasing and together cover an entire cycle. The first element of such a subsequence is \(q\) minus \(p\) units less than its predecessor. All other elements in the subsequence are \(p\) units greater than its predecessor, i.e. the first element is a local minimum. The last element of the subsequence is greater than its successor, i.e. its a local maximum. This implies that you get the difference between an element in a subsequence and an element in the next subsequences by multiplying the number of steps between the two elements by \(p\) and deduct \(q\). As the element with index zero always is zero, this implies that any element can be computed by multiplying its index by \(p\) and subtract \(q\) multiplied by the number of subsequences after the first one until the subsequence of the element. The index of the first element in a subsequence is here denoted \(\xli\) whereas the last is denoted \(\xhi\) where \(i\) specifies the subsequence. (We will explain the number scheme of \(i\) later on.) These indexes are here referred to as lower and upper borders and the corresponding elements are referred to as lower and upper border elements, denoted \(u(\xli)\) and \(u(\xhi)\) respectively. All lower border elements are less than the difference. The assumption that the difference and modulus are coprimes, implies that the integers in the interval \([0, q)\) exist exactly once in a cycle, i.e. two equal elements within a cycle must be the same element. This implies that the number of lower border elements in a cycle equals the difference, i.e. the number of subsequences equals the difference. The upper border elements, on the other hand, are not less than the modulus minus the difference. The element with index zero will always equal zero which implies that it will be a lower border element and in fact the smallest lower border element. We number the corresponding subsequence one and denote the index of the lower border \(\xlone\) to indicate that it is the lower border of the first subsequence. The last upper border element equals \(q\) minus \(p\), i.e. it is the smallest one.

Next let us look at an example of an increasing sequence. The function \(\mod{5x}{13}\) defines the sequence:
\[0, 5, 10, 2, 7, 12, 4, 9, 1, 6, 11, 3, 8, ...\]
The elements above complete a cycle. Note that the modulus is 13 and that the integers in the interval \([0, 13)\) exist exactly once in the cycle. This cycle consists of five (the difference) increasing subsequences , where the elements increase five units compared to its predecessor. The lower border elements of these subsequences consist of the integers in the interval \([0, 5)\) and these are eight units (modulus minus difference) less than their predecessor. The upper border elements consist of the integers in the interval \([8, 13)\) and the last one is the smallest one equal to the modulus minus the difference, i.e. eight.

Now we will take a look at the sibling sequence defined by the function \(\mod{8x}{13}\) which gives the sequence:\\
\[8, 3, 11, 6, 1, 9, 4, 12, 7, 2, 10, 5, 0, ...\]
Comparing carefully the sibling sequences above you will see that the decreasing one is the reverse of the increasing one. This is true in general for siblings. However, the start of the cycle of the decreasing sequence must be chosen with care. For increasing sequences we decided that the first subsequence of a cycle starts at index zero. To make the decreasing sequence the reverse of the increasing sibling the first subsequence should start at index one. The last element in a cycle will then have index \(q\) whereas the last element of the increasing sibling has index \(q\) minus one. Considering the fact that the sum of the differences of sibling sequences equals the modulus and that the decreasing sequence is the reverse of the increasing sibling, it is trivial to realize that the cycle of a decreasing sequence consists of \(q\) minus \(p\) decreasing subsequences where the first element of such a subsequence is \(p\) units greater than its predecessor, i.e. it is a local maximum. All other elements in the subsequence are \(q\) minus \(p\) units less than its predecessor. The last element of the subsequence is less than its successor, i.e. it is a local minimum. This implies that you get the difference between an element in a subsequence and an element in the next subsequences by multiplying the number of steps between the two elements by \((q - p)\) and subtract this product from \(q\). As the element with index zero always is zero, any element in the sequence can be computed by multiplying \(q\) by the subsequence number of the element and subtract \(q\) minus \(p\) multiplied by the index of the element. The lower border elements of these subsequences consist of all integers in the interval \([p, q)\) and the first one is the smallest one. The upper border elements consist of the integers in the interval \([0, q - p)\) and the last one is the smallest one (equal to zero). The lower borders of sibling sequences are the same except the first one which is zero for the increasing sequence and one for the decreasing one. The upper borders of sibling sequences are also the same except the last one which is \(q\) minus one for the increasing sequence and \(q\) the decreasing one. From the fact that an ARM sequence is the reverse of its sibling we can also conclude that \(u(-x)\) equals \(\us(x)\). This since the cycle starting at \(\xlone\) is preceded by another cycle which reversed becomes a cycle of the sibling sequence. Another characteristic of ARM sequences, which we will use often here is the fact that the sum of an element and the element of the sibling sequence,
with the same index, is always equal to the modulus unless the index is a multiple of the modulus. If the index is a multiple of the modulus then both elements are equal to zero. For instance, the elements with index 2 in the sibling sequences above are 10 and 3 and the elements with index 4 are 7 and 6. In both cases the sum equals the modulus, i.e. 13.

Now we will have a look at the specializations where the difference equals half the modulus. The assumption that the difference and modulus are coprimes actually implies that there is only one specialization which meets this criteria, namely the one given by:
\[p = 1 \text{ and } q = 2\]
These parameter values give the following sequence which actually can be regarded both as an increasing and decreasing sequence:
\[0,1,0,1,0,1...\]
Below follows a formal description of what we concluded so far about ARM sequences.
\begin{frameit}
\begin{defi}\label{BA19}
\leavevmode
\begin{itemize}
\item
If \(2p < q\) then the sequence \(\useq\) is called increasing.
\item
If \(2p > q\) then the sequence \(\useq\) is called decreasing.
\end{itemize}
If two ARM sequences have the same modulus and the sum of their differences equals the modulus then these are called siblings. One of the siblings will be increasing and the other one decreasing. The sibling of \(u\) is denoted \(\us\).
\end{defi}
\begin{lemm}\label{BA20}\ \\
\\
Increasing sequences:
\begin{itemize}
\item
Within a cycle of \(\useq\) there are \(p\) local minima and \(p\) local maxima, which indexes are denoted \(\xli\) and \(\xhi\) respectively.
\item
\(\xlone=0, \xhp=q-1\)
\item
\(\xlip = \xhi + 1\)
\item
\(u(\xlip) = u(\xhi) - q +p\)
\item
\(u(\xli) <  p\)
\item
\(u(\xhi) \geq q-p\)
\item
\(\xli < x < \xhi \implies
p \leq u(x) < q - p\)
\item
\(\xli \leq x , x + z \leq \xhi \implies
u(x + z) = u(x) + zp \text{ where } z\inZ\)
\item
\(u(\xlone) = 0 < u(\xlj)\;\forall\; j \neq 1\)
\item
\(u(\xhp) = q - p < u(\xhj)\;\forall\; j \neq p\)
\end{itemize}
Decreasing sequences:
\begin{itemize}
\item
Within a cycle of \(\useq\) there are \(q - p\) local maxima and \(q - p\) local minima, which indexes are denoted \(\xli\) and \(\xhi\) respectively.
\item
\(\xlone=1, \xhqmp=q\)
\item
\(\xlip = \xhi + 1\)
\item
\(u(\xlip) = u(\xhi) + p\)
\item
\(u(\xli) \geq  p\)
\item
\(u(\xhi) < q-p\)
\item
\(\xli < x < \xhi \implies
q - p \leq u(x) < p\)
\item
\(\xli \leq x , x + z \leq \xhi \implies
u(x + z) = u(x) - z(q - p) \text{ where } z\inZ\)
\item
\(u(\xlone) = p < u(\xlj)\;\forall\; j \neq 1\)
\item
\(u(\xhqmp) = 0 < u(\xhj)\;\forall\; j \neq q - p\)
\end{itemize}
Lower borders of siblings are the same except for the first ones. Upper borders of siblings are the same except the last ones.
\end{lemm}
\begin{lemm}\label{BA14}
\(2p = q \iff p =1 \text{ and } q = 2\)
\end{lemm}
\begin{lemm}\label{BA22}
Given \(\abs{x-\xt}<q:\;u(x)=u(\xt)\iff x=\xt\)
\end{lemm}
\begin{lemm}\label{BA16}
\begin{align*}
x = zq & \iff
u(x)=0 \iff \us(x)=0 \text{ where }z\inZ\\
u(x) \neq 0 & \iff \us(x) \neq 0
\iff u(x) = q - \us(x) = q - u(-x)
\end{align*}
\end{lemm}
\begin{lemm}\label{BA29}
\begin{align*}
2 p < q: \quad &
\xli\leq x \leq\xhi<x+z\leq\xhip\implies u(x + z)= u(x) + zp - q\\
2 p > q: \quad & \xli \leq x \leq \xhi < x + z \leq \xhip \implies u(x+z) = u(x) - z(q - p) + q\\
& \text{where }z\inN
\end{align*}
\end{lemm}
\begin{lemm}\label{BA34}
\begin{align*}
2 p < q: & \quad
u(x) = px - ( i - 1) q \text{ where } \xli \leq x \leq \xhi\\
2 p > q: & \quad
u(x)= iq - (q - p) x \text{ where } \xli \leq x \leq \xhi
\end{align*}
\end{lemm}
\end{frameit}
The number of elements in a subsequence is here called the length of this subsequence. The length of subsequence number \(i\) is denoted \(\li\). The lengths of a specific sequence vary with one unit. For increasing sequences, the lower border element of the longer subsequences are less than a limit, whereas they are greater than a limit for decreasing sequences. If the sequence is increasing and the modulus is greater than one, then the first lower border element, equal to zero, is always less than the limit, i.e. the first subsequence is always long. If the sequence is decreasing, then the first lower border element, equal to \(p\), is always less than the limit, i.e. the first subsequence is short. This implies that \(\xhone\) equals the short length of the sequence for both increasing and decreasing sequences since \(\xlone\) equals zero for increasing sequences and one for decreasing ones.
\begin{frameit}
\begin{defi}\label{BA3}
\(\li = \xhi - \xli + 1\)
\end{defi}
\begin{lemm}\label{BA28}
\begin{align*}
2 p < q: \quad
&\begin{aligned}[t]
u(\xli)<\mod{q}{p}\implies &\li=\floorfrac{q}{p}+1\\
u(\xli)\geq\mod{q}{p}\implies &\li = \floorfrac{q}{p}\\
p > 1 \implies &\lone = \floorfrac{q}{p} + 1
\end{aligned}\\
2 p > q: \quad
&\begin{aligned}[t]
u(\xli)\geq q - \mod{q}{q - p}\implies &\li = \floorfrac{q}{q - p} +1\\
u(\xli) < q - \mod{q}{q -p}\implies &\li=\floorfrac{q}{q - p}
\end{aligned}
\end{align*}
\end{lemm}
\begin{lemm}\label{BA37} Given \(p > 1\)
\[\xhone = \floorfrac{q}{p}\]
\end{lemm}
\end{frameit}
Another characteristic of ARM sequences, that we use in this article, is that the function \(u\) is quasi-additive.
\begin{frameit}
\begin{lemm}\label{BA8}
\begin{gather*}
u(x+y)=
\begin{cases}
u(x)+u(y) &\text{ if }u(x)+u(y) <q\\ 
u(x)+u(y)-q &\text{ if }u(x)+u(y)\geq q\\
\end{cases}
\end{gather*}
\end{lemm}
\begin{lemm}\label{BA9}
\begin{gather*}
u(x-y)=
\begin{cases}
u(x)-u(y) &\text{ if }u(x)\geq u(y)\\ 
u(x)-u(y)+q &\text{ if }u(x) < u(y)\\
\end{cases}
\end{gather*}
\end{lemm}
\end{frameit}
Next we will discuss the characteristic of ARM sequences defining the algorithm presented here. The characteristic allowing us to design such an efficient algorithm. The local minima of an ARM sequence form another ARM sequence. For instance, the lower border elements of the increasing ARM sequence above gives the sequence:
\[0, 2, 4, 1, 3, ...\]
This is the ARM sequence defined by the function \(\mod{2x}{5}\).  The upper border elements of the decreasing sibling form the reversed sequence defined by \(\mod{3x}{5}\):
\[3, 1, 4, 2, 0\]
The sequence consisting of the local minima of an ARM sequence is here called the border sequence of this ARM sequence. The border sequence of an increasing ARM sequence consists of its lower border elements, i.e. all non-negative integers less than the difference. The border sequence of the decreasing sibling consists of its upper border elements which are the same as the lower border elements of the increasing sibling but the order is reversed. This means that the border sequence of the decreasing sibling is the sibling of the border sequence of the increasing sibling. Since the border sequence of an ARM sequence is an ARM sequence, it has a border sequence, which is an ARM sequence having a border sequence and so on. It continues like this until an atomic sequence is reached. In other words, inside an ARM sequence hides another ARM sequence and inside that one another one and so on, i.e. ARM sequences are like Russian dolls. This creates a sequence of ARM sequences where all, but the first one, are the border sequence of the previous one. This sequence of sequences is here called a border sequence sequence. We denote the function defining the jth sequence in this sequence \(\uj\), its difference \(\vbj\) and its modulus \(\vj\). This implies that \(\uone, \vbone\) and \(\vone\) equals \(u, p\) and \(q\) respectively. The jth sequence in the border sequence sequence of the sibling sequence is denoted \(\ujs\). This notation is actually ambiguous since it could also mean the sibling sequence of \(\uj\). Luckily these are the same. (However, this is not true for the last sequence in a border sequence sequence if the second last sequence has the difference one and modulus two. In this case the sum of the differences of \(\uj\) and \(\ujs\) for the last sequences is not equal to the modulus.) The parameters \(\vbj\) and \(\vj\) of a border sequence sequence form a sequence of pairs where the jth pair is denoted \((\vbj, \vj)\). This sequence of pairs is here referred to as the diff-mod sequence of an ARM sequence as it consists of the differences and moduli of the sequences in a border sequence sequence. The number of definable pairs in a diff-mod sequence is finite and the last pair, denoted \((\vbt, \vt)\), will always be either \((1, 1)\) or \((0, 1)\). Both of these pairs correspond to a sequence whose cycle consists only of the integer zero. All sequences in a border sequence sequence, but the last one, meet the criteria of \cref{BA4}. The second last one also meets additional criteria. These are given below together with a formal summary of what we stated above and formulas for computing the local minima of an ARM sequence.
\begin{frameit}
\begin{defi}\label{IN7}\ \\
The sequence of functions \((\uj)\) is defined by:
\begin{align*}
\uj(i) = \mod{\vbj i}{\vj}\text{ where }\left\{
\begin{aligned}
\vbone = p, \vone = q &\\
\\
\text{Given }\vjp > 0:\\
\\
2\vbj\leq\vj:\quad & \vjp=\vbj\\
& \vbjp=\vjp-\mod{\vj}{\vjp}\\
\\
2\vbj>\vj:\quad & \vjp=\vj-\vbj\\
& \vbjp=\mod{\vj}{\vjp}
\end{aligned}\right.
\end{align*}

The sequence of sequences defined by \((\uj)\) is here called the border sequence sequence of \(u\).

The sequence of the pairs \((\vbj, \vj)\) is here called the diff-mod sequence of \(u\).

The sequence of functions \((\ujs)\), where \(\vbones = \vone - \vbone\) and  
\(\vones = \vone\), defines the border sequence sequence of the sibling sequence of \(u\).
\end{defi}
\begin{lemm}\label{BS9}
\leavevmode
\begin{itemize}
\item
Either (1,1) or (0,1) is a pair in all diff-mod sequences. This element is denoted (\vbt,\vt).
\item
\(j < \ot \implies 0 < \vbj < \vj \text{ and } 0 < \vjp < \vj\)
\item
\(j \leq \ot \implies gcd(\vbj, \vj) = 1\)
\item
\((\vbt,\vt) = (1,1) \implies \vbtm = 1\)\\
\((\vbt,\vt) = (0,1) \implies \vtm = \vbtm + 1\text{ and }\vbtm > 1\)
\item 
(\vbtp,\vtp) is undefined.
\end{itemize}
\end{lemm}
\begin{lemm}\label{BS11}
\begin{align*}\left.
\begin{aligned}
2\vbj\leq\vj:\quad & \uj (\ilkp)=\ujp(k)\\
\\
2\vbj>\vj:\quad & \uj (\ihk)=\ujp(k)\\
\end{aligned}\right\}j<\ot
\end{align*}
\end{lemm}
\begin{lemm}\label{FFA5}
\begin{align*}
&\vjs = \vj\\
\\
&\text{Given }2\vbjm \neq \vjm: \quad \vbjs = \vj - \vbj
\end{align*}
\end{lemm}
\end{frameit}
In this article, we also use the inverse function of \(u\), denoted \(\um\). As any other inverse function it undoes the action of the original function, i.e. in this case \(u\), but only for argument values within the interval \([0, q)\).
\begin{frameit}
\begin{defi}\label{FEB4}
\leavevmode
\begin{itemize}
\item
\(\pm\) denotes the multiplicative inverse of p modulo q  where \(0<\pm<q\).
\item
\(\um\) is called the inverse function of \(u\).
\item
\(\um (x)=\mod{\pm x}{q}\)
\end{itemize}
\end{defi}
\begin{lemm}\label{FEB5} Given that \(0 \leq x < q\)\\
\[\um (u(x)) = x\]
\end{lemm}
\end{frameit}
As already mentioned a subproblem of the algorithm presented here is to find the first element in an ARM sequence not greater than a specific limit (see \cref{FF7}). To find this element we use the fact that this element will be one of the local minima of the ARM sequence if the limit is not greater than the greatest local minima. This is the case if the modulus of the border sequence of the ARM sequence is greater than the limit and then the problem can be transformed into the problem of finding the first element of the border sequence less than the limit. This transformation can be repeated as long as the limit is not greater than the greatest local minima, i.e. we can use the border sequence sequence of the ARM sequence to solve this problem. Below we formulate this formally. Recall that local minima of increasing ARM sequences are lower border elements whereas local minima of decreasing sequences are upper border elements.
\begin{frameit}
\begin{defi}\label{FFE1}\ \\
\(\ibj\) is the smallest integer \(i > 0\) such that \(\uj(i) \leq L\)
\end{defi}
\begin{lemm}\label{FFE3} Given \(L < \vjp \)
\begin{align*}
\begin{aligned}
2\vbj < \vj: &\quad \ibj = \ilibjpp\\
2\vbj > \vj: &\quad \ibj = \ihibjp\\
\end{aligned}\\
\\
\uone(\ibone) = \ujp(\ibjp)
\end{align*}
\end{lemm}
\end{frameit}
\begin{proveit}{FFE3}\ \\
\\
We start by assuming that \(2\vbj < \vj\). In this case, according \cref{BA20},
 \(\ibj\) equals \(\ilitp\) since \(\ibj\) is the smallest integer \(i\) greater than zero such that:
\[\uj(i) \leq L < \vjp = \vbj\]
This implies that \(\itilde > 0\) since \(\ibj > 0 = \ilone\) and that \(\itilde\) is the smallest integer \(k > 0\) such that:
\[\uj(\ilkp) = \ujp(k) \leq L\quad\intref{BS11}\]
In other words, \(\itilde\) equals \(\ibjp\) and \(\uj(\ibj) = \ujp(\ibjp)\).\\
\\
If \(2\vbj > \vj\) then \(\ibj\) equals \(\ihit\) since \(\ibj\) is the smallest integer \(i\) greater than zero such that:
\[\uj(i) \leq L < \vjp = \vj - \vbj\]
This implies that \(\itilde\) is the smallest integer \(k > 0\) such that:
\[\uj(\ihk) = \ujp(k) \leq L\]
In other words, \(\itilde\) equals \(\ibjp\) and \(\uj(\ibj) = \ujp(\ibjp)\).\\
\\
As \(\uj(\ibj) = \ujp(\ibjp)\) independent of the relative size of \(\vbj\) and \(\vj\) we can conclude that:
\[\uone(\ibone) = \ujp(\ibjp)\]
\end{proveit}
Another subproblem of the algorithm presented here is to find the smallest positive integer \(x\), denoted \(\xb\), such that the ratio \(u(x)/x\) is less than or equal to a non-negative real number, denoted \(L\). Next in this section, we will present and prove an algorithm solving this problem. This algorithm is, just like the algorithm described previously, based on that the solution to this problem can be computed by means of closed form formulas when a specific condition apply and if this condition does not apply then this problem can be transformed into an equivalent problem of the border sequence of \(u\) and this transformation can be repeated until we have formulated a problem where the condition is satisfied, i.e. we will again use the border sequence sequence of \(u\) to solve this problem. The number of sequences in this border sequence sequence, up to and including the first sequence which satisfies the condition, is denoted by \(\os\). In this case the transformation of the problem from one sequence in the border sequence sequence to the next one will involve computing a new limit. The limit for the jth sequence will be denoted \(\Lj\). The condition we want to obtain is: 
\[\vbj = 1 \text{ and/or }\vbj \leq \Lj\]

If \(\vbone \leq \Lone\) then \(\xb\) equals one. If \(\vbone = 1 > \Lone\) then \(\useq\) becomes:
\[1, 2, ..., \vone - 1, 0, 1, ...\]
Since \(\uone = u\) and \(\vone = q\) this implies that the ratio \(u(x)/x\) is one for \(0 < x < q\) and zero when \(x\) equals \(q\), i.e. \(\xb\) equals \(q\). Continuing the analysis of this problem we will assume that none of these conditions apply, which implies that \(\os > 1\). 

Below follows a formal description of what we discussed  above, i.e. the assumptions we make, the problem we want to solve and how a new limit is computed when transforming the problem to the next sequence in the border sequence sequence of \(u\). We will explain the computation of a new limit in detail later on. As we will use the border sequence sequence of \(u\) to solve this problem, we define it in terms of this sequence. Note that \(\xb\) equals \(\ibone\) and \(L\) equals \(\Lone\) in this definition. Finally we prove a lemma guaranteeing that \(\os\) is defined and that \(\Lj\) is also defined and not less than zero for \(j\) not greater than \(\os\).
\begin{frameit}
\begin{assu}\label{FRB1}\ \\
In this section from this point onward, unless expressly stated otherwise:
 \[\vbone > 1 \text{ and } \vbone > \Lone \text{ where } \Lone \text{ is a positive real number}\]
\end{assu}
\begin{defi}\label{FRB2}
\leavevmode
\begin{itemize}
\item
\(\ibj\) is the smallest integer \(i > 0\) such that:
\begin{align*}
\frac{\uj (i)}{i} \leq \Lj \text{ where }
\left\{\begin{aligned}
2\vbj<\vj:\quad & \Ljp=\frac{\vj\Lj}{\vjp-\Lj}\\
\\
2\vbj>\vj:\quad & \Ljp=\frac{\vj\Lj}{\vjp+\Lj}\\
\end{aligned}\right.
\end{align*}
\item
\(\os\) is the smallest positive integer \(j\) such that \(\vbj = 1\) and/or \(\vbj \leq \Lj\).
\end{itemize}
\end{defi}
\begin{lemm}\label{FRB3}
\[
\Lj \geq 0 \text{ if }j \leq \os \text{ and } \os \leq \ot
\]
\end{lemm}
\end{frameit}
\begin{proveit}{FRB3}\ \\
We start with proving that \(\Lj\) is defined and not less than zero when \(j\) is not greater than neither \(\os\) nor \(\ot\). Then we will prove that \(\os\) is not greater than \(\ot\). 

The first proof we do by induction and assume that:
\[\Lj > 0 \text{ and } j < \os, \ot\]
This guarantees that \(\vj\) and \(\vjp\) are defined. For the case where \(2\vbj\) is greater than \(\vj\) it is trivial to realize that \(\Ljp\) is defined and greater than zero. For the case where \(2\vbj\) is less than \(\vj\) this is the case as long as:
\[\vjp - \Lj = \vbj - \Lj > 0\]
That this is the case follows from the definition of \(\os\). Noting that our assumption is fulfilled for \(j\) equal to one finalizes our proof by induction.

Next we will show that \(\os\) is not greater than \(\ot\). Recall that the last pair in a diff-mod sequence \((\vbt, \vt)\) is either \((1, 1)\) or   \((0, 1)\). If the last pair is \((1, 1)\) then \(\vbtm\) equals one. Therefore, we can conclude that there will be an integer \(j\) less than or equal to \(\ot\) such that \(\vbj\) equals one or \(\vbj \leq \Lj\). However, an observant reader might also have noticed that \(\Ljp\) is not defined when \(2\vbj\) equals \(\vj\).
However, we can conclude that
\(\os\) is defined also when there is a \(j\) such that \(2\vbj\) equals \(\vj\), even though \(\Ljp\) is not defined in this case. This as \(\vbj\) will equal one according to \cref{BA14} which implies that \(\os\) is less or equal to \(j\). All in all, \(\os\) is always defined.
\end{proveit}
Many of the upcoming lemmas assume that \(\vbj > \Lj\). This implies that:
\[j<\os\leq\ot\]
According to \cref{BS9}, this implies that \(\uj\) fulfills the assumptions made for ARM sequences in this article, i.e. all properties of \(u\) are valid for \(\uj\). Next we will prove a lemma to be used in the next one.
\begin{frameit}
\begin{lemm}\label{FRB4}
\begin{align*}
2p < q: \quad& u \left(\xhi+\floorfrac{q}{p}\right) = u (\xhi)-\mod{q}{p}=u (\xhip)\text{ or }u (\xhip-1)\\
2p > q: \quad & u \left(\xhi+\floorfrac{q}{q - p}\right) = u (\xhi)+\mod{q}{q - p}=u (\xhip)\text{ or }u (\xhip-1)
\end{align*}
\end{lemm}
\end{frameit}
\begin{proveit}{FRB4}\ \\
We prove this lemma for increasing sequences. The proof for decreasing sequences is analogous to this one.

The length of a subsequence is according to \cref{BA28}:
\[
\floorfrac{q}{p} \text{ or }\floorfrac{q}{p}+1
\]
From the fact that \(\xhi+\ldip=\xhip\) follows then that:
\[
\xhi+\floorfrac{q}{p}
= \xhip\text{ or }\xhip-1\\
\]
We can then finalize this proof by means of \cref{BA29}:
\begin{align*}
u(\xli+\floorfrac{q}{p})= &u(\xli)+\floorfrac{q}{p} p-q\\
\intref{INT10}= &u(\xli)+\frac{q-\mod{q}{p}}{p}p-q\\
= &u(\xli)-\mod{q}{p}
\end{align*}
\end{proveit}
Next we will show how we can use the border sequence sequence of \(u\) to limit the number of candidates which need to be considered to find \(\ibj\).
\begin{frameit}
\begin{lemm}\label{FRB5}
Given that \(\vbj > \Lj\)
\begin{align*}
\intertext{\bullet\(2\vbj<\vj \implies \ibj\) equals \(\ilk\) where \(1 < k \leq \vjp + 1\)}
\intertext{\bullet 2\(\vbj>\vj\)}
&\bullet\frac{\uj (\ihone)}{\ihone} \leq \Lj \implies
\uj(\ibj) = \vj-\vjp\ceilfrac{\vj}{\vjp+\Lj}\\
&\bullet\frac{\uj(\ihone)}{\ihone} > \Lj \implies \ibj \text{ equals }\ihk \text{ where } 1 < k \leq \vjp
\end{align*}
\end{lemm}
\end{frameit}
\begin{proveit}{FRB5}\ \\
\bullet\(2\vbj < \vj\):\\
First we will show that the ratio \(\uj(i)/i\) is ascending within a subsequence of \(\uj\) except within the first subsequence where the ratio is constant. Assuming that \(\ilk \leq i < \ihk\):
\begin{align*}
\frac{\uj (i+1)}{i+1} & = \frac{\uj (i)+\vbj}{i+1}\cdot\frac{i}{\uj (i)}\cdot\frac{\uj (i)}{i}\quad\intref{BA20}\\
& = \frac{\uj (i) i+\vbj i}{\uj (i)i+\uj (i)}\cdot\frac{\uj (i)}{i}\\
\intref{BA34} & = \frac{\uj (i) i+\vbj i}{\uj (i)i+\vbj i-(k-1)\vj}\cdot\frac{\uj (i)}{i}\\
& \geq \frac{\uj (i)}{i}
\end{align*}
\begin{align*}
\intertext{From this we can conclude that \(\ibj\) is greater than \(\ihone\) since:}
\frac{\uj(i)}{i} \geq \frac{\uj(1)}{1} = \vbj >\Lj \text{ where } \ilone \leq i \leq \ihone
\intertext{We can also conclude that \(\ibj\) will equal \(\ilk\) for some k since:}
\frac{\uj(i)}{i} \geq \frac{\uj(\ilk)}{\ilk}\text{ where } \ilk \leq i \leq \ihk
\intertext{All in all, we can conclude that \(1 < k \leq \vbj + 1 = \vjp + 1\) since:}
\uj (\ilvjpp)/\ilvjpp = 0 \leq \Lj
\intertext{\bullet\(2\vbj>\vj:\)
\newline
\newline
In this case it is trivial to realize that the ratio \(\uj(i)/i\) is descending within a subsequence of \(\uj\), noting that \(\uj\) is descending within a subsequence. Based merely on this characteristic, we cannot draw many conclusions about \(\ibj\) so we need to analyse this case further.\newline
\newline
We start with the case where \(\ibj \leq \ihone\) which implies that:}
\frac{\uj (\ihone)}{\ihone} \leq \Lj
\intertext{When \(i\leq\ihone\) we get:}
\intref{BA34}\quad\frac{\uj(i)}{i} = \frac{\vj-\vjp i}{i}
\intertext{This implies that the ratio \(\uj(i)/i\) is smaller than  or equal to \(\Lj\) when:}
i \geq \frac{\vj}{\vjp+\Lj}
\intertext{Hence follows that:}
\ibj = \ceilfrac{\vj}{\vjp+\Lj}
\intertext{Finally, we can conclude that:}
\uj(\ibj) = \vj - \vjp\ibj 
=  \vj - \vjp\ceilfrac{\vj}{\vjp+\Lj}\\
\\
\intertext{Next we will deal with the case \(\ibj > \ihone\). \newline
\newline
According to \cref{FRB4} holds that:}
\uj\left(\ihk+\floorfrac{\vj}{\vjp}\right) = \uj(\ihkp)\text{ or } \uj(\ihkp-1)
\intertext{We will show that:}
\frac{\uj \left(\ihk+\floorfrac{\vj}{\vjp}\right)}{\ihk+\floorfrac{\vj}{\vjp}}\geq\frac{\uj (\ihk)}{\ihk}
\intertext{Hence, we can conclude that \(\ibj\) will equal \(\ihk\) for some k where \(1 < k \leq \vjp\) since:}
\uj (\ihvjp)/\ihvjp = 0 \leq \Lj
\end{align*}
\begin{align*}
\uj \left(\ihk+\floorfrac{\vj}{\vjp}\right)= & \uj(\ihk)+\mod{\vj}{\vjp}\\
= & \uj (\ihk)+\vbjp\\
\\
\ihk+\floorfrac{\vj}{\vjp}=&\frac{\vjp\ihk+\vj-\mod{\vj}{\vjp}}{\vjp}\quad\intref{INT10}\\
=&\frac{\vjp\ihk+\vj-\vbjp}{\vjp}\\
\\
\frac{\uj \left(\ihk+\floorfrac{\vj}{\vjp}\right)}{\ihk+\floorfrac{\vj}{\vjp}}-\frac{\uj (\ihk)}{\ihk}= & \frac{\vbjp(\vjp\ihk+\uj (\ihk))-\vj\uj (\ihk)}{(\vjp\ihk+\vj-\vbjp)\ihk}\\
\intref{BA34}= & \frac{\vbjp\vj k-\vj\uj (\ihk)}{(\vjp\ihk+\vj-\vbjp)\ihk}\\
\intref{BS11} = &\frac{\vj(\vbjp k-\ujp (k))}{(\vjp\ihk+\vj-\vbjp)\ihk}\\
\geq & 0 \text{ iff }\vbjp k-\ujp (k)\geq 0
\end{align*}\\
From \cref{BS11} follows that:
\[
\frac{\uj(\ihone)}{\ihone}
= \frac{\ujp(1)}{\ihone}
= \frac{\vbjp}{\ihone} 
> \Lj \geq 0 
\]
This implies that \(\vbjp\) is greater than zero. As \(2\vbj\) is greater than \(\vj\), \(\vbjp\) will be zero if \(j\) equals \(\ot - 1\), according to \cref{BS9}, i.e. \( j+1\) is less than \(\ot\) which in turn implies that \(\ujp\) fulfills the assumptions made for ARM sequences in this article.
\begin{align*}
\intertext{To finalize this proof we will deal with three different cases separately.}
\intertext{If \(2\vbjp=\vjp\) then \(\vbjp=1\) and \(\vjp=2\) according to \cref{BA14} which implies that:}
\vbjp k-\ujp(k) = & k-\mod{k}{2}\geq k-1\geq 0\\
\intertext{If \(2\vbjp<\vjp\) and \(\kll\leq k\leq \khl\):}
\vbjp k-\ujp(k) = & \vbjp k - (\vbjp k - \vjp (l-1))\quad\intref{BA34}\\
= & \vjp (l-1)) \geq 0\\
\intertext{If \(2\vbjp>\vjp\) and \(\kll\leq k\leq \khl\):}
\vbjp k-\ujp (k) = & \vbjp k - (\vjp l - (\vjp - \vbjp) k)\\
= & \vjp (k-l)) \geq 0
\end{align*}
\end{proveit}
When \(2\vbj < \vj\) and \(\vbj > \Lj\) follows from \cref{FRB5} that \(\ibj\) equals \(\ilkp\) where k is the smallest positive integer k such that:  
\[\frac{\uj(\ilkp)}{\ilkp} \leq \Lj\]
When \(2\vbj > \vj\) and
\(\uj(\ihone)/\ihone > \Lj\) follows that \(\ibj\) equals \(\ihk\) where \(k\) is the smallest positive integer such that:  
\[\frac{\uj(\ihk)}{\ihk} \leq \Lj\]
Next we will express this in terms of \(\ujp\) and \(\Ljp\).
\begin{frameit}\begin{lemm}\label{FRB6}
Given that \(\vbj > \Lj\)
\leavevmode
\begin{itemize}
\item
\(2\vbj<\vj\)\\
\[\frac{\uj (\ilkp)}{\ilkp}\leq \Lj\iff\frac{\ujp (k)}{k} \leq \Ljp\text{ where }k > 0\]
\item
\(2\vbj>\vj\)\\
\[\frac{\uj(\ihk)}{\ihk} \leq \Lj\iff\frac{\ujp (k)}{k} \leq \Ljp\text{ where }k > 0\]
\end{itemize}
\end{lemm}\end{frameit}
\begin{proveit}{FRB6}\ \\
When \(2\vbj < \vj\) then follows from \cref{BS11} that:
\[
\ujp(k) = \uj(\ilkp)
= \ilkp \vjp - \vj k \quad \intref{BA34}
\]
This gives us that:  
\[
\frac{\uj (\ilkp)}{\ilkp} \leq \Lj
\iff
\frac{\ujp (k)}{k} \leq \frac{\Lj\vj}{\vjp-\Lj} = \Ljp
\]
The proof for decreasing ARM sequences is made in an analogous way.
\end{proveit}
Now when we can express the condition \(\uj(i)/i \leq \Lj\) in terms of \(\ujp\) and \(\Ljp\) we will use this to relate \(\ibj\) and \(\ibjp\) when specific conditions apply.
\begin{frameit}\begin{lemm}\label{FRB7}
\leavevmode
\begin{itemize}
\item
\(2\vbj<\vj\text{ and }\vbj>\Lj\)\\
\[\ibj=\ilibjpp\]
\item
\(2\vbj>\vj\text{ and }\vbjp>\Ljp\)\\
\[\ibj=\ihibjp\]
\end{itemize}
\end{lemm}\end{frameit}
\begin{proveit}{FRB7}\ \\
We prove this lemma for the case where \(2\vbj > \vj\). The proof for the case where \(2\vbj<\vj\) is made in an equivalent manner.

From \cref{FRB6} follows when \(\vbjp>\Ljp\) that :
\[
\frac{\ujp (1)}{1}=\vbjp>\Ljp\implies\frac{\uj (\ihone)}{\ihone}>\Lj
\]
From \cref{FRB5} follows that \(\ibj=\ihk\) where k is the smallest positive integer such that:
\[\frac{\uj (\ihk)}{\ihk} \leq \Lj\]
This also implies that k is the smallest positive integer such that:
\[\frac{\ujp (k)}{k} \leq \Ljp\]
Hence \(k = \ibjp\) and \(\ibj=\ihibjp\).
\end{proveit}
By means of the relationship between \(\ibj\) and \(\ibjp\), provided by \cref{FRB7}, we will now show how \(\uone(\ibone\)) can be computed by means of the border sequence sequence of \(u\).
\begin{frameit}
\begin{lemm}\label{FRB8}
\begin{align*}
\vbos \leq \Los: & \quad\uone(\ibone)=\uosm (\ibosm)\\
\vbos = 1 > \Los: &\quad\uone(\ibone) = \uos (\ibos)
\end{align*}
\end{lemm}
\end{frameit}
\begin{proveit}{FRB8}\ \\
We start with the case when \(\vbos \leq \Los\). In this case it is trivial to realize that this lemma is correct when \(\os\) equals two. We now continue this proof by means of induction and assume that:
\[
\uone(\ibone) = \ujm(\ibjm)\text{ where } 1 < j < \os \leq \ot
\]
Note that this assumption implies that \(\os > 2\) and that it is clearly correct for \(j = 2\). When \(2\vbjm<\vjm\) follows from \cref{FRB7}  that \(\ibjm=\ilibjp\) since \(\vbjm > \Ljm\). This implies that:
\[
\uone(\ibone) 
=\ujm (\ibjm)
=\ujm (\ilibjp)
=\uj (\ibj)\quad\intref{BS11}
\]
When \(2\vbjm > \vjm\) then \(\ibjm=\ihibj\) since \(\vbj>\Lj\). This implies that:
\[
\uone(\ibone) 
= \ujm (\ibjm) 
= \ujm(\ihibj) 
= \uj (\ibj)
\]
We can exclude the case \(2\vbjm = \vjm\) since this would imply according to \cref{BA14} that \(\vbjm\) equals one which in turn would imply that \(j\) minus one equals \(\os\).
Finally, with \(j = \os - 1\) we get:
\[\uone(\ibone)=\uosm(\ibosm)\]
The case where \(\vbos = 1 > \Los\) can be proven in almost exactly the same way.
\end{proveit}
Now we are ready to show how \(\uone(\ibone)\) can be computed from the diff-mod sequence of \(u\).
\begin{frameit}
\begin{theo}\label{FRB9}
Given \(\vbos = 1 > \Los\)
\[\uone(\ibone) = 0\]
\end{theo}
\begin{theo}\label{FRB10}
Given \(\vbos \leq \Los\)
\leavevmode
\begin{itemize}
\item
\(2\vbosm<\vosm:\quad \uone(\ibone)=\vbos\)
\item
\(2\vbosm>\vosm:\quad \uone(\ibone)=\vosm-\vos\ceilfrac{\vosm}{\vos+\Losm}\)
\item
\(2\vbosm\neq\vosm\)
\end{itemize}
\end{theo}\end{frameit}
\begin{proveit}{FRB9}\ \\
The sequence defined by \(\uos(i)\) becomes in this case:
\[1, 2, ..., \vos - 1, 0, 1, ...\]
This implies that the ratio \(\uos (i)/i\) is one for \(0 < i < \vos\) and zero when \(i\) equals \(\vos\). This gives us:
\[\intref{FRB8} \quad \uone (\ibone) = \uos (\ibos) = \uos (\vos) = 0\]
\end{proveit}
\begin{proveit}{FRB10}
\begin{align*}
\intertext{When  \(2\vbosm<\vosm\) follows from
\cref{FRB5} that \(\ibosm=\ilk\) where \(k > 1\) since \(\vbosm > \Losm\) and from \cref{FRB6} follows that \(k = 2\) since:}
\frac{\uos(1)}{1} = \vbos \leq \Los \implies \frac{\uosm(\iltwo)}{\iltwo} \leq \Losm
\intertext{Thereby, we can conclude that:}
\intref{FRB8}\quad \uone(\ibone) = \uosm(\ibosm) = \uosm(\iltwo) = \uos(1) = \vbos\quad\intref{BS11}
\intertext{When \(2\vbosm>\vosm\) follows that:}
\frac{\uos(1)}{1}= \vbos \leq \Los \implies \frac{\uosm (\ihone)}{\ihone} \leq \Losm
\intertext{From \cref{FRB5} follows then:}
\uone(\ibone)=\uosm(\ibosm) = \vosm - \vos\ceilfrac{\vosm}{\vos+\Losm}
\end{align*}
We can exclude the case \(2\vbosm = \vosm\) altogther since \(\vbosm\) will equal one in this case according to \cref{BA14} which is not consistent with the definition of \(\os\).
\end{proveit}
Now when \cref{FRB10} gives us an efficient way to compute \(\uone(\ibone)\) from the diff-mod sequence of \(u\) we only need to apply \cref{FEB5} on the result in order to compute \(\ibone\), which equals \(\xb\). However, the case where \(\uone(\ibone)\) equals zero must be handled differently. Computing the index by means of \(\um\) will give the result zero. However in both cases the index of this element is \(q\).

Last in this section we will analyse the time complexity of the two algorithms we have presented above. Both of them computes the diff-mod sequence starting with the pair \((p, q)\) and then the desired result is given by means of closed form formulas based on specific pairs in this sequence. This implies that the time complexity of these algorithms equal the time complexity for computing the diff-mod sequence. We will show that the worst case time complexity for computing this sequence is the same as the Euclidean algorithm with positive remainders. 

The Euclidean algorithm finds the greatest common divisor of two integers. If we assume that these integers are \(p\) and \(q\), then according to Donald E. Knuth \cite{Knuth1981} the worst case time complexity of the Euclidean algorithm is \(O(\log(min(p, q)))\), i.e. it is a logarithmic function of \(p\) in our case as we assume that \(p\) is less than \(q\). This algorithm can be expressed by means of the following sequence:
\[
\omipp = \mod{\omi}{\omip} \text{ where } \omone = p,\; \omtwo = q \text{ and } \omk = 0
\]
The greatest common divisor of \(p\) and \(q\) then equals \(\omkm\).

Each pair in the diff-mod sequence can be computed by means of a constant number of elementary operations. Therefore, the time complexity for computing the entire diff-mod sequence is decided by the number of pairs in this sequence. We will now show that the number of pairs in the diff-mod sequence, starting with the pair \((p, q)\), are less than the number of items in the sequence computed by the Euclidean algorithm when finding the greatest common divisor of \(p\) and \(q\), \(\omiseq\). This will prove that computing these sequences has the same time complexity. 

We start by assuming that the pair \((\vbj, \vj)\) in this diff-mod sequence equals two consecutive items in \(\omiseq\):
\begin{align*}
\vj = \omi\\
\vbj = \omip
\end{align*}
We refer to this condition as condition A which obviously is fulfilled when \(j\) equals one. We will now investigate how \((\vbjp, \vjp)\) can be derived when condition A is fulfilled and \(2\vbj\) is greater than \(\vj\). When analysing this, we will use the well-known fact that:
\[
\mod{a}{b} = a - b \text{ if }2b > a > b
\]
In this case \(2\vbj > \vj\)  implies that \(2\omip > \omi\) which gives us:
\[
\omipp = \mod{\omi}{\omip} = \omi - \omip
\]
\begin{align*}
\vjp = &\vj - \vbj\\
= & \omi - \omip\\
= & \mod{\omi}{\omip}\\
= &\omipp\\
\vbjp = &\mod{\vj}{\vjp}\\
= & \mod{\omi}{\omipp}\\
= & \mod{\omi - \omipp}{\omipp}\\
= & \mod{\omip}{\omipp}\\
= & \omippp
\end{align*}
This implies that \((\vbjp, \vjp)\) equals two consecutive items in \(\omiseq\), i.e. condition A is fulfilled.

Next we will analyse how \((\vbjp, \vjp)\) can be derived when condition A is fulfilled and \(2\vbj\) is less than or equal to \(\vj\):
\begin{align*}
\vjp = &\vbj = \omip\\
\vbjp = &\vjp - \mod{\vj}{\vjp}\\
= &\omip - \mod{\omi}{\omip}\\
= &\omip - \omipp
\end{align*}
In this situation \(\vjp\) equals one item in \(\omiseq\) and \(\vbjp\) equals the difference between this item and the next one. We refer to this condition as condition B.

We will now investigate how \((\vbjpp, \vjpp)\) can be derived when condition B is fulfilled and we start with the case where 
\(2\vbjp\) is strictly less than \(\vjp\) which gives us that:
\[
2(\omip - \omipp) < \omip
\implies 2\omipp > \omip >\omipp\\
\]
\begin{align*}
\vjpp = &\vbjp = \omip - \omipp\\
= &\mod{\omip}{\omipp}\\
= &\omippp\\
\vbjpp = &\vjpp - \mod{\vjp}{\vjpp}\\
= &\omippp - \mod{\omip}{\omippp}\\
= &\omippp - \mod{\omip - \omippp}{\omippp}\\ 
= &\omippp - \mod{\omipp}{\omippp}\\
= &\omippp - \omipppp
\end{align*}
This implies that condition B is fulfilled.
\\

Next we will investigate how \((\vbjpp, \vjpp)\) can be derived when condition B is fulfilled and \(2\vbjp\) is strictly greater than \(\vjp\) which gives us that:
\begin{align*}
\vjpp = \vjp - \vbjp = \omip - (\omip - \omipp) = \omipp\\
\vbjpp = \mod{\vjp}{\vjpp} = \mod{\omip}{\omipp} = \omippp
\end{align*}
This implied that condition A is fulfilled.\\

Now remains to consider how \((\vbjpp, \vjpp)\) can be derived when condition B is fulfilled and \(2\vbjp\) equals \(\vjp\). In this case, \(\vbjp\) equals one and \(\vjp\) equals two according to \cref{BA14}, i.e. we have reached the second last pair in the diff-mod sequence. From the relationship between \(\vbjp, \vjp, \omip\) and \(\omipp\) given by condition B we can conclude that \(\omip\) equals two and \(\omipp\) equals one which in turn gives that \(\omippp\) equals zero, i.e. the last item in \(\omiseq\). Since \((\vbjpp, \vjpp)\) is the last pair in the diff-mod sequence and equal to \((1, 1)\) according to \cref{BS9} this implies that:
\begin{align*}
\vjpp = \omipp\\
\vbjpp = \omipp - \omippp
\end{align*}

Hence, we have proven that all pairs in the diff-mod sequence can be derived from items in \(\omiseq\) as long as we have not run out of items in this sequence. Since \(p\) and \(q\) are coprimes, the two last items in \(\omiseq\) will be one followed by zero. Looking at the relationships between the pairs in the diff-mod sequence and the items in \(\omiseq\) and noting that \(\vj\) is nonzero for all \(j\) less than or equal to \(\ot\), it is not difficult to realize that the two last items in \(\omiseq\) will result in the pair \((0, 1)\), or \((1, 1)\) i.e. the last pair in the diff-mod sequence. It is also trivial to show that the two last items in \(\omiseq\) are the only subsequent items in this sequence which can result in the last pair in the diff-mod sequence. Furthermore, we have shown that every time we compute a new pair in the diff-mod sequence, we involve an item in \(\omiseq\), which we have not involved before. This implies that the number of pairs in the diff-mod sequence is less than or equal to the number of items in \(\omiseq\). Hence, we have proven that the algorithms presented above have the same worst case time complexity as the Euclidean algorithm.
\section{The smallest integer in the codomain of \(\oG\) per residue class}\label{FM}
In this section, we lay the foundation for the algorithm, solving the Frobenius problem in three variables, presented in this article. Below we define this problem formally. In addition, we state an assumption about the set \(A\) determining the Frobenius number, denoted \(g(A)\), i.e. the number we are searching for. Due to a well-known theorem, discovered by Johnson \cite{Johnson1960}, we can make this assumption without loss of generality.
\begin{frameit}
\begin{defi}\label{IN2}
\ \\
Given the set of integers \(A=\{\aone,\atwo,\athree\}\) where \(1 < \aone < \atwo < \athree\) the Frobenius problem in three variables is to find the greatest integer \(g(A)\) which is not in the codomain of \(\oG\) defined by:
\[\oG(X)=\aone\xone + \atwo\xtwo + \athree\xthree \text{ where }\xi\text{ are non-negative integers for all } i\]
\end{defi}
\begin{assu}\label{IN12}
The three integers of the set \(A\) are pairwise coprimes.
\end{assu}
\end{frameit}
The algorithm presented in this article is based on the following theorem for computing the Frobenius number, which is also the foundation of the algorithm, finding the same number, invented by Tripathi \cite{Tripathi2017}.
\begin{frameit}
\begin{defi}\label{IN4}
\leavevmode
\begin{itemize}
\item
\amtwo\ is the multiplicative inverse of \atwo\ modulo \aone\ where \(0 < \amtwo <\aone\).
\item
\(\azero = \mod{-\amtwo\athree}{\aone}\)
\item
\(F(n,r) = \atwo\mod{\azero n-r}{\aone} + \athree n\)
\end{itemize}
\end{defi}
Note that the assumption that \aone\ and \atwo\ are coprimes guarantees the existence of \amtwo.
\begin{theo}\label{IN5}
Tripathi's theorem
\[g(A)=
\underset{0<r<\aone}{max}\bigl(\underset{0 \leq n<\aone}{min}F(n,r)\bigl)-\aone\]
\end{theo}
\end{frameit}
Tripathi's theorem, is based on that the following expression gives us the smallest integer in the codomain of \(\oG\) belonging to a specific residue class modulo \(\aone\) and that the residue class is distinct for each integer value of \(r\) in the interval \([0, \aone)\):
\[\underset{0<n<\aone}{min}F(n,r)\]
As \cref{IN5} (Tripathi's theorem) states, the greatest of these minima minus \(\aone\) then gives us the Frobenius number. 

In this section, we will derive a sequence from which all these minima can be found. Below we define this sequence and prove a theorem stating how this is done. We also prove a few lemmas concerning this sequence which we will use in the continued analysis.
\begin{frameit}
\begin{defi}\label{FM1}
The sequence \niseq\ is defined by the following conditions:
\begin{itemize}
\item
\(h(n) = \mod{\azero n}{\aone}\)
\item
\(f(n) = \atwo h(n) + \athree n\)
\item
\(f(\nim)<f(\ni)\leq f(n)\;\forall\;n\text{ where } h(n)>h(\nim)\)
\item
h(\ni)\textgreater h(\nim)
\item
f(\ni)\textless\aone\atwo
\item
\nzero=0
\item
\nm\ is the last element in this sequence.
\end{itemize}
\end{defi}
\begin{lemm}\label{FM24}
\begin{align*}
f(n)\neq \aone\atwo\\
\athree+\atwo\azero \neq \aone\atwo
\end{align*}
\end{lemm}
\begin{lemm}\label{FM27}
\[\ni < \aone\]
\end{lemm}
\begin{theo}\label{FM2}
\begin{gather*}
\underset{n}{min}\;F(n,r)=
\begin{cases}
f(\ni)-\atwo r &\text{if } h(\nim)<r\leq h(\ni)\\ 
\aone\atwo-\atwo r &\text{if } h(\nm) < r < \aone
\end{cases}\\
\text{where } 0 < i \leq m
\end{gather*}
\end{theo}
\end{frameit}
\begin{proveit}{FM24} Proof by contradiction. \\
We assume that:
\[
f(n) = \atwo h(n) + \athree n = \aone\atwo
\]
This would imply that \atwo\ divides \(n\), as gcd(\atwo,\athree) = 1, which is absurd since \(n < \aone < \atwo\).

In addition:
\[f(1) = \athree + \atwo\azero\]
\end{proveit}
\begin{proveit}{FM27}\ \\
This lemma follows trivially from the fact that \(f(\ni)\) is less than \(\aone\atwo\) and that:
\[f(n) \geq \athree n > \atwo n\]
\end{proveit}
\begin{proveit}{FM2}
\begin{align*}
F(n, r) = & \atwo\mod{\azero n -r}{\aone} + \athree n\\
\intref{IN9} = & \atwo \mod{h(n) - r}{\aone} + \athree n
\intertext{\bullet\(h(n) \geq r \implies 0 \leq h(n) -r < \aone\)}
F(n, r) = & \atwo h(n) - \atwo r + \athree n \quad\intref{IN8}\\
= & f(n) - \atwo r
\intertext{\bullet \(h(n) < r \implies 0 < h(n) - r + \aone < \aone\)}
F(n, r) = & \atwo \mod{h(n) - r + \aone}{\aone} + \athree n\\
= & f(n) + \aone\atwo - \atwo r
\end{align*}
By realizing that the minimum of \(F(n,r)\) for the case \(h(n) < r\) is given by setting n to zero we get the following result:
\[
\underset{0<n<\aone}{min}F(n,r) = min
\Bigl(\underset{h(n)\geq r}{min}f(n),\aone\atwo\Bigr)-\atwo r
\]
(Note that \cref{FM24} guarantees that there is no risk of a tie when choosing the minima of \(f(n)\) and \aone\atwo.)

This implies that provided that there is an integer \(\none\) such that \(f(\none)\) is less than \(\aone\atwo\)
and \(min\;f(n)\) equals \(f(\none)\) for \(h(n)\) greater than 0 then:
\[
\underset{n}{min}\;F(n, r) = f(\none) - \atwo r\text{ for } 0<r\leq h(\none)
\]
Provided that there is yet another integer \(\ntwo\) such that \(f(\ntwo)\) is less than \(\aone\atwo\) and \(min\;f(n)\) equals \(f(\ntwo)\) for \(h(n)\) greater than \(h(\none)\) then the minima for \(h(\none)<r\leq h(\ntwo)\) equals:
\[f(\ntwo) - \atwo r\] Continuing along these lines we derive a sequence of integers \(\none, \ntwo, ... \nm\), where \(\nm\) is the last remaining integer \(n\) for which \(f(n)\) is less than \(\aone\atwo\) which implies that:
\[
h(\nm)=\underset{i}{max}\;h(\ni)\]
From this sequence we can calculate all minima of \(F(n, r)\) for a specific r where \(0<r\leq h(\nm)\). Note that there is no minima defined for \(r=0\). This since the goal is to find the max of \(\underset{n}{min}\; F(n,r)\) for \(0<r<\aone\).

The remaining minima for \(h(\nm)<r<\aone\) is given by:
\[\aone\atwo - \atwo r\]

However, before moving on, we need to ask ourselves whether the sequence \niseq\ is uniquely defined. When selecting the numbers of this sequence, is there a possibility of ties, i.e. can a pair of different integers \(n\) and \(\nt\) exist such that \(f(n)\) equals \(f(\nt)\)? To prove that this cannot happen we assume without loss of generality that \(n<\nt\). That \(f(n)\) is less than \(f(\nt)\) if \(h(n)\leq h(\nt)\) follows trivially. For the case \(h(n)\geq h(\nt)\) \cref{BA9} gives:
\[
f(\nt)-f(n) = f(\oD)-\aone\atwo\neq 0 \text{ where } \oD = \nt-n
\]
\end{proveit}
Next we will analyse which integers the sequence \niseq\ consists of. To figure this out we will analyse the behaviour of \(f(n)\). Recall that \(h(n)\) is an ARM sequence. Analysing the behaviour of \(f(n)\), we will use the properties of this sequence type extensively. However, to do that we must first prove that \(h\)  fulfills \cref{BA4}.
\begin{frameit}
\begin{lemm}\label{FM22}
\(gcd(\aone,\azero) = 1\text{ and }0 < \azero < \aone\)
\end{lemm}
\end{frameit}
\begin{proveit}{FM22}\ \\
Proving this lemma, we will use the well-known lemmas below where \(\am\) is the multiplicative inverse of \(a\) modulo \(b\):
\begin{align*}
gcd(a, b) = 1 \implies & gcd(\am,b)=1\\
gcd(a,b) = gcd(a,c) = 1 \implies  & gcd(a,bc) = 1
\end{align*}
\begin{align*}
gcd(\amtwo, \aone) = 1 = & gcd(\athree, \aone)\\
= & gcd(\amtwo\athree, \aone)\\
= & gcd(-\amtwo\athree, \aone)\\
\intref{IN10}\quad = & gcd(\mod{-\amtwo\athree}{\aone}, \aone)\\
= & gcd(\azero,\aone)
\end{align*}
From \cref{IN11}, recalling that \aone\ is greater than one, follows that:
\[\azero = \mod{-\amtwo\athree}{\aone} > 0\]
\end{proveit}
Recall that ARM sequences are increasing or decreasing within intervals defined by lower and upper borders. We will denote these borders \(\nli\) and \(\nhi\) respectively for the sequence defined by \(h\). Next we will prove a lemma telling whether \(f(n)\) is increasing or decreasing within these intervals.
\begin{frameit}\begin{lemm}\label{FM3} Given that \(\nli\leq n<\nhi:\)
\begin{gather*}
\intertext{\bullet 2\azero\textless\aone}
f(n+1)>f(n)\\
\intertext{\bullet 2\azero\textgreater\aone}
f(n+1)>f(n)\\
\iff\\
\athree+\atwo\azero>\aone\atwo
\end{gather*}
\end{lemm}\end{frameit}
\begin{proveit}{FM3}\ \\
That \(f\) is increasing within a subsequence of \(h\) when \(h\) is increasing follows trivially. Whether \(f\) is increasing or decreasing when \(h\) is decreasing depends on the relative size of \(\athree + \atwo\azero\) and \(\aone\atwo\) since:
\begin{align*}
f(n+1)-f(n)= & \athree +\atwo(h(n+1) - h(n))\\
= &\athree +\atwo(h(n) - (\aone-\azero)- h(n))\\
= & \athree +\atwo\azero - \aone\atwo
\end{align*}
\end{proveit}
Knowing under which circumstances \(f(n)\) is increasing and decreasing, we are ready to show which integers \niseq\ consists of for two special cases paving the way for the analysis of the remaining case. The first special case is the case where the sum \(\athree+\atwo\azero\) is greater than \(\aone\atwo\).
\begin{frameit}\begin{lemm}\label{FM4}
\[\athree+\atwo\azero>\aone\atwo\implies\niseq=(0)\]
\end{lemm}
\begin{lemm}\label{FM25}
\[
2\azero = \aone \implies \athree+\atwo\azero > \aone\atwo
\]
\end{lemm}
\end{frameit}
\begin{proveit}{FM4}\ \\
It is trivial to realize that zero is part of the sequence \niseq\ and, from the fact that \(f(1)\) equals \(\athree + \atwo\azero\), that one is not if \(\athree+\atwo\azero\) is greater than \(\aone\atwo\). Next we will show that none of the integers greater than one is part of this sequence. Showing that we will consider the following three cases separately: \(2\azero<\aone\), \(2\azero\ = \aone\) and  \(2\azero>\aone\).

For increasing sequences \(f\) is increasing within a subsequence of \(h\) according to \cref{FM3}. This implies that \(f(n)\) is greater than \(\aone\atwo\) for all \(n\) in the first subsequence greater than zero. We will now prove that \(f(\nli)\) is greater than \(\aone\atwo\) for all \(i\) greater than one, which implies that the only integer in the sequence \(\niseq\) is zero. The proof goes as follows:
\begin{align*}
f(\nlip)-\aone\atwo= &\athree\nlip+\atwo h(\nlip)-\aone\atwo\\
\intref{BA34}= &\frac{\athree(h(\nlip)+i\aone)}{\azero}+\atwo h(\nlip)-\aone\atwo\\
= &\frac{(\athree+\atwo\azero) h(\nlip)+\aone(i\athree-\atwo\azero)}{\azero}\\
\biggl\{&\athree>\atwo(\aone-\azero)>\atwo\azero\biggr\}\\
> &\frac{(\athree+\atwo\azero) h(\nlip)+\aone\atwo\azero(i-1)}{\azero}\\
> &0\quad\forall\;i>0
\end{align*}
In the computation above and in the continued analysis, \azero\ appears as denominator. Therefore, we have to ensure that  \azero\ is not equal to zero which luckily follows from \cref{FM22}.\\
\\
For decreasing sequences \(f\) is increasing within a subsequence of \(h\) iff \(\athree+\atwo\azero\) is greater than \(\aone\atwo\). Furthermore, \(f(\nli)\) for \(i\) greater than one is greater than \(f(\nlone)\) since  \(h(\nlone)\) is less than \(h(\nli)\) according to \cref{BA20}. As \(f(\nlone)\) is greater than \(\aone\atwo\), follows that the sequence \(\niseq\) consists of only zero in this case as well.

Finally we have the case 2\azero = \aone\ which implies that \(\azero\) equals one and \(\aone\) equals two according to \cref{BA14}. Since \(\ni\) is less than \(\aone\) according to \cref{FM27}, \(\niseq\) consists only of zero also in this case.
\end{proveit}
\begin{proveit}{FM25}\ \\
\(2\azero = \aone\) implies that \(\azero\ = 1\) and \(\aone\ = 2\) according to \cref{BA14}\ which in turn implies that:
\[
\athree+\atwo\azero=\athree+\atwo>2\atwo=\aone\atwo
\]
\end{proveit}
In the remainder of this section we will analyse which integers \niseq\ consists of when \(\athree + \azero\aone\) is less than \(\aone\atwo\). (Note that \(\athree + \azero\aone\) cannot be equal to \(\aone\atwo\) according to \cref{FM24}.) We will as above often consider the two cases \(2\azero<\aone\) and \(2\azero>\aone\) separately. However, we will not have to consider the case \(2\azero=\aone\) as this implies that \(\athree + \atwo\azero\) is greater than \(\aone\atwo\) according to \cref{FM25}.
\begin{frameit}
\begin{assu}\label{FM23}\ \\
In this section from this point onwards, unless expressly stated otherwise:
\[
\athree + \atwo\azero < \aone\atwo
\] 
\end{assu}
\end{frameit}
The next special case we will address is the one where \(h(n)\) only has one subsequence. Addressing this case we will introduce a constant, denoted \nb, which will play a major role in the algorithm presented here.
\begin{frameit}
\begin{defi}\label{IN13}\ \\
\nb\ is the smallest integer \(n\geq 0\) such that \(f(n+1)>\aone\atwo\).
\end{defi}
\begin{lemm}\label{FM6}
\begin{align*}
2\azero<\aone:\quad & \azero = 1 \implies \niseq = (0, 1, 2, ..., \nb)\text{ where }\nb < \nhone\\
2\azero>\aone:\quad & \aone - \azero = 1 \implies \athree+\atwo\azero>\aone\atwo\\
\end{align*}
\end{lemm}
\end{frameit}
\begin{proveit}{FM6}\ \\
From \cref{BA20} follows for the case \(2\azero<\aone\):
\[
\nhone = \aone - 1
\]
Using \cref{BA20} and recalling that \athree\ is greater than \atwo, it is easy to show that:
\[
f(\nhone) = \athree\nhone + \atwo\nhone > 2\atwo(\aone-1) \geq \aone\atwo
\]
This implies that \(\nb\) is less than \(\nhone\). It also implies that \(f(n)\) is greater than \aone\atwo, if \(n\) is greater than \(\nb\) since \(f\) is increasing within a subsequence of \(h\). This also implies that \niseq\ equals:
\[
(0, 1, 2,... , \nb)
\]

For the case \(2\azero>\aone\) we get that \(\athree +\atwo\azero - \aone\atwo = \athree - \atwo(\aone-\azero)>0\).
\end{proveit}
Note that the sequence \niseq\ consists of all integers in the interval \([0, \nb]\) also for the special case where \(\athree + \atwo\azero\) is greater than \aone\atwo\ since \nb\ equals zero in this case (see \cref{FM4}). In the remainder of this section we will show that this sequence consists of these numbers also for the remaining cases, although they in general do not appear in ascending order. We will do this by analysing the characteristics of \(\nb\). If \(h\) is increasing and \(f(\nhi)\) is less than \(\aone\atwo\) then \(f(n)\) is less than \(\aone\atwo\) for all integers \(n\) in the interval \([\nli, \nhi]\) since \(f\) is increasing in this interval according to \cref{FM3}. Therefore, it is interesting to know for which integers \(i\) \(f(\nhi)\) is less than \(\aone\atwo\) when analysing \(\nb\). As we will show further down, we can get \(h(\nhi)\) from the border sequence of \(\hs\). For the analogue reason it is interesting to know for which integer \(i\) \(f(\nli)\) is less than \(\aone\atwo\) when \(h\) is decreasing. We will further down show that \(h(\nli)\) also can be obtained from the border sequence of \(\hs\) and the continued analysis in this section will be based on this sequence. We denote the function defining this sequence \(e\) and define it below together with a lemma stating that \(e\) defines  the border sequence of \(\hs\).
\begin{frameit}
\begin{defi}\label{FM7}
\begin{align*}
e(i)=\mod{\oab i}{\oa} \text{ where }\left\{
\begin{aligned}
2\azero<\aone:\quad
&\oa=\azero\\
&\oab=\mod{\aone}{\oa}\\
\\
2\azero>\aone:\quad
&\oa=\aone-\azero\\
&\oab=\oa-\mod{\aone}{\oa}
\end{aligned}\right.
\end{align*}
\end{defi}
\begin{lemm}\label{FM29}
\begin{align*}
2\azero<\aone:& \quad
\hs(\nhi)= e(i)
\\
2\azero>\aone:& \quad\hs(\nlip)= e(i)
\end{align*}
\end{lemm}
\end{frameit}
\begin{proveit}{FM29}\ \\
The proof of this lemma follows from \cref{BS11} noting that the difference of the sibling sequence equals \(\aone\) minus \(\azero\).
\end{proveit}
Excluding the cases handled by \cref{FM6}, we can assume that \(\oa\) is greater than one which implies that \(e\) fulfills the criteria of \cref{BA4} according to \cref{BS9}. This implies in turn that all characteristics of ARM sequences presented here are valid for the sequence defined by \(e\). Below, we formally state our assumption about \(\oa\). In addition, we prove a lemma stating how the border values of \(h\) can be obtained from \(e\). Then we prove another lemma stating that when \(h\) is increasing then \(f(\nhi)\) is less than \(\aone\atwo\) if the ratio \(e(i)/i\) is less than a constant derived from \(A\), denoted \(\oo\). We also prove that the same condition decides wheather \(f(\nli)\) is less than \(\aone\atwo\) when \(h\) is decreasing. Finally, we prove  that the constant \(\oo\) is greater than one. A fact we will use later on.
\begin{frameit}
\begin{assu}\label{FM26}\ \\
In this section from this point onward, unless expressly stated otherwise:
\[\oa>1\]
\end{assu}
\begin{lemm}\label{FM8}
\begin{align*}
\left.\begin{aligned}2\azero<\aone:\quad &\begin{aligned}[t]
h(\nlip)=&\oa-e(i)\\
h(\nhi)=&\aone-e(i)\\
\end{aligned}\\
\\
2\azero>\aone:\quad &\begin{aligned}[t]
h(\nlip)=&\aone-e(i)\\
h(\nhi)=&\oa-e(i)\\
\end{aligned}\end{aligned}\right\}0<i<\oa
\end{align*}
\end{lemm}
\begin{defi}\label{FM10}
\begin{align*}
\oo = \frac{\aone\athree}{\ob}
\text{ where }\left\{
\begin{aligned}
2\azero<\aone:\quad\ob=\atwo\oa+\athree\\
2\azero>\aone:\quad\ob=\atwo\oa-\athree
\end{aligned}\right.\\
\end{align*}
\end{defi}
\begin{lemm}\label{FM9}
\begin{gather*}
\begin{aligned}
2\azero<\aone:\quad & f(\nhi)<\aone\atwo\\
2\azero>\aone:\quad & f(\nlip)<\aone\atwo\\
\end{aligned}
\iff
\frac{e(i)}{i} > \oo
\text{ where }0<i<\oa
\end{gather*}
\begin{align*}
\frac{e(i)}{i} & \neq \oo\;\forall \;i\\
\oab & \neq \oo
\end{align*}
\end{lemm}
\begin{lemm} \label{FM11}
\[\oo > 1 \]
\end{lemm}
\end{frameit}
\begin{proveit}{FM8}
When \(2\azero < \aone\) follows from \cref{BA16}:
\[
h(\nhi) = \aone - \hs(\nhi) = 
\aone - e(i) \quad \intref{FM29}
\]
From \cref{BA20} follows:
\[
h(\nlip) = h(\nhi) - \aone + \azero = \oa - e(i)
\]
The proof for the case \(2\azero>\aone\) goes along the same lines.
\end{proveit}
\begin{proveit}{FM9}\ \\
We prove this lemma for the case \(2\azero > \aone\). The case \(2\azero < \aone\) is proven in an analogous manner.
\begin{align*}
f(\nlip)-\aone\atwo= &\athree\nlip+\atwo h(\nlip)-\aone\atwo\\
\intref{BA34}= &\frac{\athree((i+1)\aone - h(\nlip))}{\aone-\azero}+\atwo h(\nlip)-\aone\atwo\\
= &\frac{(\atwo\oa-\athree)(h(\nlip)-\aone)+i\aone\athree}{\oa}\\
\intref{FM8}= &\frac{-\ob e(i)+i\aone\athree}{\oa}
\intertext{This allows us to conclude that:}
&\frac{e(i)}{i}>\frac{\aone\athree}{\ob}=\oo\iff f(\nlip)<\aone\atwo
\end{align*}
In addition, as \(f(n) \neq \aone\atwo\) according to \cref{FM24}, we can conclude that:
\[\frac{e(i)}{i}\neq\oo\]
Setting \(i\) equal to one, we get:
\[\frac{e(1)}{1}= \oab\neq\oo\]
\end{proveit}
\begin{proveit}{FM11}\ \\
For the case \(2\azero < \aone\) we get:
\[
\oo = \frac{\aone\athree}{\athree + \atwo\azero} >
\frac{\aone\atwo}{\athree + \atwo\azero} >
\frac{\aone\atwo}{\aone\atwo} = 1
\]
For the case \(2\azero > \aone\) we note that:
\[\atwo(\aone - \azero) - \athree = \aone\atwo - (\athree + \atwo\azero) > 0
\]
This gives us:
\[
\oo = 
\frac{\aone\athree}{\atwo(\aone - \azero) - \athree} >
\frac{\aone\athree}{\atwo(\aone - \azero) + \athree} >
\frac{\aone\athree}{\atwo\azero + \athree} > 1
\]
\end{proveit}
According to \cref{FM9} the first local maxima of \(f\), such that the value of \(f\) is greater than \(\aone\atwo\), is given by the smallest integer \(i\) such that \(e(i)/i\) is less than \(\oo\). (Note that \(e(i)/i\) can not be equal to \(\oo\) according to \cref{FM9}.) If we denote this integer \(\ib\) then this local maximum is at \(\nhib\) for increasing sequences and at \(\nlib\) for decresing ones. For decreasing sequences it is easy to realize that \(\nlib\) is not only the first lower border such  that \(f(\nli)\) is greater than \(\aone\atwo\) but also the first integer \(n\) such that \(f(n)\) is greater than \(\aone\atwo\), since \(f\) is decreasing in the interval \([\nlip, \nhip]\). For increasing sequences we can show that \(f(\nhi)\) is greater than \(f(n)\) for all \(n\) in the interval \([\nlip, \nhip)\). This gives us two cases when \(h\) is increasing. Either \(n\), such that \(f(n)\) is greater than \(\aone\atwo\), is not greater than \(\nhone\) or it equals \(\nhib\). Next we define \(\ib\) formally and prove a theorem expressing \nb\ in terms of \(\ib\). Note that this theorem holds for all values of \(\oa\). It will, besides helping us to show that \niseq\ consists of all integers in the interval \([0, \nb]\), also form the foundation of the algorithm finding \(\nb\), described in \cref{FN}. We also show that \(\ib\) is less than \(\oa\), i.e. the number of subsequences of \(h\). This implies that \(\nb\) is less than \(\aone\), which it must be to be part of \niseq\ according to \cref{FM27}.
\begin{frameit}
\begin{defi}\label{FM12}
\ib\ is the smallest integer i\textgreater 0 such that:
\[\frac{e(i)}{i}<\oo\]
\end{defi}
\begin{theo}\label{FM13}
\begin{align*}
\intertext{Given that 2\azero\textless\aone}
\oab<\oo &\quad\left\{\begin{aligned}
1 \leq \nb=\ceilfrac{\aone\atwo}{\ob} - 1<\nhone\\
f(n)>\aone\atwo\text{ for }\nb<n\leq\nhone
\end{aligned}\right.\\
\oab>\oo &\quad\nb=\nhib-1
\intertext{Given that 2\azero\textgreater\aone}
\oab<\oo &\quad \nb = \nhone = \flooraoneoa\\
\oab>\oo &\quad\nb=\nhib
\end{align*}
\end{theo}
\textbf{Note} that this theorem is valid for all values of \(\oa\).
\begin{lemm}\label{FM14}
\[\ib < \oa\]
\end{lemm}
\end{frameit}
\begin{proveit}{FM13}\ \\
We start with the case \(2\aone < \azero\). From \cref{FM6} follows when \(\oa\) is equal to one that \nb\ is less than \(\nhone\) and that:
\[\oab = \mod{\aone}{\oa} = 0 \]
This implies that \(\oab\) is less than \(\oo\) as \(\oo\) is greater than one according to \cref{FM11}. From \cref{FM9} follows when \
\(\oa\) is greater than one that:
\[
f(\nhone) > \aone\atwo \iff \frac{e(1)}{1} = \oab < \oo
\iff \ib = 1
\]
All in all, \(\nb\) is less than \(\nhone\) iff \(\oab\) is less than \(\oo\) in which case \(f(n)<\aone\atwo\) if: 
\[n<\frac{\aone\atwo}{\athree + \atwo\azero}\text{ since }f(n) = \athree n +\atwo\azero n\text{ for }n\leq\nhone\quad \intref{BA34}\]
This implies that:
\[
\nb = \ceilfrac{\aone\atwo}{\ob} - 1 \geq 1
\]

In addition, we can conclude that \(f(n)>\aone\atwo\) if \(\nb<n\leq\nhone\).

We will now turn to the case \(\oab>\oo\) which implies that \(\ib>1\) (note that \(\oab\neq\oo\) according to \cref{FM9}). Hence, \(\nhibm\) is defined. As \(f(\nhi)\) is less than \(\aone\atwo\) if \(i\) is less than \(\ib\) we can conclude that \(f(n)\) is less than \(\aone\atwo\) if \(n\) is not greater than \(\nhibm\). We will now show that \(f(\nhib-1)\) is less than \(\aone\atwo\) which implies that \(f(n)\) is less than \(\aone\atwo\) if \(n\) is not greater than \(\nhib-1\). Hence \(\nb\) equals \(\nhib-1\) since \(f(\nhib)\) is greater than \(\aone\atwo\).
\begin{align*}
f(\nhibm+\flooraoneazero)= &\athree(\nhibm+\flooraoneazero)+\atwo h(\nhibm+\flooraoneazero)\\
\intref{INT10}=& \athree(\nhibm+\frac{\aone-\mod{\aone}{\azero}}{\azero})+\atwo (h(\nhibm)-\mod{\aone}{\azero})\\
=& \athree\nhibm+\atwo h(\nhibm)+\frac{\aone\athree-(\athree+\atwo\azero)\mod{\aone}{\azero}}{\azero}\\
=& f(\nhibm)+\frac{\aone\athree-\ob\oab}{\oa}\\
&\Biggl\{\oab>\oo=\frac{\aone\athree}{\ob}\implies \aone\athree-\ob\oab<0\Biggr\}\\
&< f(\nhibm)<\aone\atwo
\end{align*}
The fact that \(\nhibm+\flooraoneazero\) equals \(\nhib\) or \(\nhib-1\) implies that it is equal to the latter since \(f(\nhib)>\aone\atwo\) and we can conclude \(f(\nhib-1)<\aone\atwo\).

Now we turn to the case \(2\aone>\azero\). The fact that \( f(1) = f(\nlone)\) is less than \(\aone\atwo\), that \(f(\nlip)\) is less than  \(\aone\atwo\) for \(i\) less than \(\ib\) according to \cref{FM9} and that \(f(n)\) is decreasing within a subsequence of \(h\), implies that \(f(n)\) is less than \(\aone\atwo\) if \(n\) is less than \(\nlibp\) whereas \(f(\nlibp)\) is greater than \(\aone\atwo\) which proves that \(\nb\) equals \(\nhib\). If \(\oab\) is less than \(\oo\) then \ib\ equals one and \cref{BA37} gives that: 
\[\nb = \nhone = \flooraoneoa\]
\end{proveit}
\begin{proveit}{FM14}
\[\frac{e(\oa - 1)}{\oa - 1}
\leq \frac{\oa - 1}{\oa - 1} = 1 
< \oo \quad \intref{FM11}\]
\end{proveit}
Next we will show that there is no integer greater than \nb\ in
the sequence \niseq. An integer \(\nt\) is not part of this sequence iff \(f(\nt)>\aone\atwo\) and/or there exists another integer \(\nu\) such that \(f(\nu) < f(\nt)\) and \(h(\nu) > h(\nt)\), which implies that \(\nu < \nt\). The next lemma states when two such elements exist, having the same position in two different subsequences of \(h\). We start by showing that if the first subsequence has the greatest lower border element then all elements in this subsequence are greater than the element with the same position in the second subsequence. Hence, it is relevant to compare the value of \(f\) at these elements. In addition, we show that when \(h\) is increasing then an element in the first subsequence is greater than an element with a lower position in the second sequence but less than an element with higher position. When \(h\) is decreasing the opposite is true. We will use this fact later on. Note that when \(h\) is increasing then the first subsequence is not longer than the second subsequence (see \cref{BA28}) since the lower border element of the first subsequence is greater than the lower border element of the second one. When \(h\) is decreasing then the first subsequence actually can be longer.
\begin{frameit}
\begin{lemm}\label{FM28}Given \(h(\nlj) > h(\nli)\), \(\nlj + \oD \leq \nhj\)  and \(\nli + \oDt \leq \nhi\)\\
\begin{align*}
2\azero < \aone:\quad
&\begin{aligned}[t]
h(\nlj+\oD) & > h(\nli+\oDt)\text{ if }\oD\geq\oDt \\
& < h(\nli+\oDt)\text{ if }\oD<\oDt\\
\end{aligned}\\
2\azero > \aone:\quad
&\begin{aligned}[t]
h(\nlj+\oD) & > h(\nli+\oDt)\text{ if }\oD\leq\oDt \\
&<h(\nli+\oDt)\text{ if }\oD>\oDt\\
\end{aligned}
\end{align*}
\end{lemm}
\begin{lemm}\label{FM15}
\ \\
\(\text{Given }\nljp<\nlip\text{ and } h(\nljp)>h(\nlip)\)
\begin{gather*}
\left.\begin{gathered}
f(\nljp+\oD)<f(\nlip+\oD)\\
\iff\\
\frac{e(i-j)}{i-j} < \oo\\
\end{gathered}\right\}
\begin{gathered}
0<i,j<\oa\\
\oD\geq 0
\end{gathered}
\\
\begin{aligned}
\text{where}\quad & \nljp+\oD\leq\nhjp\text{ if }2\azero<\aone\\
&\nlip+\oD\leq\nhip\text{ if }
2\azero>\aone&
\\
\end{aligned}\\
\end{gather*}
\end{lemm}\end{frameit}
\begin{proveit}{FM28}\ \\
When \(2\azero < \aone\) then \cref{BA20} gives us that:
\[h(\nlj + \oD) - h(\nli + \oDt) = h(\nlj) - h(\nli) + (\oD - \oDt)\azero\]
If \(\oD \geq \oDt\) then:
\[h(\nlj + \oD) - h(\nli + \oDt) 
> (\oD - \oDt) \azero \geq 0\]
If \(\oD < \oDt\) then \cref{BA19} gives:
\[h(\nlj + \oD) - h(\nli + \oDt)
< \azero - \azero = 0\]
When \(2\azero > \aone\) then we get:
\[
h(\nlj+\oD) - h(\nli + \oDt) = h(\nlj) - h(\nli) + (\oDt-\oD)(\aone - \azero)
\]
If \(\oD \leq \oDt\) then:
\[ h(\nlj+\oD) - h(\nli + \oDt)
> (\oDt-\oD)(\aone - \azero) \geq 0\]
If \(\oDt < \oD\) then:
\[h(\nlj+\oD) - h(\nli + \oDt)
< \aone - \azero - (\aone - \azero) = 0\]
\end{proveit}
\begin{proveit}{FM15}\ \\
We prove this lemma for the case \(2\azero < \aone\). The case \(2\azero > \aone\) is proven in an analogous manner. Note that according to \cref{BA28}, the length of subsequence i is longer than or equal to subsequence j since \(h(\nljp)\) is greater than \(h(\nlip)\) which implies that \(\nlip+\oD\) is not greater than \(\nhip\). This gives us:
\begin{align*}
f(\nljp+\oD)-f(\nlip+\oD)=& \athree(\nljp-\nlip)+\atwo(h(\nljp+\oD)-h(\nlip+\oD))\\
\intref{BA34},\intref{BA20}
 = &\athree\frac{h(\nljp)+j\aone-(h(\nlip)+i\aone)
}{\oa}\\
&+\atwo(h(\nljp)+\oD\oa-(h(\nlip)+\oD\oa))\\
=&\frac{(\atwo\oa+\athree)(h(\nljp)-h(\nlip))-(i-j)\aone\athree}{\oa}\\
\intref{FM8}= &\frac{\ob(e(i)-e(j))-(i-j)\aone\athree}{\oa}\\
&\left\{h(\nljp)>h(\nlip)\implies e(i)>e(j)\right\}\\
\intref{BA9}= &\frac{\ob e(i-j)-(i-j)\aone\athree}{\oa}\\
\\
\intertext{Noting that \(i-j>0\)
since \(\nljp<\nlip\) we can conclude that:}
\frac{e(i-j)}{i-j} < \frac{\aone\athree}{\ob}=\oo &\iff f(\nljp+\oD)<f(\nlip+\oD)
\end{align*}
\end{proveit}
Showing that there is no integer greater than \nb\ in the sequence \niseq, a comparison between two elements, having the same position in two different subsequence of \(h\), will never be relevant for the first subsequence, since \(h(\nlone)\) is less than \(h(\nli)\) for all \(i \neq 1\)(see \cref{BA20}). When \(h\) is increasing it instead makes sense to compare \(\nli + \oD\) with \(1 + \oD\) as \(h(\nli + \oD)\) is less than \(h(1+\oD)\). For equivalent reasons it makes sense to compare \(\nlip + 1 +\oD\) and \(1+\oD\) when \(h\) is decreasing.
\begin{frameit}\begin{lemm}\label{FM16}
\begin{gather*}
\left.\begin{aligned}
2\azero<\aone:\quad & f(1+\oD)
 < f(\nlip+\oD)\\
&\text{ where }\oD < \flooraoneoa\\
\\
2\azero>\aone:\quad & f(1 + \oD) < f(\nlip + 1 + \oD)\\
&\text{ where }\nlip+1+\oD\leq\nhip\\
\end{aligned}\right\} \begin{aligned}
0<i<\oa\\
\oD\geq 0
\end{aligned}\\
\iff\\
\frac{e(i)}{i} < \oo\\
\end{gather*}
\end{lemm}\end{frameit}
\begin{proveit}{FM16}\ \\
We prove this lemma for the case \(2\azero < \aone\). The case \(2\azero > \aone\) is proven in an analogous manner.
\begin{align*}
f(1+\oD)-f(\nlip+\oD)= &\athree(1-\nlip)+\atwo(h(1+\oD)-h(\nlip+\oD))\\
\mathclap{\left\{0\leq\oD<\flooraoneoa\implies\left\{\begin{aligned} & 1+\oD\leq\nhone\quad\intref{BA28}\\
& \nlip+\oD\leq\nhip\end{aligned}\right.\right\}}\\
\intref{BA34}= &\athree\left(1-\frac{h(\nlip) + i \aone}{\oa}\right)\\
& + \atwo(h(1)+\oD\oa-(h(\nlip)+\oD\oa))\\
=&\frac{(\atwo\oa+\athree)(\oa-h(\nlip))-i\aone\athree}{\oa}\\
\intref{FM8} = & \frac{\ob e(i)-i\aone\athree}{\oa}\\
\end{align*}
From this we can conclude that:
\[
\frac{e(i)}{i} < \frac{\aone\athree}{\ob} = \oo \iff f(1+\oD) < f(\nlip+\oD)
\]
\end{proveit}
Considering carefully the implications of \cref{FM15}, we realize that \(\nlip + \oD\) is less than \(\nhip\) in case the length of subsequence \(i + 1\) is one element longer than subsequence \(j+ 1\). However, this is only possible when \(h\) is increasing, since \(h(\nljp) > h(\nlip)\). The implications of \cref{FM16} are actually the same. The next lemma fills this gap by showing when \(f(\nhi) > f(\nhj)\). This lemma is easily proven following the same line of thought as \cref{FM15}. Therefore, the proof is left out here.
\begin{frameit}\begin{lemm}\label{FM17}
\ \\
\(\text{Given }\nhj<\nhi\text{ and } h(\nhj)>h(\nhi)\)
\begin{gather*}
\left.\begin{gathered}
f(\nhj) < f(\nhi)\\
\iff\\
\frac{e(i-j)}{i-j} < \oo\\
\end{gathered}\right\}0 < i,j < \oa
\end{gather*}
\end{lemm}\end{frameit}
The lemmas and theorems proven so far in this section form a good foundation for showing that all integers greater than \nb\ are not part of the sequence \niseq. However, there is missing one lemma gluing it all together and that is the lemma below.
\begin{frameit}\begin{lemm}\label{FM18}
Given \(0<i<\oa\) and  \(0<\ib+i<\oa\)
\begin{align*}
e(\ib)+e(i)\geq\oa\implies & \frac{e(\ib+i)}{\ib+i}<\oo\\
e(\ib)+e(i)<\oa\implies &\left\{\begin{aligned} h(\nlibip)< & h(\nlip)\\
h(\nhibi)< & h(\nhi)\end{aligned}\right.
\end{align*}
\end{lemm}\end{frameit}
\begin{proveit}{FM18}\ \\
If \(e(\ib)+e(i)\geq\oa\) then 
\(e(\ib+i)=e(\ib)+e(i)-\oa<e(\ib)\) according to \cref{BA8}. This implies that:
\[\frac{e(\ib+i)}{\ib+i}< \frac{e(\ib)}{\ib}<\oo\]
According to \cref{BA16}, \(
e(\ib) > 0\) since \(0 < \ib < \oa\) according to \cref{FM14}. If \(2\azero\) is less than \(\aone\) and the sum of \(e(\ib)\) and \(e(i)\) is less than \(\oa\) then this gives us:
\begin{align*}
h(\nlibip)= &\oa-e(\ib+i)\quad\intref{FM8}\\
= &\oa-(e(\ib)+e(i))\\
< & \oa-e(i)= h(\nlip)
\end{align*}
That \(h(\nhibi) < h(\nhi)\) in this case is proven in an analogous manner. The proof for the case \(2\azero > \aone\) is equivalent to the proof for \(2\azero < \aone\).
\end{proveit}
Now we are completely ready to show that an integer greater than \nb\ is not part of the sequence \niseq .

\begin{frameit}\begin{lemm}\label{FM19}
\[n > \nb \implies \ni \neq n\  \forall\ i\]
\end{lemm}\end{frameit}
\begin{proveit}{FM19}\ \\
We prove this lemma for the case \(2\azero < \aone\). The case \(2\azero > \aone\) is proven in an analogous manner. Recall that an integer \(\nt\) is not part of the sequence \niseq\ iff \(f(\nt)>\aone\atwo\) and/or there exists another integer \(\nu\) such that \(f(\nu)\) is less than \(f(n)\) and \(h(\nu)\) is greater than \(h(\nt)\). We start by considering the integers which can be expressed as:
\[
\nhibi \text{ where } 0 \leq i < \oa \text{ and } 0 < \ib+i \leq \oa
\]
When \(i\) equals zero we get \(\nhib\) which we can exclude from the sequence as \(f(\nhib)\) is greater than \(\aone\atwo\) according to \cref{FM9} since \(e(\ib)/\ib\) is less than \(\oo\). When \(\ib+i\) equals \(\oa\) we get \(\nhoa\) which we also can exclude since:
\begin{align*}
f(\nhoa)= &\athree\nhoa+\atwo h(\nhoa)\\
\intref{BA20} \quad = &\athree(\aone-1)+\atwo(\aone-\azero)\\
> &\atwo(\aone-1)+\atwo\\
= &\aone\atwo
\end{align*}
Now remains to consider:
\[
\nhibi \text{ where } 0<i<\oa \text{ and } 0<\ib+i<\oa
\]
If \(e(\ib)+e(i)\geq\oa\) then \(e(\ib+i)/(\ib+i)<\oo\) according to \cref{FM18} which implies that \(f(\nhibi)\) is greater than \(\aone\atwo\). On the other hand, if \(e(\ib)+e(i)\) is less than \(\oa\) then \(h(\nhibi)\) is less than \(h(\nhi)\). Furthermore, \(f(\nhibi)\) is greater than \(f(\nhi)\) according to \cref{FM17}. All in all, we can conclude that the following integers are not part of the sequence \niseq:
\[
\nhibi \text{ where } 0 \leq i<\oa \text{ and } 0 < \ib+i \leq \oa
\]
Next we will consider all integers which can be expressed as:
\[
\nlibip + \oD \text{ where } 0 \leq \oD < \flooraoneoa \text{, } 0 \leq i < \oa \text{ and } 0 < \ib+i < \oa
\]
Note that \cref{FM28} gives:
\[
h(\nlibip+\oD) < h(\nlone + 1 +\oD) = h(1+\oD)
\]
If \(i=0\) we get \(\nlibp + \oD\) which can be exclude from \(\niseq\) since:
\[
f(\nlibp+\oD) > f(1+\oD) \quad \intref{FM16}
\]
Next we will consider the case where \(0< i<\oa\). If \(e(\ib)+e(i)\geq\oa\) then:
\[
e(\ib+i)/(\ib+i) < \oo
\]
This implies that:
\[
f(\nlibip+\oD) > f(1+\oD)
\]
On the other hand, if \(e(\ib) +  e(i) < \oa\) then \(h(\nlibip) < h(\nlip)\) which implies that:
\[
f(\nlibip+\oD) > f(\nlip+\oD) \quad \intref{FM15}
\]
Note that \(\oD < \flooraoneoa\) implies that \(\nlip+\oD\leq\nhip\). All in all, we can conclude that the following integers are not part of the sequence \niseq:
\[
\nlibip + \oD \text{ where } 0 \leq \oD < \flooraoneoa, 0 \leq i<\oa \text{ and } 0<\ib+i<\oa
\]
To sum it up, all integers we have excluded from \niseq\ jointly comprise all integers \(n\) such that:
\[
\nhib \leq n \leq \nhoa = \aone-1
\]
If \(\oab\) is greater than \(\oo\) then these comprise all integers greater than \nb\ according to \cref{FM13}. If \(\oab\) is less than \(\oo\) then we can also exclude the following integers \(n\) from \niseq\ since \(f(n)\) is greater than \(\aone\atwo\) for these:
\[
\nb < n \leq \nhone = \nhib
\]
This implies that we can exclude all integers greater than \(\nb\) from \niseq\ also in this case. Recalling that \(\oab\neq\oo\), we can conclude that no integer greater than \(\nb\) is part of \niseq\ when \(2\azero\) is less than \(\aone\).
\end{proveit}
Finally, to prove that the sequence \niseq\ consisits of the integers in the interval \([0, \nb]\) and only these integers, we show that all of these are part of the sequence.
\begin{frameit}\begin{lemm}\label{FM20}
\[n\leq\nb\implies\exists\;\ni=n\]
\end{lemm}\end{frameit}
\begin{proveit}{FM20}\ \\
An integer \(\nt\) is included in the sequence \niseq\ iff \(f(\nt)\) is less than \(\aone\atwo\) and there is no integer \(\nu\) such \(h(\nu)\) is greater than \(h(\nt)\) and \(f(\nu)\) is less than \(f(\nt)\). We assume that \(h(\nu)\) is greater than \(h(\nt)\) and that \(\nt\) is not greater than \(\nb\). That \(f(\nt)\) is less than \(\aone\atwo\) follows from the definition of \(\nb\) (see \cref{IN13}). Next we will show that \(f(\nu)\) is greater than \(f(\nt)\) for all positive integers less than \aone\ which proves this lemma. 

If \(\nu\) is greater than \(\nt\) then \(f(\nu)\) is greater than \(f(\nt)\) as \(h(\nu)\) is greater than \(h(\nt)\). Moving forward we assume that \(\nu\) is less than \(\nt\) and that:
\begin{align*}
\nu=\nli + \oDu \leq \nhi \text{ where } \oDu \geq 0\\
\nt = \nlj + \oDt \leq \nhj \text{ where } \oDt \geq 0
\end{align*}
The fact that \(\nt \leq \nb\) implies that \(j\leq\ib\) (see \cref{FM13}) and the assumption that \(\nu\) is less than \(\nt\) implies that \(i\) is not greater than \(j\). However, if \(i\) equals \(j\) then \(h\) is decreasing according to \cref{BA19} since \(h(\nu)\) is greater than \(h(\nt)\) which implies that \(f(\nu)\) is greater than \(f(\nt)\) according to \cref{FM3}. Moving forward we assume that \(i\) is less than \(j\).

We start with the case \(2\azero < \aone\) which we will divide into two subcases: one where \(h(\nli)\) is less than \(h(\nlj)\) and one where \(h(\nli)\) is greater. We start with the former. From the fact that \(h(\nu)\) is greater than \(h(\nt)\) follows that:
\[
\oDt<\oDu\leq\flooraoneoa=\nhone \quad\intref{BA37}
\]
This implies that:
\[\oDt<\flooraoneoa\]
This implies that \(f(\nlj+\oDt) <f(1+\oDt)\) according to \cref{FM16} since: \[0<i<j\leq\ib<\oa\implies 0<j-1<\ib<\oa\]
Since \(h(\nlone) \leq h(\nli)\) follows that \(h(\nlone + \oDu) \leq h(\nli+\oDu)\) which gives that:
\[
f(\nt) = f(\nlj + \oDt) < f(1 + \oDt) \leq f(\nlone + \oDu) \leq f(\nli+\oDu) = f(\nu)
\]
Next we look at the case \(h(\nli)>h(\nlj)\). This condition implies that \(i\) is greater than one. From this we can conclude that \(f(\nlj + \oDu)\) is less than \(f(\nli + \oDu)\) according to \cref{FM15} since: 
\[j-i \leq \ib-i < \ib\]
Furthermore, \(\oDu \geq \oDt\) since \(h(\nu) > h(\nt)\) which implies that:
\[
f(\nt) = f(\nlj+\oDt) \leq 
f(\nlj+\oDu) <
f(\nli + \oDu) = f(\nu)
\]
Now we continue with the case \(2\azero > \aone\) and start with the subcase where \(h(\nli)\) is less than \(h(\nlj)\) which implies that \(\oDu\) is less than \(\oDt\) as \(h(\nu)\) is greater than \(h(\nt)\) which in turn implies that 
\(\nlj+1+\oDu\leq\nhj\). Furthermore, since \(h(\nlone)\) is not greater than \(h(\nli)\) for any \(i\), we can conclude that \(h(\nlone + \oDu) \) is not greater than \(h(\nli+\oDu)\) for any \(i\). This allows us to conclude that:
\[
f(\nt) = f(\nlj+\oDt) \leq f(\nlj+1+\oDu) < f(1+\oDu) =
f(\nlone + \oDu) \leq
f(\nli+\oDu) = f(\nu)
\]
Left to consider is the subcase where \(h(\nli)\) is greater than \(h(\nlj)\) which implies that \(i\) is greater than one and \(\oDu\) is not greater than \(\oDt\) which allow us to conclude that:
\[
f(\nt) = f(\nlj+\oDt) < f(\nli+\oDt) \leq f(\nli+\oDu) = f(\nu)
\]
\end{proveit}
\begin{frameit}
\begin{theo}\label{FM21}
\ \\
\niseq\ consists of all integers in the interval \([0,\nb]\) where \(\nb\) is greater than zero. These integers appear exactly once and there are no other elements in the sequence. Hence, \(m\) equals \nb. The elements appear in ascending order based on the value of \(h\) at the element.
\end{theo}
\textbf{Note} that this theorem is valid for all values of \(\oa\).
\end{frameit}
\begin{proveit}{FM21}\ \\
From \cref{IN13} follows that \(\nb > 0\) since:
\[f(1) = \athree + \atwo\azero < \aone\atwo\]
According to \cref{FM6}, \niseq\ equals \((0, 1, 2, ..., \nb)\) if \(\oa=1\) and \cref{FM19,,FM20} jointly gives that \niseq\ consists of the integers in the interval \([0, \nb]\) if \(\oa>1\) and no other elements. From \cref{FM1} follows that the elements appear in ascending order based on the value of \(h\) at the element. Noting that the first element in the sequence has index zero, it is trivial to realize that \(m\) equals \nb.
\end{proveit}
\section{Finding \(\nb\)}\label{FN}
As will be shown in \cref{FF}, the Frobenius number can be computed without computing all elements in the sequence \ninbseq\ by cherry picking specific ones. However, the computations in that section are based on knowing what \nb\ is. Therefore, in this section we will design an efficient algorithm finding \(\nb\). However, first we will conclude that an algorithm for finding \nb\ is not needed in two cases. When the sum \(\athree + \atwo\azero\) is greater than \(\aone\atwo\), then \nb\ equals zero according to \cref{FM4}. We can also disregard the case where \(\oab\) is less than \(\oo\) from the discussion since \cref{FM13} provides closed form formulas for computing \nb\ in this case. All in all, we can make the assumption below when trying to find \nb\ for the remaining cases, recalling that \(\athree + \atwo\azero\) can not be equal to \(\aone\atwo\) according to \cref{FM24} and that \(\oab\) can not be equal to \(\oo\) according to \cref{FM9}.
\begin{frameit}
\begin{assu}\label{FN1}
In this entire section we will assume that:
\begin{align*}
\athree + \atwo\azero < \aone\atwo\\
\oab > \oo
\end{align*}
\end{assu}
\end{frameit}
Recall that \nb\ can be found by means of \cref{FM13} if \(\ib\) is known. The integer \(\ibj\) defined by \cref{FRB2} is defined as the first element in an ARM sequence such that the ratio of the element and its index is not greater than a constant. \ib\ is the smallest positive integer \(i\) such that the ratio \(e(i)/i\) is strictly less than the constant \(\oo\). However, from \cref{FM9} follows that \(e(i)/i\) can not equal \(\oo\) for any \(i\) so we will find \(\ib\) using the algorithm finding \(\ibj\). This algorithm assumes that \(\oab\) is greater than one and the constant \(\oo\). This assumption is fullfilled since \cref{FM11} and \cref{FN1} gives:
\[\oab > \oo > 1\]
(Note that this also implies that \(e\) meets the criteria of \cref{BA4} according to \cref{BS9}, i.e. all characteristics of ARM sequences presented here are valid for the sequence defined by \(e\).)

Following the algorithm for computing \(\ibj\) we get \(e(\ib)\) by either \cref{FRB9} or \cref{FRB10}. However, \cref{FRB9} implies that \(\ib\) equals \(\oa\) which is not the case according to \cref{FM14}, i.e. we can assume that \(e\) always is given by \cref{FRB10}. Next we finish this short section by showing how \cref{FRB10} and \cref{FM13} combined allows us to compute \(\nb\) from
the diff-mod sequence of \(e\).
\begin{frameit}
\begin{defi}\label{FN2}\ \\
\leavevmode
\begin{itemize}
\item
The pair of integers \((\oabone, \oaone)\) where \(\oabone=\oab\) and \(\oaone=\oa\) defines a diff-mod sequence (see \cref{IN7}). 
\item
\(\os\) is the smallest positive integer \(j\) such that \(\oabj\leq\ooj\) where:
\begin{align*}
2\oabj<\oaj:\quad & \oojp=\frac{\oaj\ooj}{\oajp-\ooj}\\
\\
2\oabj>\oaj:\quad & \oojp=\frac{\oaj\ooj}{\oajp+\ooj}\\
\\
&\ooone=\oo
\end{align*}
\end{itemize}
\end{defi}
\begin{theo}\label{FN11}
\leavevmode
\begin{itemize}
\item
\(2\oabosm<\oaosm:\quad \nb= \hms (\oabos+\azero)\)
\item
\(2\oabosm>\oaosm:\quad
\nb=\hms \left(\oaosm-\oaos\ceilfrac{\oaosm}{\oaos+\ooosm}+\azero\right)\)
\end{itemize}
\textbf{Note }that \(2\oabosm\neq\oaosm\)
\end{theo}\end{frameit}
\begin{proveit}{FN11}\ \\
Recall that \cref{FN1} implies that \(\oa>1\) in this entire section, which means that all lemmas and theorems in \cref{FM} made after \cref{FM26} are valid.

When \(2\azero < \aone\) and \(\oab > \oo\) then \(\nb = \nhib-1\) according to \cref{FM13}. Noting that \(\ib\) is greater than zero and less than \(\oa\) according to \cref{FM12,FM14} respectively we can conclude that:
\[
\intref{BA20} \quad h(\nhib-1) = h(\nhib) - \azero = \aone-e(\ib) - \azero \quad \intref{FM8}
\]
Since \(0 \leq \nhib - 1 < \aone\) we can get \(\nb\) from the inverse function of \(h\):
\begin{align*}
\nb = \nhib-1 = & \hm (\aone - (e(\ib) + \azero)) \quad \intref{FEB5}\\
= & \hm(- (e(\ib) + \azero))\\
\intref{BA16} \quad = & \hms(e(\ib)+\azero)
\end{align*}
The same result is obtained for the case \(2\azero > \aone\) in an equivalent manner and the case \(2\azero = \aone\) we can disregard altogether since \(\athree + \atwo\azero\) is greater than \(\aone\atwo\) according to \cref{FM25} in this case. This proof is then completed by replacing \(e(\ib)\) by means of \cref{FRB10}.
\end{proveit}
\section{Formulas for the Frobenius number}\label{FF}
Equipped with the efficient algorithm for computing \nb, described in \cref{FN}, we are in this section ready to derive formulas for computing the Frobenius number \(g(A)\). In total we will derive six different formulas for the same number of mutually exclusive cases which jointly cover all possible cases. Let us start with the first one.
\begin{frameit}
\begin{theo}\label{FF1}
Given that \(\athree + \atwo\azero > \aone\atwo\)
\[g(A) = \aone\atwo - \atwo -\aone\]
\end{theo}
\end{frameit}
\begin{proveit}{FF1}\ \\
When \(\athree + \atwo\azero > \aone\atwo\) follows from \cref{FM4}\ that:
\[\nb = 0\]
This implies according to \cref{FM2}\ that:
\[\underset{n}{min}\;F(n,r) = \aone\atwo-\atwo r \text{ if } r > 0\]
Hence by means of \cref{IN5} we can conclude that:
\begin{align*}
g(A) & = \underset{0<r<\aone}{max}\bigl(\underset{0<n<\aone}{min}F(n,r)\bigl)-\aone\\
& = \underset{0<r<\aone}{max}(\aone\atwo - \atwo r)-\aone\\
& = \aone\atwo - \atwo -\aone
\end{align*}
\end{proveit}
Moving forward we assume that the criterion for the case above is not fulfilled. (Note that \(\athree + \atwo\azero\) is not equal to \(\aone\atwo\)
according to \cref{FM24} and that this assumption implies that \(2\azero\) is different from \(\aone\) according to \cref{FM25}.) Then we take a closer look at \cref{IN5} (Tripathi's theorem) to see how we can cherry pick values of \(r\) to compute the Frobenius number. Then we use this result to derive the formula for the next case.
\begin{frameit}
\textbf{\Cref{FM23}}.\\
In this section from this point onward, unless expressly stated otherwise:
\[\athree + \atwo\azero < \aone\atwo\]
\begin{defi}\label{FF3}
\(\oDi=h(\ni)-h(\nim)
\text{ where }0<i\leq \nb\)
\end{defi}
\begin{lemm}\label{FF4}
\[g(A)=max\Bigl(\underset{i}{max}\;\athree \ni +\atwo(\oDi - 1),\atwo(\aone-h(\nnb)-1)\Bigl)-\aone\text{ where }0<i\leq \nb\]
\end{lemm}
\begin{theo}\label{FF5}
Given \(2\azero < \aone\) and \(\oab < \oo\)
\begin{align*}
g(A) = max( & \athree\nb + \atwo(\azero - 1),\\
& \atwo(\aone - \nb\azero -1))
 - \aone
\end{align*}
\end{theo}
\end{frameit}
\begin{proveit}{FF4}\ \\
Recall that \cref{IN5} states that:
\[
g(A)=\underset{0<r<a1}{max}\bigl(\underset{0 < n < \aone}{min}F(n,r)\bigl)-\aone
\]
Next we pick, for specific intervals, the r-value which is a candidate to give the maximum of this expressions.\\

\bullet\(h(\nim) < r \leq h(\ni)\text{ where } 0 < i \leq \nb\)
\begin{align*}
\underset{r}{max}\bigl(\underset{n}{min}F(n,r)\bigl) &=\underset{r}{max}(f(\ni)-\atwo r)\quad\intref{FM2}\\
&=f(\ni)-\atwo(h(\nim)+1)\\
&=\athree n+\atwo h(\ni)-\atwo(h(\nim)+1)\\
&=\athree n+\atwo(\oDi-1)\\
\intertext{\(\bullet r>h(\nnb)\)}
\underset{r}{max}\bigl(\underset{n}{min}F(n,r)\bigl) &=\underset{r}{max}(\aone\atwo-\atwo r)\\
&=\aone\atwo-\atwo (h(\nnb)+1)\\
&=\atwo (\aone-h(\nnb)-1)\\
\end{align*}
\end{proveit}
\begin{proveit}{FF5}\ \\
In this case \(\nb < \none\) according to \cref{FM13} which implies that we get \(\ni\) by adding one to \(\nim\) since:
\[h(n) = n\azero\text{ when }0 < n < \none \quad\intref{BA20}\]
From this follows that \(\oDi = \azero\) for all \(i\) which in turn implies that:
\[\underset{i}{max}\;\athree \ni+\atwo(\oDi-1) = \athree\nb + \atwo(\azero -1) \]
This theorem then follows from \cref{FF4}\ noting that:
\[h(\nnb) = \nb\azero\]
\end{proveit}
According to \cref{FF4} there are two parameters that we have to consider when computing \(g(A)\), namely \(\ni\) and \(\oDi\). The greater \(\ni\) and \(\oDi\) are for a specific \(i\), the more likely it is that \(g(A)\) is given by the
following expression:
\[
\athree \ni+\atwo(\oDi-1) -\aone
\]
Therefore, we will now spend some time analysing how the size of \(\ni\) and \(\oDi\) are correlated and below we prove a lemma which will be instrumental for doing this. However, the proof of this lemma needs another lemma so we start by proving that one.
\begin{frameit}
\begin{lemm}\label{FF34}
\[0\leq h(n) + \oD < \aone \implies h(n) + \oD = h(n + \hm (\oD))\]
\end{lemm}
\begin{defi}\label{FF6}
\leavevmode
\begin{itemize}
\item
\(\oDdi\text{ is the smallest integer }\oD>0\text{ such that }\hm (\oD)\leq\ni.\)
\item
\(\oDbi\text{ is the smallest integer }\oD>0\text{ such that }\hms (\oD) \leq \nb - \ni.\)
\end{itemize}
\end{defi}
\begin{lemm}\label{FF7}
\begin{align*}
\oDi= & min(\oDdi,\oDbi)\\
\ni= &\left\{\begin{aligned}
& \nim+\hm (\oDdi)>\nim\text{ if }\oDi=\oDdi\\
& \nim - \hms (\oDbi) < \nim \text{ if } \oDi = \oDbi\\
\end{aligned}\right.
\end{align*}
\end{lemm}
\end{frameit}
\begin{proveit}{FF34}
\begin{align*}
h(n + \hm(\oD)) = & \mod{\azero(n + \mod{\azerom \oD}{\aone}}{\aone}\\
\intref{IN9}\quad = & \mod{\mod{\azero n}{\aone} + \mod{\azero\azerom}{\aone}\oD}{\aone}\\
= & \mod{h(n) + \oD}{\aone}\\
\intref{IN8}\quad = & h(n) + \oD
\end{align*}
\end{proveit}
\begin{proveit}{FF7}\ \\
Assume that \(\oD = h(\ni) - h(n)\) for some n such that \( 0 \leq n < \aone\) and \(\oD > 0\). From \cref{FF34} follows that:
\[
h(n) = h(\ni + \hm (-\oD)) \text{ since } 0 \leq h(n) = h(\ni) - \oD < \aone
\]
In addition, from \cref{BA16} follows that:
\[
h(n) = h(\ni + \aone - \hm (\oD)) =
h(\ni - \hm (\oD)) =
h(\ni + \hms (\oD))
\]
If \(\hm(\oD) \leq \ni\) then according to \cref{BA22} follows that:
\[
0\leq n=\ni-\hm(\oD) < \nb
\text{ since } 0 \leq \ni-\hm(\oD) < \aone
\]
If \(\ni < \hm(\oD) < \ni-\nb + \aone\) then follows that:
\[
\nb < n = \ni-\hm (\oD)+\aone<\aone \text{ since }
\nb < \ni - \hm(\oD) + \aone < \aone
\]
If \(\hm(\oD) \geq \ni - \nb + \aone\) then \(\hms(\oD) \leq \nb - \ni\). From this follows that:
\[
0 < n = \ni + \hms(\oD) \leq \nb \text{ since } 0 < \ni + \hms(\oD) \leq \nb
\]
According to \cref{FM21} this implies that \(n\) is an element in \ninbseq\ for \(\oD\) such that \(\hm(\oD)\) is not greater than \(\ni\) and for \(\oD\) such that \(\hms(\oD)\) is not greater than \(\nb - \ni\) and \(n\) then eqauls \(\ni - \hm(\oD)\) and \(\ni + \hms(\oD)\) respectively. For \(\oD\) such that none of these conditions are fullfilled, \(n\) is not an element in \ninbseq. From \cref{FM1} follows that \(n\) is \(\nim\) for the smallest \(\oD\) fulfilling one of these conditions and \(\oD\) then equals \(\oDi\).
\end{proveit}
The lemma above implies that \(\oDbi\) increases or stays the same when \(\ni\) increases since \(\oDbi\) is the index of the first element in an ARM sequence lower than a limit which decreases when \(\ni\) increases. Equivalent reasoning yields that \(\oDdi\) decreases or stays the same when \(\ni\) increases. This in turn implies that if \(\nk\) is less than \(\ni\) and \(\oDi\) equals \(\oDbi\) then \(\oDk\) equals \(\oDbk\). On the other hand, if \(\oDk\) equals \(\oDdk\) then \(\oDi\) equals \(\oDdi\). Below we prove and formulate this formally. We also show that \(\oDi\) equals \(\oDdi\) when \(\ni\) equals \(\nb\). In addition, we show that \(\oDdi\) and \(\oDbi\) never are equal for a specific \(i\), i.e. \(\oDi\) is equal to exactly one of them.
\begin{frameit}
\begin{lemm}\label{FF8}
\begin{align*}
\nk>\ni\implies &\oDbk\geq\oDbi\\
&\oDdi\geq\oDdk\\
&\oDbk<\oDdk\implies \oDbi<\oDdi\\
&\oDdi<\oDbi\implies \oDdk<\oDbk\\
\end{align*}
\end{lemm}
\begin{lemm}\label{FF9}
\[\ni=\nb\implies\oDi=\oDdi\]
\end{lemm}
\begin{lemm}\label{FF10}
\[\oDdi \neq \oDbi\]
\end{lemm}
\end{frameit}
\begin{proveit}{FF8}\ \\
\(\oDbk\) is the smallest positive integer \(\oD\) such that \(\hms(\oD)\) is not greater than \(\nb\) minus \(\nk\). \(\oDbi\) is also the smallest positive integer \(\oD\) such that \(\hms(\oD)\) is not greater than \(\nb\) minus \(\ni\) and this limit is higher as \(\ni\) is smaller than \(\nk\). This implies that \(\oDbk\) is greater or equal to \(\oDbi\). Similar reasoning leads to the conclusion that \(\oDdi\) is greater or equal to \(\oDdk\).

If \(\oDbk < \oDdk\) then follows that:
\[
\oDbi \leq \oDbk < \oDdk \leq \oDdi
\]
If on the other hand \(\oDdi<\oDbi\) then follows that: 
\[
\oDdk\leq\oDdi<\oDbi\leq\oDbk
\]\
\end{proveit}
\begin{proveit}{FF9}\ \\
When \(\ni = \nb\) then \(\ni > \nim\) which implies that \(\oDi = \oDdi\) according to \cref{FF7}.
\end{proveit}
\begin{proveit}{FF10}
\[
\hm(\oDdi) \leq \ni \leq \nb - \hms(\oDbi) = \hm(\oDbi) + \nb - \aone < \hm(\oDbi) \implies \oDdi \neq \oDbi
\]
\end{proveit}
What conclusions can we draw from the analysis made so far about the change of \(\oDi\) when \(\ni\) increases? Let us assume that \(\oDi\) equals \(\oDbi\) when \(\ni\) is within the interval \([1, k)\) and that \(\oDi\) equals \(\oDdi\) when \(\ni\) equals \(k\). From \cref{FF8} follows that \(\oDi\) will increase or stay the same when \(\ni\) increases within the interval \([1, k)\). Futhermore, from this lemma follows that \(\oDi\) equals \(\oDdi\) for all \(\ni\) within the interval \([k, \nb]\) and that \(\oDi\) will decrease or stay the same when \(\ni\) increases within this interval. Furthermore, \cref{FF9} guarentees that there will be
an integer like \(k\) in the interval \([1, \nb]\) if \(\oDi\) equals \(\oDbi\) when \(\ni\) equals one, i.e. there is a break point within this interval, where \(\oDi\) equals \(\oDbi\) when \(\ni\) is less than this break point and equals \(\oDdi\) when it is greater or equal. In addition, \cref{FF10} ensures there will not be any value of \(\ni\) in this interval where there is a tie between \(\oDbi\) and \(\oDdi\). On the other hand, if \(\oDi\) equals \(\oDdi\) when \(\ni\) equals one then \(\oDi\) equals \(\oDdi\) for all \(\ni\) in the interval \([1, \nb]\) and \(\oDi\) will decrease or stay the same when \(\ni\) increases within this interval. We have actually already encountered such a case, namely when \(2\azero\) is less than \(\aone\) and \(\alpha\) is less than \(\oo\). The next lemma proves that this is also the only case where \(\oDi\) equals \(\oDdi\) when \(\ni\) equals one.
\begin{frameit}
\begin{lemm}\label{FF11}
\begin{gather*}
\oDi=\oDdi\text{ when }\ni=1\\
\iff\\
2\azero<\aone \text{ and } \oab<\oo
\end{gather*}
\end{lemm}
\end{frameit}
\begin{proveit}{FF11}\ \\

\bullet Assume that \(\oDi=\oDdi\) when \(\ni=1\).\\

From \cref{FF7} follows that \(\nim < \ni = 1\) since \(\oDi\) equals \(\oDdi\). This implies that: 
\[\nim = 0 = \nzero \implies i = 1\]
This in turn implies that there is no \(k\) such that:
\[0 = h(\nzero) < h(\nk) < h(\none) = \azero\]
When \(2\azero > \aone\) then \(\nb \geq \nhone\) according to \cref{FM13} which implies that \(2\azero\) can not be greater than \(\aone\) since according to \cref{BS11} follows that:
\[ 
h(\nhone) = \mod{\aone}{\aone - \azero} < \aone - \azero < \azero
\]
When \(2\azero < \aone\) and \(\oab > \oo\) then \(\nb > \nhone\) which implies that \(\nltwo\) is one of the elements of \ninbseq\ and \(h(\nltwo) < \azero\) according to \cref{BA20}.

All in all, this implies that the only option remaining is that \(2\azero\) is less than \(\aone\) and \(\oab\) is less than \(\oo\) (Note that \(\oab\neq\oo\) according to \cref{FM9}).\\
\\
\bullet Assume that \(2\azero < \aone\) and \(\oab<\oo\).\\
\[
\ni \leq \nb <\nhone \implies  h(\ni)=\ni\azero\quad\intref{BA34}
\]
This implies that \(\none\) equals one which in turn implies that \(\oDone\) equals \(\oDdone\) according to \cref{FF7} since \(\nzero\) equals zero which is less than \(\none\).
\end{proveit}
Below we prove the formula for computing \(g(A)\) for the next case. However, first we show that, for this case and the remaining cases, we can simplify \cref{FF4} since \(g(A)\) will never be given by:
\[
\atwo(\aone-h(\nnb)-1)-\aone
\]
\begin{frameit}
\begin{lemm}\label{FF12}\ \\
Given that \(2\azero > \aone\) or \(\oab > \oo\)
\[g(A) = \underset{i}{max}\;\athree \ni +\atwo(\oDi - 1) \text{ where } 0 < i \leq \nb\]
\end{lemm}
\begin{theo}\label{FF13}
Given \(2\azero > \aone\) and \(\oab < \oo\)
\begin{align*}
g(A) = max(& \athree\nb + \atwo(\mod{\aone}{\oa} - 1),\\
&\athree(\nb - 1) + \atwo(\oa - 1)) - \aone
\end{align*}
\end{theo}
\end{frameit}
\begin{proveit}{FF12}\ \\
Assume that \(\oD = \aone - h(n)\text{ where } 0 \leq n < \aone\). From \cref{FEB5} follows that:
\[
n = \hm(\aone - \oD) = \hms(\oD) \quad \intref{BA16}
\]
Assuming that \(\oDt\) is the smallest integer \(\oD\) such that \(\hms(\oD) \leq \nb\) we get that:
\[
\nnb = \hms(\oDt)
\]
This implies that \(
h(\nnb) = \aone - \oDt\). As \(2\azero > \aone\) or \(\oab > \oo\), we can according to  \cref{FF11} conclude that there is an \(i\) such that \(\oDi\) equals \(\oDbi\). Recall that \(\oDbi\) is the smallest positive integer \(\oD\) such that:
\[\hms(\oD) \leq \nb -\ni < \nb\]
This implies that:
\[\oDi \geq \oDt = \aone - h(\nnb)\]
Hence, we can conclude that:
\[\athree\ni + \atwo(\oDi - 1) > \atwo(\oDi - 1) \geq \atwo(\aone - h(\nnb) - 1)\]
This lemma then follows from \cref{FF4}.
\end{proveit}
\begin{proveit}{FF13}\ \\
From \cref{FM13} follows that \(\nb = \nhone\).  This implies that:
\[
\intref{BA34}\quad h(\ni) = \aone - \oa \ni
\]
From this we can conclude that
\(\none = \nhone = \nb\) which gives us that:
\begin{align*}
\oDone = h(\nhone) - h(0)
= \mod{\aone}{\oa} &\quad\intref{BS11}\\
\oDi = \oa \text{ for }i > 1&
\end{align*}
This means that:
\[
\underset{\ni < \nb}{max}\;\athree \ni +\atwo(\oDi - 1) = 
\athree(\nb - 1) + \atwo(\oa - 1))
\]
This theorem then follows from \cref{FF12}.
\end{proveit}
Recall that \(\oDbi\) is the smallest positive integer \(\oD\) such that \(\hms(\oD)\) is not greater than the limit \(\nb\) minus \(\ni\) and that \(\oDdi\) is the smallest positive integer \(\oD\) such that \(\hm(\oD)\) is not greater than the limit \(\ni\). In both cases we would like to find the first element in an ARM sequence not greater than a limit. To do this we can use the fact that this element will be one of the local minima of the sequence if the limit is not greater than the greatest local minima. This is the case if the modulus of the border sequence of the ARM sequence is greater than the limit and then the problem can be transformed into finding the first element of the border sequence less than the limit. This transformation can be repeated as long as the limit is not greater than the greatest local minima, i.e. we can again use the border sequence sequence of the ARM sequence to solve this problem. This is the essence of \cref{FFE3}. We will use this lemma to compute \(\oDbi\) and \(\oDdi\). In our case, the corresponding sequence is defined by \(\hm\) or \(\hms\) and the limit is \(\ni\) or \(\nb\) minus \(\ni\). Since none of these limits are greater than \(\nb\), we can transform these problems into finding the first element not greater than the limit in the first sequence of the border sequence sequence of \(\hm\) or \(\hms\), where \(\nb\) is greater than the greatest local minima in the sequence. We denote the sequence number of this sequence \(\op\). (An observant reader might ask himself what happens with \cref{FFE3} if \(2\vbj\) equals  \(\vj\). In this case, \(\vbj\) equals one according to \cref{BA14}, which in turn implies that \(\vjp\) equals one which is less than or equal to our limit \(\nb\) according to \cref{FM21}, i.e. this case is not relevant.)

Below we define the counterparts of \(\oDbi\) and \(\oDdi\) for each sequence in the border sequence sequence of \(\hm\) and \(\hms\) and denote these \(\odbij\) and \(\oddij\) where \(j\) indicate the number of the sequence in the border sequence sequence. We also show that \(\op\) is less than \(\ot\), implying that \(\op\) is always defined. Finally, since we are actually not looking for the first element in the sequence, defined by \(\hm\) or \(\hms\), but the index of this element, we show how \(\oDdi\) and \(\oDbi\) can be derived from \(\hop(\oddi)\) and \(\hops(\odbi)\) respectively.
\begin{frameit}
\begin{defi}\label{FF14}
\leavevmode
\begin{itemize}
\item
\(\hj(\od) = \mod{\ovbj\od}{\ovj}\) where \(\ovbone = \azerom\) and \(\ovone = \aone\) defines a boarder sequence sequence.
\item
\(\ovopp \leq \nb < \ovop\)
\item
\(\oddij\) is the smallest integer \(\od > 0\) such that \(\hj(\od) \leq \ni\).
\item
\(\odbij\) is the smallest integer \(\od > 0\) such that \(\hjs(\od) \leq \nb - \ni\).
\item
\(\odij = min(\oddij, \odbij\))
\item
\(\oddi = \oddiop\)
\item
\(\odbi = \odbiop\)
\item
\(\odi = min(\oddi, \odbi)\)
\end{itemize}
\end{defi}
\begin{lemm}\label{FF15}
\[\op < \ot\]
\end{lemm}
\begin{lemm}\label{FF17}
\begin{align*}
\oDdi = & h(\hop(\oddi))\\
\oDbi = & \hs(\hops(\odbi))
\end{align*}
\end{lemm}
\end{frameit}
\begin{proveit}{FF15}\ \\
From \cref{FM13} follows that \(\nb \geq 1\) since \(\athree + \atwo\azero < \aone\atwo\). From the fact that \(\ovjp\) is less than \(\ovj\) when \(j\) is less than \(\tau\) (see \cref{BS9}) and that \(\ovot\) equals one follows then that \(\psi\) is always uniquely defined and less than \(\tau\).
\end{proveit}
\begin{proveit}{FF17}\ \\
Since \(\ni \leq \nb < \ovop\) follows from \cref{FF6} and \cref{FFE3} that:
\[
\hm(\oDdi) 
= \hone(\oddione)
= \hop(\oddiop)
= \hop(\oddi)
\]
From \cref{FEB5} follows that:
\[\oDdi = h(\hop(\oddi))\]
In the same way we get:
\[\oDbi = \hs(\hops(\odbi))\]
\end{proveit}
The relationship between \(\oDdi\) and \(\oddi\) and between \(\oDbi\) and 
\(\odbi\) provided by \cref{FF17} allows us to replace the analysis of the correlation between \(\ni\) and \(\oDi\) with the corresponding analysis of \(\ni\) and \(\odi\). However, to be able to do that we first need to prove some propositions. We need to show that the relative size of \(\oddi\) and \(\odbi\) reflects the relative size of \(\oDdi\) and \(\oDbi\). In addition, we must show that the relative size of \(\odi\) for two different values of \(i\) reflects the relative size of \(\oDi\) for the same values of \(i\). We also have to show that \(\oddi\) is not equal to \(\odbi\) for any \(i\) meaning that \(\odi\) is always equal to exactly one of these. Below we prove all three propositions mentioned above. However, first we show how \(\oddij\) and \(\odbij\) can be expressed in terms of \(\oddijp\) and \(\odbijp\) respectively. We will use this proposition to prove the other ones. Having shown this, we have shown that the analysis of the correlation between \(\ni\) and \(\oDi\) can be based on the corresponding analysis of \(\ni\) and \(\odi\). In addition, we have shown that the conclusions which we have already made about the correlation between \(\ni\) and \(\oDi\) also holds for \(\ni\) and \(\odi\), i.e. that \(\odi\) equals \(\odbi\) when \(\ni\) equals one, unless \(2\azero\) is less than \(\aone\) and \(\oab\) is less than \(\oo\), and as \(\ni\) increases \(\odi\) will remain equal to \(\odbi\) until a break point is reached and thereafter \(\odi\) will equal \(\oddi\). Before this break point \(\odi\) will increase or stay the same when \(\ni\) increases and after \(\odi\) will decrease or stay the same.
\begin{frameit}
\begin{lemm}\label{FF16} Given \(\nb < \ovjp \)
\begin{align*}
2\ovbj < \ovj:\quad &
\begin{aligned}[t] \oddij =& \odloddijpp\\
\odbij =& \odhodbijp
\end{aligned}\\
2\ovbj > \ovj:\quad &
\begin{aligned}[t] \oddij =& \odhoddijp\\
\odbij =& \odlodbijpp
\end{aligned}
\end{align*}
\end{lemm}
\begin{lemm}\label{FF18}
\begin{align*}
\oddi < \odbi \implies & \oDdi < \oDbi\\
\oddi > \odbi \implies & \oDdi > \oDbi
\end{align*}
\end{lemm}
\begin{lemm}\label{FF19}
\[
\odi > \odk \implies \oDi > \oDk
\]
\end{lemm}
\begin{lemm}\label{FF20}
\[\oddi \neq \odbi\]
\end{lemm}
\end{frameit}
\begin{proveit}{FF16}\ \\
This lemma follows from \cref{FF6} and \cref{FFE3}, noting that \(\ni\) and \(\nb\) minus \(\ni\) are not greater than \(\nb\) which in turn is less than \(\ovjp\).
\end{proveit}
\begin{proveit}{FF18}\ \\
We assume that \(0 < j < \psi\) which implies that \(\nb < \ovjp\) and start with the case where \(2\ovbj\) is less than \(\ovj\).

If \(\oddijp < \odbijp\) then:
\begin{align*}
\intref{FF16}\quad\oddij = \odloddijpp \leq \odlodbijp < \odhodbijp = \odbij
\intertext{If \(\oddijp > \odbijp\) then:}
\oddij = \odloddijpp > \odhoddijp > \odhodbijp = \odbij
\end{align*}
For the case where \(2\ovbj > \ovj\) it is easy to show in an analogous manner that:
\begin{align*}
\oddijp < \odbijp \implies \oddij < \odbij\\
\oddijp > \odbijp \implies \oddij > \odbij 
\end{align*}

Now follows a proof by induction where we assume that:
\[
\oddij < \odbij \implies \oDdi < \oDbi
\]
From this assumption follows that:
\[
\oddijp < \odbijp \implies \oddij < \odbij \implies \oDdi < \oDbi
\]
The base case where \(j = 1\) is given by the fact that \(\oddione\) equals \(\oDdi\) and \(\odbione\) equals \(\oDbi\) which means that:
\[
\oddione < \odbione \implies \oDdi < \oDbi
\]
Finally the case \(j = \op - 1\)  gives:
\[
\oddi = \oddiop < \odbiop = \odbi \implies \oDdi < \oDbi
\]
In an analogous manner it is easy to show that:
\[
\oddi > \odbi \implies \oDdi > \oDbi
\]
\end{proveit}
\begin{proveit}{FF19}\ \\
We assume that \(\odijp > \odkjp\) and \(j < \op\) which implies that \(\nb < \ovjp\).
We start with the case where \(2\ovbj\) is less than \(\ovj\).

If \(\odijp = \oddijp\) and \(\odkjp = \odbkjp\) we can by looking at the proof of \cref{FF18} conclude that:
\begin{align*}
\oddijp < \odbijp \implies & \oddij < \odbij \\
\odbkjp < \oddkjp \implies & \odbkj < \oddkj
\end{align*}
From \cref{FF16} then follows that:
\[
\odij = \oddij = \odloddijpp > \odlodbkjpp > \odhodbkjp = \odbkj = \odkj
\]
On the other hand, if \(\odijp = \odbijp\) and \(\odkjp = \oddkjp\) then:
\[
\odij = \odbij = \odhodbijp > 
\odlodbijp \geq \odloddkjpp = \oddkj = \odkj
\]
It is trivial to show that same condition apply for the two remaining cases, i.e. the case where \(\odijp\) equals \(\oddijp\) and \(\odkjp\) equals \(\oddkjp\) and the case where \(\odijp\) equals  \(\odbijp\) and \(\odkjp\) equals \(\odbkjp\). All in all, we have shown that:
\[\odijp > \odkjp \implies \odij > \odkj\]
In an analogous manner we can also easily show that this condition applies for the case where \(2\ovbj\) is greater than \(\ovj\). The finalization of this proof then easily follows by induction in the same manner as \cref{FF18}.
\end{proveit}
\begin{proveit}{FF20}\ \\
The following shows us that \(\oddi \neq \odbi\):
\[
\hop(\oddi) \leq \ni \leq \nb - \hops(\odbi) = \hop(\odbi) + \nb - \ovop < \hop(\odbi)
\]
\end{proveit}
As we have already found formulas giving \(g(A)\) for all possible cases when \(\oab\) is less than \(\oo\), we can assume that \(\oab\) is greater than \(\oo\) for the remaining ones. (Recall that \(\oa\) is not equal to \(\oo\) according to \cref{FM9}.) Below we analyse which conclusions we can make about the diff-mod sequence of \(\hm\) assuming this. Having done that we have done the necessary groundwork for deriving the formula for the next case, which we do thereafter.
\begin{frameit}
\begin{lemm}\label{FF21}
\[
\ovbop = 1 \implies 2\azero < \aone \text{ and } \oab < \oo
\]
\end{lemm}
\begin{lemm}\label{FF22}
\begin{gather*}
\oDi=\aone - \azero \text{ when }\ni=1\\
\implies\\
2\azero > \aone \text{ and } \oab<\oo
\end{gather*}
\end{lemm}
\begin{lemm}\label{FF23}
\[\ovbop = \ovop - 1 \implies \oab < \oo\]
\end{lemm}
\begin{lemm}\label{FF24}
\begin{gather*}
\oab > \oo \implies 
\left\{\begin{aligned}
\ovbop > 1\\
\ovop - \ovbop > 1\\
\ovopp > \ovbopp > 0\\
2\ovbop \neq \ovop\\
\end{aligned}\right.
\end{gather*}
\end{lemm}
\begin{theo}\label{FF25}
Given that \(\oab > \oo\) and \(\nb = \ovop -1\)
\begin{align*}
g(A) = max(&\athree \nb + \atwo(h(\ovbop) - 1),\\
&\athree(\ovbop -1) + \atwo (\hs(\ovop - \ovbop) - 1))
- \aone
\end{align*}
\end{theo}
\end{frameit}
\begin{proveit}{FF21}\ \\
If \(\ovbop = 1\) then \(\odi = \oddi = 1\) when \(\ni = 1\) since:
\[
\hop(1) = \ovbop = 1 = \ni
\]
According to \cref{FF18} this implies that \(\oDi = \oDdi\) when \(\ni = 1\), which in turn implies that \(2\azero\) is less than \(\aone\) and \(\oab\) is less \(\oo\) according to \cref{FF11}.
\end{proveit}
\begin{proveit}{FF22}\ \\
We assume that \(\nk = 1\). Given that \(h\) is increasing and \(\oab\) is less than \(\oo\) then \nb\ is less than \(\nhone\) according to \cref{FM13}. If 
\(n\) is less than \(\nhone\) then follows from \cref{BA34} that:
\[h(n) = n \azero\]
This implies that \(\none = 1 = \nk\) which gives that:
\[
\oDk = h(1) - h(0)
= \azero < \aone - \azero
\]
Given that \(h\) is increasing and \(\oab\) is greater than \(\oo\) then \(\nb \geq \nhtwo - 1\).
\begin{align*} 
\oDk = & h(\nk) - h(\nkm)\\
\intref{BA20}\quad \leq & h(1) - h(\nltwo)\\
< &\azero
\end{align*}
Given that \(h\) is decreasing and \(\oab\) is greater than \(\oo\) then \(\nb \geq \nhtwo\). This implies that:
\begin{align*}
\oDk = & h(\nk) - h(\nkm)\\
\leq & h(\nlone) - h(\nltwo + 1)\\
= & \azero - (h(\nltwo) - (\aone - \azero))\\
= & \aone - h(\nltwo)\\
< & \aone - \azero
\end{align*}
Hence we can conclude that the only remaining case is when \(h\) is decreasing and \(\oab\) is less than \(\oo\).
\end{proveit}
\begin{proveit}{FF23}\ \\
If \(\nb = 1\) then \(\nb \leq \nhone\) which implies that \(\oab < \oo\) according to \cref{FM13}.\\ 
(For instance \(\aone = 5, \atwo = 8, \athree = 9\) gives \(\nb = 1\) and \(\ovbop = 1\) and \(\ovop = 2\).)

If \(\nb > 1\) then \(\ni = 1\) gives:
\[
\nb - \ni \geq 2 - 1 = \ovop - \ovbop = \hops(1)\quad\intref{FFA5}
\]
This implies that \(\odi = \odbi = 1\) which gives:
\begin{align*}
\intref{FF17}\quad\oDi = &\hs(\hops(\odbi))\\
= & \hs(\hops(1))\\
= & \hs(\ovop - \ovbop)\\
= & \hs(1)\\
= & \aone - \azero
\end{align*}
This implies that \(\oab < \oo\) according to \cref{FF22}.
\end{proveit}
\begin{proveit}{FF24}\ \\
From \cref{BS9} follows that \(\ovop > \ovbop > 0\) since \(\op < \ot\) according to \cref{FF15}. Since \(\ovbop \) is not equal to one according to \cref{FF21}, this implies that:
\[\ovbop > 1\]
The fact that \(\ovbop \neq \ovop - 1\) according to \cref{FF23}, implies that:
\[\ovop - \ovbop > 1\]
The fact that \(\ovbop \neq 1\) and that \(\ovbop \neq \ovop - 1\), implies that \(\op + 1\) is less than \(\tau\) according to \cref{BS9}. From the same lemma follows then that:
\[\ovopp > \ovbopp > 0\]
Finally, since \(\ovbop = 1\) if \(2\ovbop = \ovop\), according to \cref{BA14}, follows that:
\[2\ovbop \neq \ovop\]
\end{proveit}
\begin{proveit}{FF25}\ \\
Note that \(\nb = \ovop -1 > \ovbop\) and \(\ovbop - 1 > 0\) according to \cref{FF24}.

\bullet \(\ni \leq \ovbop -1\):
\[
\nb - \ni 
\geq \ovop - 1 - (\ovbop - 1) 
= \ovop - \ovbop
= \hops(1)\quad\intref{FFA5}
\]
This implies that \(\odi = \odbi = 1\) which gives:
\[
\oDi = \hs(\hops(1)) 
= \hs(\ovop - \ovbop)
\]
This yields that:
\[
\underset{\ni < \ovbop}{max}\;\athree \ni +\atwo(\oDi - 1) = \athree(\ovbop -1) + \atwo (\hs(\ovop - \ovbop) - 1))
\]
\bullet \(\ni \geq \ovbop\):

\(\odi = \oddi = 1\) since \(\hop(1)  = \ovbop \leq \ni\) which gives:
\[
\oDi = h(\hop(1)) = h(\ovbop)
\]
This implies that:
\[
\underset{\ni \geq \ovbop}{max}\;\athree \ni +\atwo(\oDi - 1) = 
\athree \nb + \atwo(h(\ovbop) - 1)
\]
This theorem is then given by \cref{FF12}.
\end{proveit}
The next formula involves a constant, here named \(\oed\), derived from the diff-mod sequence of \(\hm\). Below we prove two lemmas concerning this constant and then we derive the formula for this case.
\begin{frameit}
\begin{defi}\label{FF26}
\[
\oed = \floorfrac{\nb - \ovopp + 1 + \ovbopp}{\ovopp}
\]
\end{defi}
\begin{lemm}\label{FF27}
Given \(\oab > \oo\), \(\nb < \ovop - 1\) and \(2\ovbop < \ovop\)\\
\[ 
0 \leq \oed < \odhone 
\text{ where } \odhone \text{ is the first upper border of } \hop
\]
\end{lemm}
\begin{lemm}\label{FF28}
\[
\oed = 0 \implies \ovopp - \ovbopp - 1 > 0
\]
\end{lemm}
\begin{theo}\label{FF29}
Given that \(\oab > \oo\), \(\nb < \ovop -1\) and \(2\ovbop < \ovop\):\\
\begin{align*}
g(A) = max(&\athree \nb + \atwo(h(\ovbop) - 1),\\
&\athree(\ovbop -1) + \atwo (\hs(\oed \ovbop - \ovbopp) - 1))
- \aone
\end{align*}
\end{theo}
\end{frameit}
\begin{proveit}{FF27}\ \\
From \cref{FF26} follows that:
\begin{gather*}
\oed \leq \frac{\nb - \ovopp + 1 + \ovbopp}{\ovopp} < \oed + 1\\
\oed \ovopp + \ovopp - \ovbopp - 1 \leq 
\nb < 
(\oed+1) \ovopp + \ovopp - \ovbopp - 1
\end{gather*}
This implies that:
\begin{align*}
\oed \ovopp + \ovopp - \ovbopp - 1 \leq 
\nb < \ovop -1\\
\intref{IN7} \quad \oed \ovbop < \ovop + \ovbopp - \ovopp\\
\intref{FFA5}\quad\oed \ovbop < \ovop - \hopps(1)\\
\intref{BS11}\quad\oed \ovbop < \ovop - \hops(\odhone)\\
\intref{BA16}\quad \oed \ovbop < \hop(\odhone)\\
\intref{BA34}\quad \oed \ovbop < \odhone \ovbop\\
\oed < \odhone
\end{align*}
Note that the preconditions of \cref{FFA5} are met since \cref{FF24} guarantees that \(2\ovbop\) is not equal to \(\ovop\).

In addition:
\begin{align*}
\frac{\nb - \ovopp + 1 + \ovbopp}{\ovopp} \geq \frac{1 +\ovbopp}{\ovopp} > 0 \implies \oed \geq 0
\end{align*}
\end{proveit}
\begin{proveit}{FF28}\ \\
When \(\oed = 0\) \cref{FF26} gives that:
\[
\frac{\nb - \ovopp + 1 + \ovbopp}{\ovopp} < 1
\]
This lemma then follows from the fact that \(\nb \geq \ovopp\).
\end{proveit}
\begin{proveit}{FF29}\ \\
Note that from \cref{FF14} and \cref{FF24} follows that:
\[
\nb \geq \ovopp > \ovbopp > 0
\]
\bullet \(\ni \leq \ovbopp\):

In this case \(\oddi > \odhone\) since 
\(\hop(\odltwo) <
\hop(\od)\;\forall\;\od\in[1, \odhone]\) (See \cref{BA19}) and:
\[
\ni \leq \ovbopp 
= \hopp(1) 
= \hop(\odltwo)\quad\intref{BS11}
\]
This in turn implies that \(\odi = \odbi \leq \odhone\) since:
\begin{align*}
\nb - \ni \geq &\ovopp - \ovbopp\\ 
\intref{FFA5}\quad = &\hopps(1) = \hops(\odhone)
\end{align*}
\bullet \(\ovbopp \leq \ni  < \ovbop\):

Note that from \cref{FF24} follows that:
\begin{align*}
\nb \geq \ovopp = \ovbop > 1\\
\ovbop = \ovopp > \ovbopp > 0
\end{align*}
In this case \(\oddi = \odltwo\) since:
\begin{align*}
\ni \geq \ovbopp = \hop(\odltwo)\\
\ni < \ovbop = \hop(1) < \hop(\od)\;\forall\;\od\in(1, \odhone]
\end{align*}
\bullet \(\ni = \ovbop -1\)  and \(\oed \geq 1\):

From \cref{FF26} follows that:
\begin{gather*}
\oed \leq \frac{\nb - \ovopp + 1 + \ovbopp}{\ovopp} < \oed + 1\\
\oed \ovopp + \ovopp - \ovbopp - 1 \leq 
\nb < 
(\oed+1) \ovopp + \ovopp - \ovbopp - 1
\end{gather*}
This gives us:
\begin{align*}
\nb - \ni + \geq &
\oed \ovopp + \ovopp - \ovbopp - 1 - (\ovbop -1)\\
= & \hopps(1) + (\oed - 1) \ovopp\\
= & \hops(\odhone) + (\oed - 1) \ovopp\\
= & \hops(\odhone + 1 - \oed)\quad\intref{BA20}\\
\\
\nb - \ni < &
(\oed + 1) \ovopp + \ovopp - \ovbopp - 1 - (\ovbop -1)\\
= & \hops(\odhone) + \oed\ovopp\\
= & \hops(\odhone - \oed)\quad\intref{BA20}
\end{align*}
This implies that \(\odi = \odbi = \odhone + 1 - \oed\) since:
\[
\intref{FF27} \quad 1 < \odhone + 1 - \oed \leq \odhone\text{ and }\oddi = \odltwo
\]
From \cref{FF17} follows that:
\begin{align*}
\oDi = & \hs(\hops(\odhone + 1 - \oed))\\
= & \hs(\hops(\odhone) + (\oed - 1)\ovopp))\\
= & \hs(\hopps(1) + (\oed - 1)\ovopp))\\
= & \hs(\ovopp - \ovbopp + (\oed - 1)\ovopp))\\
= & \hs(\oed\ovbop - \ovbopp)
\end{align*}
As \(\oDi\) is increasing or static when \(\ni\) increases up till \(\ovbop - 1\) we can conclude that:
\[
\underset{\ni < \ovbop}{max}\;\athree \ni +\atwo(\oDi - 1) = \athree(\ovbop -1) + \atwo (\hs(\oed \ovbop - \ovbopp) - 1))
\]
\bullet \(\ni = \ovbop -1\)  and \(\oed = 0\):

From \cref{FF26} follows that:
\[
\ovopp - \ovbopp - 1 \leq \nb < 2\ovopp - \ovbopp - 1
\]
(Note that \(2\ovopp - \ovbopp - 1 > \ovopp\) according to \cref{FF28} which is important since \(\nb\) is greater than or equal to \(\ovopp\).)

This implies that \(\odbi > \odltwo\) since \(\hops(\odhone) < \hops(\odltwo)\) and:
\begin{align*}
\nb - \ni < &
2\ovopp - \ovbopp - 1 - (\ovbop - 1)\\
= &\ovopp - \ovbopp\\
= &\hopps(1)\\
= &\hops(\odhone) < \hops(\od)\;\forall\;\od\in[1, \odhone)
\end{align*}
This in turn implies that
\(\odi = \oddi = \odltwo\) which gives:
\[
\oDi = h(\hop(\odltwo)) 
= h(\ovbopp)
= \aone - h(\oed \ovbop -\ovbopp)
= \hs(\oed \ovbop -\ovbopp)
\]
Since \(\odi \leq \odhone\) when \(\ni \leq \ovbopp\) and \(\odi \leq \odltwo\) when \(\ovbopp \leq \ni  < \ovbop\) we get the same result as we obtained for \(\oed \geq 1\), namely:
\[
\underset{\ni < \ovbop}{max}\;\athree \ni +\atwo(\oDi - 1) = \athree(\ovbop -1) + \atwo (\hs(\oed \ovbop - \ovbopp) - 1))
\]
\bullet \(\ni \geq \ovbop\):

Following the same line of reasoning as in the proof of \cref{FF25} we obtain:
\[
\underset{\ni \geq \ovbop}{max}\;\athree \ni +\atwo(\oDi - 1) = 
\athree \nb + \atwo(h(\ovbop) - 1)
\]
This theorem is then given by \cref{FF12}.
\end{proveit}
Now we have come to the formula for the last case. It also involves a constant, here named \(\oeb\), derived from the diff-mod sequence of \(\hm\). Below we prove two lemmas concerning this constant and then we derive the formula for this case.
\begin{frameit}
\begin{defi}\label{FF30}
\[
\oeb = \floorfrac{\nb + 1 - \ovbopp}{\ovopp}
\]
\end{defi}
\begin{lemm}\label{FF31}
Given \(\oab > \oo\), \(\nb < \ovop - 1\) and \(2\ovbop > \ovop\)\\
\[ 0 \leq \oeb < \odhone\]
\end{lemm}
\begin{lemm}\label{FF32}
\[
\oeb = 0 \implies \ovbopp > 1
\]
\end{lemm}
\begin{theo}\label{FF33}
Given that \(\oab > \oo\), \(\nb < \ovop -1\) and \(2\ovbop > \ovop\)
\begin{align*}
\intertext{If \(\nb = \oeb \ovopp + \ovbopp - 1\) then:}
g(A) = & \athree\nb + \atwo(h(\ovbopp + (\oeb - 1)\ovopp) - 1) - \aone\\
\intertext{If \(\nb > \oeb \ovopp + \ovbopp - 1\) then:}
g(A) = & max(
\begin{aligned}[t]
&\athree \nb + \atwo(h(\ovbopp + \oeb \ovopp) - 1),\\
&\athree(\oeb \ovopp + \ovbopp - 1
) + \atwo (h(\ovbopp + (\oeb - 1)\ovopp) - 1)) - \aone
\end{aligned}
\end{align*}
\end{theo}
\end{frameit}
\begin{proveit}{FF31}\ \\

From \cref{FF30} follows that:
\[
\oeb \leq \frac{\nb + 1 - \ovbopp}{\ovopp} < \oeb + 1
\]
Since \(\nb < \ovop - 1\) this gives us:
\begin{align*}
\oeb \ovopp + \ovbopp - 1 \leq 
\nb < \ovop -1\\
\oeb \ovopp + \hopp(1) < \ovop \\
\intref{BS11}\quad\oeb \ovopp + \hop(\odhone) < \ovop \\
\intref{BA34}\quad \oeb \ovopp + \ovop - \ovopp\odhone < \ovop\\
\oeb < \odhone
\end{align*}
In addition:
\[
\frac{\nb + 1 - \ovbopp}{\ovopp} \geq \frac{\ovopp + 1 -\ovbopp}{\ovopp}
> \frac{1}{\ovopp} > 0 \implies \oeb \geq 0
\]
\end{proveit}
\begin{proveit}{FF32}\ \\
When  \(\oeb = 0\)
\cref{FF30} gives that:
\[
\frac{\nb + 1 - \ovbopp}{\ovopp} < 1
\]
This lemma then follows from the fact that \(\nb \geq \ovopp\).
\end{proveit}
\begin{proveit}{FF33}\ \\
\bullet \(\oeb = 0\):

From \cref{FF30} follows that:
\[
\frac{\nb + 1 - \ovbopp}{\ovopp} < 1
\]

From \cref{FF14} follows that:
\[
\ovopp \leq \nb < \ovopp + \ovbopp -1
\]

From \cref{FF28} follows that \(\ovbopp > 1\).

If \(\ni < \ovbopp\) then \(\oddi > \odltwo\) since:
\begin{align*}
\ni < & \ovbopp \\
= & \hopp(1) \\
\intref{BS11}\quad = & \hop(\odhone) \\
\leq &\hop(n)\;\forall\; n \in [1, \odltwo]\quad\intref{BA20}
\end{align*}
This in turn implies that \(\odi = \odbi \leq \odltwo\) since:
\begin{align*}
\nb - \ni > & \ovopp - \ovbopp\\
\intref{FFA5}\quad =& \hopps(1) = \hops(\odltwo)
\end{align*}
If \(\ni = \ovbopp - 1\) then \(\odi = \odbi = \odltwo\) since:
\begin{align*}
\nb - \ni  < & \ovopp + \ovbopp - 1 - (\ovbopp - 1)\\
= & \ovopp = \ovop - \ovbop = \hops(1) < \hops(\od)\;\forall\; \od\in(1, \odhone]
\end{align*}
For \(\ni = \ovbopp - 1\) this gives that :
\begin{align*}
\oDi = & \hs(\hops(\odltwo))\quad\intref{FF17}\\
= & \hs(\hopps(1))\\
= & \hs(\ovopp - \ovbopp))\\
= & \aone - \hs( \ovbopp - \ovopp )\quad\intref{BA16}\\
= & h( \ovbopp - \ovopp )\quad\intref{BA16}\\
= & h( \ovbopp + (\oeb -1) \ovopp )
\end{align*}
As \(\oDi\) is increasing or static when \(\ni\) increases up till \(\oeb\ovopp + \ovbopp - 1\;(\oeb = 0)\) we get:
\[
\underset{\ni < \ovbopp}{max}
\;\athree \ni +\atwo(\oDi - 1) = 
\athree(\oeb \ovopp + \ovbopp - 1) + \atwo (h(\ovbopp + (\oeb - 1)\ovopp) - 1)
\]
If \(\ovbopp \leq \ni < \ovopp + \ovbopp - 1\) then we get that \(\odi = \oddi = \odhone\) since \(\odbi\) is greater than \(\odhone\) and:
\begin{align*}
\ni \geq & \ovbopp = \hop(\odhone) \\
\\
\ni < & \ovopp + \ovbopp - 1 \\
< & \hop(\odhone) + \ovopp \\
= & \hop(\odhone - 1)\\
< & \hop(\od)\;\forall\;\od \in [1, \odhone - 1)
\end{align*}
From this follows that:
\[
\oDi = h(\hop(\odhone)) = h(\ovbopp + \oeb \ovopp)
\]
This gives that:
\[
\underset{\ni \geq \ovbopp}{max}
\;\athree \ni +\atwo(\oDi - 1) = 
\athree\nb + \atwo ( h(\ovbopp + \oeb \ovopp) - 1)
\]
This theorem for this specific case is then given by \cref{FF12}.

\bullet \(\oeb = \ovbopp = 1\) and \(\nb = \ovopp\):

From \cref{FF30} follows that:
\[
\ovopp \leq \nb < 2 \ovopp
\]
If \(\ni = 1\) then
\(\odbi > \odhone\) since:
\[
\nb - \ni = \ovopp - 1 < \ovop - \ovbop = \hops(1) < \hops(\od)\;\forall\;\od\in(1, \odhone]
\]
This implies that \(\odi = \oddi = \odhone\) since:
\[
\ni = 1 = \ovbopp = \hop(\odhone) < \hop(\od)\;\forall\;\od\in[1, \odhone)
\]
This in turn implies that \(\oab < \oo\) according to \cref{FF11}, i.e. we can disregard this case.

\bullet \(\oeb = \ovbopp = 1\) and \(\nb > \ovopp\):

From \cref{FF30} follows that:
\[
\ovopp + 1 \leq \nb < 2 \ovopp
\]
Note that according to \cref{FF24} \(\ovopp > 1\) which implies that: 
\[2\ovopp > \ovopp + 1\]
If \(\ni = 1\) then \(\odi = \odbi = 1\) since:
\[
\nb - \ni  \geq \ovopp + 1 - 1 = \ovop - \ovbop = \hops(1)
\] 
If \(1 \leq \ni \leq \ovopp\) then \(\oddi = \odhone\) since:
\begin{align*}
\ni \geq & 1 = \ovbopp = \hop(\odhone)\\
\\
\ni < & \ovopp + 1\\
= & \ovopp + \ovbopp\\
= & \hop(\odhone - 1)\\
< & \hop(\od)\;\forall\;\od\in[1, \odhone-1)
\end{align*}
If \(\ni = \ovopp\) then \(\odbi > \odhone\) since:
\[
\nb - \ni  < 2\ovopp - \ovopp = \ovop - \ovbop = \hops(1) < \hop(\od)\;\forall\;\od\in(1, \odhone]
\]

This gives that \(\odi = \oddi = \odhone\) when \(\ni = \ovopp\) which implies that:
\[
\oDi = h(\hop(\odhone)) = h(\ovbopp + (\oeb - 1)\ovopp)
\]
Since \(\odi \leq \odhone\) when
\(1 \leq \ni \leq \ovopp = \oeb \ovopp + \ovbopp - 1\), we can conclude that:
\[
\underset{\ni \leq \ovopp}{max}
\;\athree \ni +\atwo(\oDi - 1) = 
\athree(\oeb \ovopp + \ovbopp - 1) + \atwo (h(\ovbopp + (\oeb - 1)\ovopp) - 1)
\]
If \(\ni > \ovopp\) then:
\[
\ni \geq \ovopp + 1 = \ovopp + \ovbopp = \hop(\odhone -1)
\]
If \(\odhone = 2\) then \(\odi = \oddi = \odhone - 1 = 1\).\\

If \(\odhone > 2\) then as \(\odbi > \odhone\) we also get that \(\odi = \oddi = \odhone - 1\) since:
\[
\ni < 2 \ovopp < 2 \ovopp + \ovbopp = \hop(\odhone - 2) < \hop(\od)\;\forall\;\od\in[1, \odhone - 2)
\]
This implies that
when \(\ni > \ovopp\) we get:
\[
\oDi = h(\hop(\odhone - 1)) = h(\ovbopp + \ovopp) = h(\ovbopp + \oeb\ovopp)
\]
This gives us the same result as the case where \(\oeb\) equals zero.

\bullet \(\oeb \geq 1\) and \(\ovbopp > 1\) if \(\oeb = 1\):

From \cref{FF30} follows that:
\[
\oeb\ovopp + \ovbopp - 1 \leq \nb < (\oeb + 1) \ovopp + \ovbopp - 1
\]
Note that:
\begin{align*}
(\oeb - 1)\ovopp + \ovbopp - 1 
\begin{aligned}[t]
= & \ovbopp - 1 \geq 1 \text{ if }\oeb = 1\\
\geq & \ovopp > 1 \text{ if }\oeb \geq 2\\
\end{aligned}
\end{align*}
If \(\ni \leq (\oeb - 1)\ovopp + \ovbopp -1\) then \(\odi = \odbi = 1\) since:
\begin{align*}
\nb - \ni 
\geq & \oeb\ovopp + \ovbopp - 1 
-((\oeb - 1)\ovopp + \ovbopp -1)\\
= &\ovopp = \ovop - \ovbop = \hops(1)
\end{align*}
If \((\oeb - 1)\ovopp + \ovbopp \leq \ni < \oeb\ovopp + \ovbopp
\) then \(\oddi = \odhone + 1 - \oeb \leq \odhone\) since:
\begin{align*}
\ni \geq (\oeb - 1)\ovopp + \ovbopp = \hop(\odhone + 1 - \oeb)\\
\ni < \oeb\ovopp + \ovbopp = \hop(\odhone - \oeb) < \hop(\od)\;\forall\;\od\in[1, \odhone - \oeb)
\end{align*}
If \(\ni = \oeb\ovopp + \ovbopp - 1\text{ then }\odbi > \odhone\) since:
\begin{align*}
\nb - \ni 
< &(\oeb + 1)\ovopp +\ovbopp-1
- (\oeb\ovopp + \ovbopp - 1)
\\
= &\ovopp = \ovop - \ovbop = \hops(1) < \hop(\od)\;\forall\;\od\in(1, \odhone]
\end{align*}
This gives that \(\odi = \oddi = \odhone + 1 -\oeb\) when \(\ni = \oeb\ovopp + \ovbopp - 1\) which in turn implies that:
\[
\oDi = h(\hop(\odhone + 1 - \oeb)) = h(\ovbopp + (\oeb-1)\ovopp)
\]
Since \(\odhone + 1 - \oeb > 1\) according to \cref{FF31} we can conclude that:
\[
\underset{\ni \leq \oeb\ovopp + \ovbopp -1}{max}
\;\athree \ni +\atwo(\oDi - 1) = 
\athree(\oeb \ovopp + \ovbopp - 1) + \atwo (h(\ovbopp + (\oeb - 1)\ovopp) - 1)
\]
If \(\nb = \oeb\ovopp + \ovbopp - 1\) we get \(g(A)\) from this max. If \(\nb\) is greater than this then for \(\ni\) greater than this we get:
\[
\ni \geq \oeb\ovopp + \ovbopp =  \hop(\odhone - \oeb)
\]
If \(\oeb = \odhone - 1\) then \(\odi = \oddi = \odhone - \oeb = 1\). Else we get the same result since:
\[
\ni \leq \nb < (\oeb + 1)\ovopp + \ovbopp = \hop(\odhone - 1 - \oeb) < \hop(\od)\;\forall\;\od\in [1, \odhone - 1 - \oeb)
\]
This implies that for \(\ni \geq \oeb\ovopp + \ovbopp\) we get:
\[
\oDi = h(\hop(\odhone - \oeb)) = h(\ovbopp + \oeb\ovopp)
\]
This theorem for this specific case is then proven in a manner equivalent to the proof for the cases above.
\end{proveit}
\section{Examples}\label{EX}
In this section, we show how to compute the Frobenius number by means of the algorithm presented here for various examples. (The entire algorithm is presented in the introduction.) However, no example is provided for the cases where:\\
\(\athree + \atwo\azero < \aone\atwo\), \(\oab > \oo\), \(\nb < \ovop -1\) and \(2\ovbop > \ovop\)

The reason for this is that the author has not been able to find any set \(A\) corresponding to this case.
\begin{exam}\label{EX1}
\(\athree + \atwo\azero > \aone\atwo\)\\
\begin{align*}
\text{Given } A = \{9, 11, 20\}\\
\azero = \mod{-\amtwo\athree}{\aone} = 8\\
\athree + \atwo\azero = 108 > \aone\atwo = 99\\
g(A) = \aone\atwo - \atwo -\aone = 79
\end{align*}
\end{exam}
\begin{exam}\label{EX2}
\(\athree + \atwo\azero < \aone\atwo\), \(\oab < \oo\) and \(2\azero < \aone\)\\
\begin{align*}
\text{Given }A = \{ 53, 55, 82\}\\
\azero = \mod{-\amtwo\athree}{\aone} = 12\\
\athree + \atwo\azero = 742 < \aone\atwo = 2915\\
2\azero = 24 < \aone = 53\\
\oa = \azero = 12\\
\ob = \atwo\oa + \athree = 742\\
\oab = \mod{\aone}{\oa} = 5 < \oo = \frac{\aone\athree}{\ob} = 5.857\\
\nb= \ceilfrac{\aone\atwo}{\ob} - 1 = 3\\
g(A) = max(
\begin{aligned}[t]
&\athree\nb + \atwo(\azero - 1),\\
&\atwo(\aone - \nb\azero -1)) - \aone
\end{aligned}\\
= max(851, 880) - 53 = 827
\end{align*}
\end{exam}
When \(A = \{5, 7, 8\}\) then \(g(A)\) is given by \(\athree\nb + \atwo(\azero - 1) - \aone\).
\begin{exam}\label{EX3}
\(\athree + \atwo\azero < \aone\atwo\), \(\oab < \oo\) and \(2\azero > \aone\)\\
\begin{align*}
\text{Given } A = \{19, 23, 28\}\\
\azero = \mod{-\amtwo\athree}{\aone} = 12\\
\athree + \atwo\azero = 304 < \aone\atwo = 437\\
2\azero = 24 > \aone = 19\\
\oa = \aone - \azero = 7\\
\ob = \atwo\oa - \athree = 133\\
\oab = \oa - \mod{\aone}{\oa} = 2 < \oo = \frac{\aone\athree}{\ob} = 4\\
\nb= \floorfrac{\aone}{\oa} = 2\\
g(A) = max(
\begin{aligned}[t]
&\athree\nb + \atwo(\mod{\aone}{\oa} - 1),\\
&\athree(\nb - 1) + \atwo(\oa - 1)) - \aone
\end{aligned}\\
= max(148, 166) - 19 = 147
\end{align*}
\end{exam}
When \(A = \{5, 6, 7\}\) then \(g(A)\) is given by \(\athree\nb + \atwo(\mod{\aone}{\oa} - 1) - \aone\).

\begin{exam}\label{EX4}
\(\athree + \atwo\azero < \aone\atwo\), \(\oab > \oo\), \(2\oabosm > \oaosm\) and \(\nb = \ovop -1\)\\
\begin{align*}
\text{Given } A = \{74, 79, 81 \}\\
\azero = \mod{-\amtwo\athree}{\aone} = 43\\
\athree + \atwo\azero = 3478 < \aone\atwo = 5846\\
2\azero = 86 > \aone = 74\\
\oa = \aone - \azero = 31\\
\ob = \atwo\oa - \athree = 2368\\
\oab = \oa - \mod{\aone}{\oa} = 19 > \oo = \frac{\aone\athree}{\ob} = 2.531\\
\end{align*}
\begin{table}[H]
\centering
\begin{tabular}{ |l|l|l|l|} 
\hline 
j & \(\oabj\) & \(\oaj\) & \(\ooj\)\\ 
\hline
1 & 19 & 31 & 2.531\\
\hline
2 & 7 & 12 & 5.4\\
\hline
3 \((\os)\) & 2 & 5 & 6.231\\
\hline
\end{tabular}
\end{table}
\begin{align*}
2\oabosm = 14 > \oaosm = 12\\
\nb=\hms \left(\oaosm-\oaos\ceilfrac{\oaosm}{\oaos+\ooosm}+\azero\right) = 11\\
\end{align*}
\begin{table}[H]
\centering
\begin{tabular}{ |l|l|l|} 
\hline 
j & \(\ovbj\) & \(\ovj\)\\ 
\hline
1 & 31 & 74\\
\hline
2 & 19 & 31\\
\hline
3 \((\op)\) & 7 & 12\\
\hline
4 & 2 & 5\\
\hline
\end{tabular}
\end{table}
\begin{align*}
\nb = & 11 = \ovop -1\\
g(A) = & max(
\begin{aligned}[t]
&\athree \nb + \atwo(h(\ovbop) - 1),\\
&\athree(\ovbop -1) + \atwo (\aone - h(\ovop - \ovbop) - 1))
- \aone
\end{aligned}\\
= & max(1207, 960) - 74 = 1133
\end{align*}
\end{exam}
When \(A = \{50, 59, 61\}\) then \(g(A)\) is given by
\[\athree(\ovbop -1) + \atwo (\aone - h(\ovop - \ovbop) - 1) - \aone\]
\begin{exam}\label{EX5}\ \\
\(\athree + \atwo\azero < \aone\atwo\), \(\oab > \oo\), \(2\oabosm < \oaosm\), \(\nb < \ovop -1\) and \(2\ovbop < \ovop\)\\
\begin{align*}
\text{Given } A = \{77, 82, 83\}\\
\azero = \mod{-\amtwo\athree}{\aone} = 45\\
\athree + \atwo\azero = 3773 < \aone\atwo = 6314\\
2\azero = 90 > \aone = 77\\
\oa = \aone - \azero = 32\\
\ob = \atwo\oa - \athree = 2541\\
\oab = \oa - \mod{\aone}{\oa} = 19 > \oo = \frac{\aone\athree}{\ob} = 2.515\\
\end{align*}
\begin{table}[H]
\centering
\begin{tabular}{ |l|l|l|l|} 
\hline 
j & \(\oabj\) & \(\oaj\) & \(\ooj\)\\ 
\hline
1 & 19 & 32 & 2.515\\
\hline
2 & 6 & 13 & 5.188\\
\hline
3 \((\os)\) & 5 & 6 & 83.0\\
\hline
\end{tabular}
\end{table}
\begin{align*}
2\oabosm = 12 < \oaosm = 13\\
\nb = \hms(\oabos + \azero) = 16
\end{align*}
\begin{table}[H]
\centering
\begin{tabular}{ |l|l|l|} 
\hline 
j & \(\ovbj\) & \(\ovj\)\\ 
\hline
1 \((\op)\) & 12 & 77\\
\hline
2 & 7 & 12\\
\hline
\end{tabular}
\end{table}
\begin{align*}
\nb = & 16 < \ovop -1 = 76\\ 2\ovbop =  & 24 < \ovop = 77\\
\oed = & \floorfrac{\nb - \ovopp + 1 + \ovbopp}{\ovopp} = 1\\
g(A) =& max(
\begin{aligned}[t]
&\athree \nb + \atwo(h(\ovbop) - 1),\\
&\athree(\ovbop -1) + \atwo (\aone - h(\oed \ovbop - \ovbopp) - 1)) - \aone\\
\end{aligned}\\
= & max(1328, 1323) - 77 = 1251
\end{align*}
When \(A = \{27, 29, 32\}\) then \(g(A)\) is given by:
\[\athree(\ovbop -1) + \atwo (\aone - h(\oed \ovbop - \ovbopp) - 1) - \aone
\]
\end{exam}

\section{Time complexity analysis}\label{TC}
In this section, we will show that the worst case time complexity of the algorithm presented here is \(O(\log \atwo)\) by comparing its time complexity with the time complexity of the Euclidean algorithm with positive remainders. (From now on we will simply write time complexity and let it be implied that it refers to the worst case.) The Euclidean algorithm finds the greatest common divisor of two integers. If these integers are \(a\) and \(b\), then according to Donald E. Knuth \cite{Knuth1981}, the time complexity of the Euclidean algorithm is \(O(\log(min(a, b)))\).

The Frobenius number of a set \(A\), consisting of three arbitrary positive integers, can with the help of Johnson's theorem (see \cref{IN3}) be computed by means of a closed form formula from the Frobenius number of a set \(A'\), consisting of pairwise coprimes. Computing closed form formulas generally takes constant time. The algorithm presented here assumes that we are starting with a set such as \(A'\). To get this set from the arbitrary set \(A\), we need to find the greatest common divisor of each pair in \(A\), i.e. the time complexity of creating \(A'\) from \(A\) is \(O(log(\atwo))\) assuming that \(\atwo\) is the second largest integer in this set.

When we have transformed the arbitrary set \(A\) into a set such as \(A'\), \(\azero\) (see \cref{IN4}) is computed based on this set. Therefore, we must
compute the multiplicative inverse of \(\atwo\) modulo \(\aone\). This can be done by means of the extended Euclidean algorithm. The time complexity for finding the multiplicative inverse of an integer \(a\) modulo \(b\) by means of this algorithm is the same as finding the greatest common divisor of these integers by means of the ordinary Euclidean algorithm, i.e. the time complexity for computing \(\azero\) is \(O(log(\aone))\). In some cases we also have to compute \(\azerom\), which takes less time since \(\azero\) is less than \(\aone\).

In some cases, we then have to compute the diff-mod sequences starting with \((\oab, \oa)\) and \((\azerom, \aone)\) respectively. In \cref{ARM} we show that the time complexity for computing a diff-mod sequence starting with the pair \((a, b)\) is \(O(log(a))\). Since \(\oab\) and \(\azerom\) are both less than \(\aone\), the time complexity for computing the diff-mod sequences we need, is less than \(O(log(\aone))\). Once these diff-mod sequences have been computed, the Frobenius number is given by means of closed form formulas based on specific pairs in these sequences.

To sum up this complexity analysis, we can conclude that the algorithm presented here consists of a constant number of steps, which time complexity is \(O(\log \atwo)\) or less, implying that the overall time complexity is that as well. Greenberg's algorithm \cite{Greenberg1988} is the fastest algorithm known today for solving the three variable case of Frobenius problem and its worst case time complexity is \(O(\log \aone)\) which one can argue is faster. However, the time complexity of both algorithms is a logarithmic function of \(A\).
\bibliography{document}

\end{document}